%% file: paper_sparse_curves.tex
\documentclass{article}
\usepackage[utf8]{inputenc}
\usepackage{arxiv}
\usepackage{authblk}

\usepackage[british]{babel}
\usepackage{amsmath}
\usepackage{amssymb}
\usepackage{amsthm}
\usepackage{dsfont}
\usepackage{hyperref}
\usepackage{float}
\usepackage{mathtools}
\usepackage[ruled, vlined]{algorithm2e}

\newtheorem{theorem}{Theorem}[section]
\newtheorem{defi}[theorem]{Definition}
\newtheorem{lemma}[theorem]{Lemma}
\newtheorem{cor}[theorem]{Corollary}

\newtheorem{remark}[theorem]{Remark}

\newenvironment{proof2}{}{\hfill$\square$}
\newcommand{\R}{\mathbb{R}}
\newcommand{\N}{\mathbb{N}}
\newcommand{\1}{\mathds{1}}

\DeclareMathOperator*{\argmax}{argmax}
\DeclareMathOperator*{\arginf}{arginf}

\title{Elastic analysis of irregularly or sparsely sampled curves}
\date{\today}

\author[*]{Lisa Steyer \thanks{Corresponding Author: Lisa.Steyer@hu-berlin.de}}
\author[*]{Almond St\"ocker}
\author[*]{Sonja Greven}
\affil[*]{Humboldt-Universit\"at zu Berlin,
School of Business and Economics,
Chair of Statistics}
\begin{document}
\maketitle
\begin{abstract}
We provide statistical analysis methods for samples of curves when the image but not the parametrisation of the curves is of interest. A parametrisation invariant analysis can be based on the elastic distance of the curves modulo warping, but existing methods have limitations in common realistic settings where curves are irregularly and potentially sparsely observed.
We provide methods and algorithms to approximate the elastic distance for such curves via interpreting them as polygons. Moreover, we propose to use spline curves for modelling smooth or polygonal Fr\'echet means of open or closed curves with respect to the elastic distance and show identifiability of the spline model modulo warping. We illustrate the use of our methods for elastic mean and distance computation  by application to two datasets. The first application clusters sparsely sampled GPS tracks based on the elastic distance and computes smooth means for each cluster to find new paths on Tempelhof field in Berlin. The second classifies irregularly sampled handwritten spirals of Parkinson's patients and controls based on the elastic distance to a mean spiral curve computed using our approach. All developed methods are implemented in the \texttt{R}-package \texttt{elasdics} and evaluated in simulations.
\end{abstract}


\input{0_introduction}
\input{1_method}
\input{2_simulation}
\input{3_application}
\input{4_discussion}

\section*{Acknowledgement}
The authors gratefully acknowledge funding by grant GR 3793/3-1 from the German research foundation (DFG). We thank the members of the Chair of Statistics who contributed to data collection on Tempelhof field, and Manuel Pfeuffer for alerting us to the Parkison's data.

\newpage
\bibliographystyle{plain}
\bibliography{paper_sparse_curves}

\newpage
\appendix
\renewcommand{\thesubsection}{\Alph{subsection}}
\renewcommand{\thesection}{\thesubsection}
\section*{Appendices}
\input{proofs_computations}
\newpage
\input{supplem_plots}

\end{document}

%% file: 0_introduction.tex
\section{Introduction}
\label{sec:intro}
\begin{figure}[ht]
\centering
\includegraphics[scale=0.6]{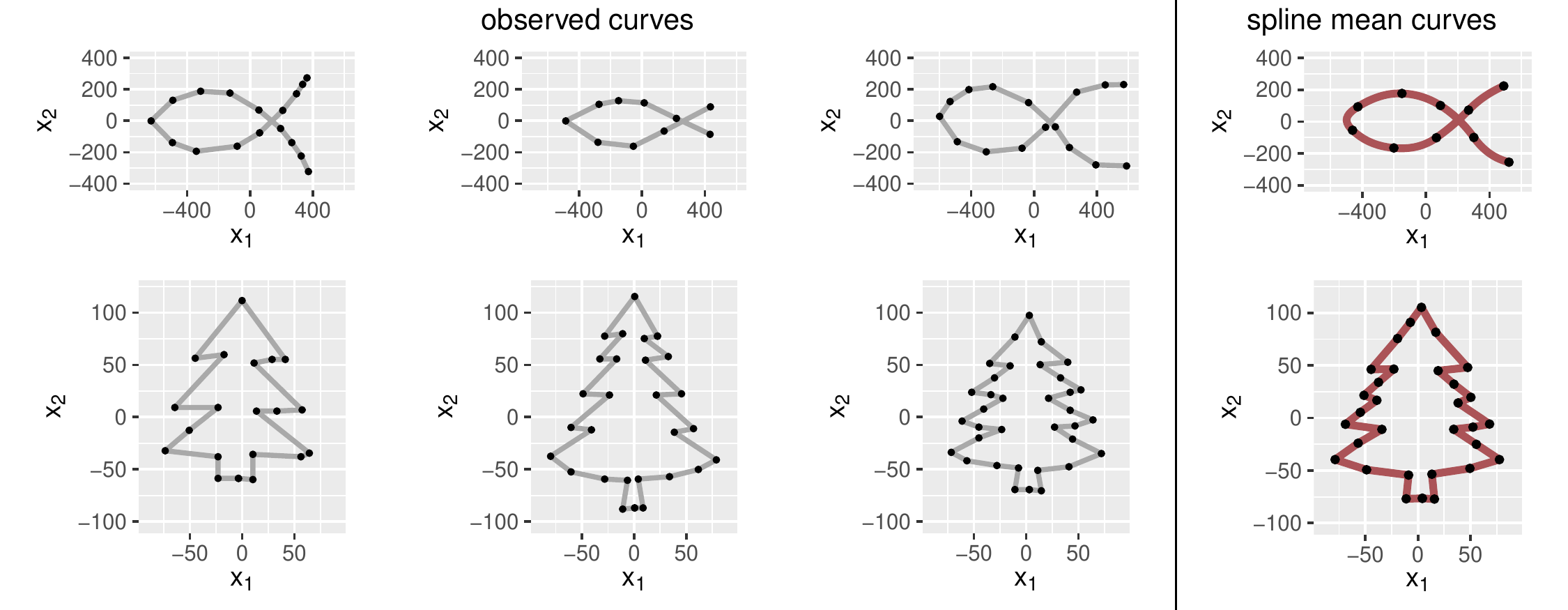}
\caption{\label{fig:sampled_curves} Two toy examples of sparsely and irregularly observed curves in $\mathbb{R}^2$ with observed points indicated as black dots (first three columns). The corresponding smooth or polygonal spline means (last column in red) are computed using our algorithms, with black dots indicating values at the model based spline knots.}
\end{figure}

Elastic analysis of curves 
$\boldsymbol{\beta}: [0,1] \to \mathbb{R}^d$, $d \in \mathbb{N}$, refers to an analysis of the curves' image without taking their parametrisation over the interval $[0, 1]$ into account. Examples for such curves in $\mathbb{R}^2$ are handwritten letters or the outline of an object. Here only the image of the curve represents the object, not the speed with which the parametrisation traverses the outline. Hence, for statistical analysis of such curves including mean computation, clustering or classification, the analysis should be invariant under different possible parametrisations.  
Ideally, the analysis should also yield an optimal alignment of different curves to allow comparison of corresponding points such as bumps and other features.
Consider for instance the handwritten symbols in Fig. \ref{fig:sampled_curves}. Here we would like the mouths of the different fish and the branches of the trees to correspond after alignment. As in this example, curves are often observed at a differing number of discrete points.
The aim of this paper is to extend elastic statistical methodology to such realistic cases where curves are irregularly and sparsely sampled. In particular, this includes suitable algorithms for alignment and distance computation for samples of such curves, as well as identifying appropriate spline model spaces for elastic 
(Fr\'echet) mean curves.
These means can be smooth curves, such as shown for the fish in Fig. \ref{fig:sampled_curves}, or polygonal curves, better suited for curves with sharp corners like the trees in Fig. \ref{fig:sampled_curves}.
To this end, we, i.a., derive a useful simplification of the warping problem when interpreting the observed curves as polygons, and show that certain first and second order splines meet the identifiability properties required 
in a modulo warping context.


The alignment problem for curves in $\mathbb{R}^d$ is closely related to the registration problem in functional data analysis (Ramsay and Silverman\cite{ramsay_silverman}), which is the alignment problem in the case $d = 1$.
For two functions $f_1$ and $f_2$, registration (also called warping) has commonly been treated as an optimisation problem $\inf_{\gamma \in \Gamma} \|f_1 - f_2 \circ \gamma\|_{L_2}$ on a suitable function space $\Gamma$ of warping functions $\gamma$. This choice is problematic as the mapping $(f_1, f_2) \mapsto \inf_{\gamma \in \Gamma} \|f_1 - f_2 \circ \gamma\|_{L_2}$ does not define a proper distance on the space of curves modulo parametrisation. 
This yields two major pitfalls: First, the mapping is not symmetric, which means aligning $f_2$ to $f_1$ will not be equivalent to aligning $f_1$ to $f_2$. Furthermore, $\inf_{\gamma \in \Gamma} \|f_1 - f_2 \circ \gamma\|_{L_2}$ can be zero even if  $f_2$ is not a warped version of $f_1$, which is related to the so-called 'pinching' problem (Marron et al.\cite{marron}). Intuitively speaking, this 'pushes' the integration mass to parts of the domain where $f_1$ and $f_2$ are close. To avoid this 'pinching' effect, a regularisation term can be added to the loss function (Ramsay and Silverman\cite{ramsay_silverman}). This is done in various dynamic time warping algorithms, where usually large values of the derivative of the warping function are penalised (Sakoe and Chiba\cite{sakoe}, Keogh and Ratanamahatana\cite{keogh}). Alternatively, one can choose a small number of basis functions for the warping or combine both approaches to use penalised basis functions (Ramsay and Li\cite{ramsay_li}).  Moreover, Bayesian approaches to modelling warping functions have been suggested (Cheng et al.\cite{cheng}, Lu et al.\cite{lu}). Recently, Matuk et al.\cite{matuk} developed such a Bayesian approach for sparse one-dimensional functions.

All of these approaches restrict the amount of warping, thus the analysis is not completely independent of the observed parametrisation. This seems more suitable for one-dimensional functions $f_1, f_2: [0,1] \to \mathbb{R}$ where one seeks to separate phase (parametrisation) and amplitude (image) but considers both as informative. 
If we analyse curves $\boldsymbol{\beta}: [0,1] \to \mathbb{R}^d$ with $d > 1$ however, we are usually only interested in the image representing the curve, which makes penalised, restricted or Bayesian approaches for the warping less suitable. In this case the object of interest is the equivalence class of the curve with respect to (w.r.t.) parametrisation, hence a proper distance on the resulting quotient space modulo warping is desirable.

To overcome the shortcomings of the usual $L_2$ distance for curve alignment, Srivastava et al.\cite{srivastava} propose an elastic distance which instead minimises the Fisher-Rao Riemannian metric. This gives a proper metric on the quotient space of absolutely continuous curves modulo parametrisation and translation. For more details on this  square-root-velocity (SRV) framework see Srivastava and Klassen\cite{srivastava_book}. They show that for two absolutely continuous curves $\boldsymbol{\beta}_1$ and $\boldsymbol{\beta}_2$, the Fisher-Rao metric can be simplified to the $L_2$-distance between the corresponding square-root-velocity (SRV) curves, which can be minimised to obtain the elastic distance.

\begin{defi}[Elastic distance and SRV transformation (Srivastava et al.\cite{srivastava})]
Let $\boldsymbol{\beta}_1, \boldsymbol{\beta}_2: [0,1] \to \mathbb{R}^d$ be absolutely continuous and $[\boldsymbol{\beta}_1]$ and $[\boldsymbol{\beta}_2]$ their respective equivalence classes modulo parametrisation and translation. Then the elastic distance between $[\boldsymbol{\beta}_1]$ and $[\boldsymbol{\beta}_2]$ is 
\begin{align} \label{def:elastic_dist}
\displaystyle d([\boldsymbol{\beta}_1], [\boldsymbol{\beta}_2]) = \inf_{\gamma_1, \gamma_2} \| (\mathbf{q}_1 \circ \gamma_1) \cdot \sqrt{\dot{\gamma_1}} - (\mathbf{q}_2 \circ \gamma_2) \cdot \sqrt{\dot{\gamma_2}} \|_{L_2},
\end{align}
with monotonically increasing, onto and differentiable warping functions $\gamma_1, \gamma_2: [0,1] \to [0,1]$ and SRV transformations $\mathbf{q}_1$ and $\mathbf{q}_2$ of $\boldsymbol{\beta}_1$ and $\boldsymbol{\beta}_2$ defined via
\begin{align} \label{def:srv_trafo}
\mathbf{q}_i(t) = \begin{cases}
\frac{\dot{\boldsymbol\beta}_i(t)}{\sqrt{\|\dot{\boldsymbol\beta}_i(t)\|}} & \text{if } \dot{\boldsymbol\beta}_i(t) \neq 0 \\
0 & \text{if } \dot{\boldsymbol\beta}_i(t) = 0
\end{cases}
\quad \text{for }i = 1,2.
\end{align}
Here, $(\mathbf{q}_i \circ \gamma_i) \cdot \sqrt{\dot{\gamma}_i}$ is the SRV transformation of the re-parametrised curve $\boldsymbol{\beta_i} \circ \gamma_i$, $i = 1,2$.
\end{defi}

Srivastava and Klassen\cite{srivastava_book} observe the following properties of the SRV transformation and the elastic distance.
\begin{remark}
\phantom{}
\begin{itemize}
\item[i)] To obtain a proper quotient space structure on the space of absolutely continuous curves, we need to consider the closure of SRV curves with respect to re-parametrisation as equivalence classes. That is for a curve $\boldsymbol{\beta}$ with SRV transformation $\mathbf{q}$, $[\boldsymbol{\beta}]$ consists of all curves whose SRV transformation is in the closure of $\{ (\mathbf{q}_i \circ \gamma) \cdot \sqrt{\dot{\gamma}}| \gamma \in \Gamma \}$, with $\Gamma$ being the set of monotonically increasing, onto and differentiable warping functions.
\item[ii)] In \eqref{def:elastic_dist}, it is in fact sufficient to align one of the curves, that is
\begin{align} \label{eq:warp_prob}
\displaystyle d([\boldsymbol{\beta}_1], [\boldsymbol{\beta}_2]) = \inf_{\gamma \in \Gamma} \| \mathbf{q}_1 - (\mathbf{q}_2 \circ \gamma) \cdot \sqrt{\dot{\gamma}} \|_{L_2},
\end{align}
with $\Gamma$ being the set of monotonically increasing, onto and differentiable warping functions $\gamma: [0,1] \to [0,1]$.
\item[iii)] Every square integrable SRV curve $\mathbf{q}$ uniquely defines an absolutely continuous curve $\boldsymbol{\beta}$ up to translation, with the back-transform given as $\boldsymbol{\beta}(t) = \boldsymbol{\beta}(0) + \int_0^t \mathbf{q}(s) \| \mathbf{q}(s) \| \ ds$.
\end{itemize}
\end{remark}

Note that any statistical analysis based on this elastic distance will be modulo translation as a result of taking derivatives. If the actual position of the curve in space is of interest as well, it has to be included separately in the analysis. On the other hand, if curves are used to model shape objects, translation invariance is a desired property. As in classical shape data analysis  (Dryden and Mardia\cite{dryden}), the analysis should then additionally be independent of the size and the orientation of the shape in space. In this paper, we solely discuss the invariance under re-parametrisation but not the invariance under rotation and scaling and give examples of GPS tracks and handwritten spirals where this elastic analysis 
is suitable. However, re-parametrisation invariance presents a key aspect of functional shape analysis (Srivastava and Klassen\cite{srivastava_book}) and may, therefore, also be viewed in this context.

A solution to the variational problem in the distance (\ref{eq:warp_prob}) is usually approximated using a dynamic programming algorithm or gradient-based optimisation (for instance in Srivastava et al.\cite{srivastava}). Both approaches discretise the warping space $\Gamma$. The dynamic programming algorithm for instance assumes a discrete grid for the domain of the warping function. An extension by Bernal et al.\cite{bernal} allows for an unequal number of points on both curves and improves computation time. Lahiri et al.\cite{lahiri} provide an algorithm to align two piecewise linear curves and show that an optimal warping exists if at least one of the curves is piecewise linear. Such an optimal warping also exists if both curves are continuously differentiable (Bruveris\cite{bruveris}).

A direct application of computing pairwise elastic distances for a sample of observed curves is their use for distance-based statistical analysis like various clustering or classification algorithms (Kurtek et al.\cite{kurtek}, Laborde et al.\cite{laborde}, Strait and Kurtek\cite{strait}).A more challenging but important task is to compute the mean of a random sample of such objects. Srivastava et al.\cite{srivastava_functional_data} suggest to approximate the Fréchet mean (which they call Karcher mean) for curves w.r.t. the elastic distance via alternating between optimising the alignment to the current mean and computing the $L_2$ mean of the SRV curves given the current alignment. Their perspective is focused on the curves as functions and, in practice, they rely on evaluating the SRV curves on a regular grid for the mean computation, which works well in the case of densely observed curves. Nevertheless, in real-world applications, we observe curves only at a finite (and often small) number of discrete points, where even the number of points might differ between curves (so-called sparse and irregular setting). 
This is the case in our example of handwritten symbols displayed in Fig. \ref{fig:sampled_curves}. Here the number of points differs randomly across curves (for the fish in the top row) or with the number of prominent features (for the Christmas trees in the bottom row). We show in examples that (elastic) methods designed for densely observed curves have limitations for such sparse settings. This problem is well-known in functional data analysis ($d=1$), where spline representations or some other smoothing method are frequently used to model sparsely and/or irregularly observed functions (e.g.\ Yao et al.\cite{yao}, Greven and Scheipl\cite{greven}).

The main contribution of this paper is to carefully introduce spline functions for modelling elastic curves in $\mathbb{R}^d$ on SRV level, extending approaches for functional data also to $d \geq 2$ and to the elastic setting. This includes piecewise constant SRV curves, which corresponds to polygonal curves, as a special case. The main advances of this work are the following points:
\begin{itemize}
\item We provide algorithms to fit elastic spline means for open and closed curves, show the proposed spline curves are identifiable via their coefficients modulo parametrisation and  discuss limitations of this identifiability.
\item We develop algorithms to align open and closed curves if at least one of them is piecewise linear, for instance a sparsely observed curve which is treated as a polygon. In the special case of open curves with both curves piecewise linear, Lahiri et al.\cite{lahiri} proposed an alternative algorithm. Our algorithm is, however, far simpler to understand and implement. We show local maximization properties of our algorithm.
\item We demonstrate how the elastic distance can be used for statistical analysis of irregularly or sparsely observed curves in two examples, involving mean computation, clustering and classification of curves. We provide an implementation of our methods in the \texttt{R}-package \texttt{elasdics}\cite{elasdics}.
\end{itemize}
Moreover, the proposed methodology for elastic spline mean estimation can be viewed as a first step towards an elastic regression analysis for sparsely observed curves when including covariates.

We structure our results as follows: In Section \ref{sec:methods}, we present algorithms to approximate the elastic distance in (\ref{eq:warp_prob}), introduce spline functions to compute a smooth representative of the Fréchet mean of observed curves and discuss identifiability properties of such elastic spline curves. In Section \ref{sec:simulation} we test the developed methods via simulation and compare our implementation to the one available in the \texttt{R} Package \texttt{fdasrvf}\cite{fdasrvf}. 
We demonstrate how the elastic distance can be used to cluster GPS tracks and compute smooth mean paths in Section \ref{sec:application}. A second example dataset comprises handwritten spirals of Parkinson's patients and a healthy control group, which we classify based on the elastic distance to the mean spiral curve computed using our method. Section \ref{sec:discussion} closes with a discussion.

%% file: 1_method.tex
\section{Elastic analysis of observed curves}
\label{sec:methods}
In practice, we observe curves in $\mathbb{R}^d$, $d \in \mathbb{N}$, not continuously but only discretely via evaluations of these curves on discrete (and potentially sparse and curve-specific) grids. An elastic analysis needs to explicitly address this point for distance and subsequently mean computation. 
We propose to treat a discretely observed curve $\boldsymbol{\beta}$ with SRV transformation $\mathbf{q}$ as a polygon parametrised with constant speed between the observed corners $\boldsymbol{\beta}(s_0), \dots, \boldsymbol{\beta}(s_m)$. In this case, the problem of finding an optimal re-parametrisation $\boldsymbol{\beta} \circ \gamma$ of $\boldsymbol{\beta}$ to another curve with SRV transformation $\mathbf{p}$ can be simplified (similar as in Lahiri et al.\cite{lahiri}). We can show that instead of solving the minimisation problem (\ref{eq:warp_prob}) over the function space of all suitable warping functions $\gamma$, we only need to solve a maximisation problem over a subset of $\mathbb{R}^{m-1}$ w.r.t. the new parametrisations $t_1 = \gamma^{-1}(s_1), \dots, t_{m - 1} = \gamma^{-1}(s_{m - 1})$ at the corners of the observed polygon. 

\begin{lemma} \label{lem:warp_prob_discrete}
Let $\boldsymbol{\beta}$ be a polygon in $\mathbb{R}^d$ with constant speed parametrisation between its corners $\boldsymbol{\beta}(s_0), \dots, \boldsymbol{\beta}(s_m)$. That means its SRV transformation $\mathbf{q}$ is piecewise constant with $\mathbf{q}|_{[s_j, s_{j+1}]} = \mathbf{q}_j \in \mathbb{R}^d$ for all $0 = 1, \dots, m - 1$.
Moreover, let $\boldsymbol{\tilde{\beta}}$ be an absolutely continuous curve with SRV transformation $\mathbf{p}$, $\|\mathbf{p}\|_\infty < \infty$. Then calculating the optimal $\gamma$ in \eqref{eq:warp_prob} to obtain the elastic distance $d([\boldsymbol{\beta}], [\boldsymbol{\tilde{\beta}}])$ is equivalent to the following problem.
\begin{align} \label{eq:warp_prob_discrete}
\textbf{Maximise } &\quad \Phi(\mathbf{t}) = \Phi(t_1, \dots, t_{m-1}) = \sum_{j = 0}^{m-1} \sqrt{ (s_{j+1} - s_j) \int_{t_j}^{t_{j+1}} \langle \mathbf{p}(t), \mathbf{q}_j \rangle^2_+ \ dt} \\
\textbf{w.r.t } &\quad 0 = t_0 \leq t_1 \leq \dots \leq  t_m = 1, \nonumber
\end{align}
where $\langle \cdot, \cdot \rangle_+$ denotes the positive part of the $d$-dimensional scalar product. For a maximiser $(t_1, \dots, t_{m-1})$ of \eqref{eq:warp_prob_discrete} there is a $\gamma: [0,1] \to [0,1]$ with $\gamma(t_j) = s_j$ for all $j = 1, \dots, m-1$ which is a minimiser of \eqref{eq:warp_prob}.
\end{lemma}

A proof can be found in Appendix \ref{app:warp_prob_discrete}. It includes an explicit construction of the minimising warping function $\gamma$ (or a minimising sequence of warping functions). Although we formulated the warping problem in \eqref{eq:warp_prob} only for open curves (or closed curves with known start and end point) we can formulate a similar criterion for closed curves, using a different set of warping functions. Here we assume $\gamma: [0,1] \to [0,1]$ such that there exists $t_0 \in [0,1]$ with
\begin{align*}
\gamma(t_0) = 0, \quad
\lim_{t \nearrow 1}\gamma(t) = \gamma(0), \quad
\lim_{t \nearrow t_0} \gamma(t) = 1,
\end{align*}
and $\gamma$ monotonically increasing and differentiable on $[0, t_0[$ and on $[t_0, 1]$. This allow us to obtain a similar result as in Lemma \ref{lem:warp_prob_discrete} for closed curves.

\begin{cor}[Optimisation problem for closed curves]
Let $\mathbf{p}$ and $\mathbf{q}$ be as in Lemma \ref{lem:warp_prob_discrete} and additionally let them be the SRV transformations of closed curves. Let $\mathbf{p}^*$ be the periodic extension of $\mathbf{p}$ to the whole real line, that is $\mathbf{p}^*(t) = \mathbf{p}(t - \lfloor t \rfloor)$ for all $t \in \mathbb{R}$. Then the optimisation problem for closed curves is equivalent to the following problem.
\begin{align} \label{eq:warp_prob_discrete_closed}
\textbf{Maximise } &\quad \Phi^*(\mathbf{t}) = \Phi^*(t_0, t_1, \dots, t_{m-1}) = \sum_{j = 0}^{m-1} \sqrt{ (s_{j+1} - s_j) \int_{t_j}^{t_{j+1}} \langle \mathbf{p^*}(t), \mathbf{q}_j \rangle^2_+ \ dt} \\
\textbf{w.r.t } &\quad t_0 \leq t_1 \leq \dots \leq  t_m = t_0 + 1. \nonumber
\end{align}
For a maximiser $(t_0, t_1, \dots, t_{m-1})$ of \eqref{eq:warp_prob_discrete_closed} there is a $\gamma: [0,1] \to [0,1]$ with $\gamma(t_j - \lfloor t_j \rfloor) = s_j$ for all $j = 0, \dots, m-1$ which is a minimiser of the corresponding warping problem for closed curves.
\end{cor}

Thus, the warping problem for open or closed curves can be simplified if one of the SRV curves is piecewise constant, no matter which form the second SRV curve $\mathbf{p}$ has.
If $\mathbf{p}$ is at least continuous, for example the SRV curve of a model-based smooth mean curve, the loss functions in \eqref{eq:warp_prob_discrete} and \eqref{eq:warp_prob_discrete_closed} are differentiable. We propose to tackle the remaining maximisation problem with a gradient descent algorithm that can handle linear constrains (for instance method \texttt{'BFGS'} in \texttt{constrOptim} from \texttt{R}-package \texttt{stats}\cite{stats}) and provide a derivation of the gradient in Appendix \ref{app:gradient}. 

In the following, Subsection \ref{subsec:elastic_polygons} provides an algorithm to compute the elastic distance if the second curve is piecewise linear, for instance an observed polygon as well. In Subsections \ref{subsec:splines} and \ref{subsec:spline_mean} we introduce spline functions to model smooth or polygonal elastic mean curves and discuss identifiability of such spline curves modulo reparametrisation in Subsection \ref{subsec:identify}.

\subsection{Elastic distance for two piecewise linear curves}
\label{subsec:elastic_polygons}
We present an algorithm that can be used to find an optimal warping function, and therefore compute the elastic distance, in the case where both curves are piecewise linear. This is relevant either because we model one of the curves as a linear spline (see Subsection \ref{subsec:splines}), or because we want to compute the elastic distance between two observed curves. The latter allows us to perform any distance-based analysis of the data such as clustering or classification, for instance.

To obtain an optimal warping for a curve with piecewise constant SRV transformation $\mathbf{q}$ to another curve with SRV transformation $\mathbf{p}$, we first notice that the maximisation in one $t_j$ direction of the loss function given in \eqref{eq:warp_prob_discrete} only depends on the current values of $t_{j-1}$ and $t_{j+1}$ for any $\mathbf{p}$. Moreover, if $\mathbf{p}$ is a piecewise constant SRV curve as well, we can even derive a closed form solution of the maximisation problem in (\ref{eq:warp_prob_discrete}) with respect to each coordinate direction $t_j \in [t_{j -1}, t_{j +1}]$ (see Appendix \ref{app:closed_form_max}). Hence we propose a coordinate wise maximisation procedure, where we iterate two steps.

\begin{algorithm}[H]
\caption{Elastic distance for two open polygons \label{algo:dist_open_curves}}
\KwIn
{
 piecewise constant SRV curves $\mathbf{p}, \mathbf{q}$;
convergence tolerance $\epsilon > 0$\;
starting values $0 \leq t_1^{(0)} \leq \dots \leq t_{m-1}^{(0)} \leq 1$
\tcp*{e.g. relative arc length.}
}
\For{$k \in \mathbb{N}$}{
    \For{$j = 1, \dots, m-1$}{
        \If{$j - k$ even}{$t_j^{(k)} = 
        \argmax_{t_j \in \left[t_{j-1}^{(k-1)}, t_{j+1}^{(k-1)}\right]}
        \Phi\mid_{\{t_{j'} = t_{j'}^{(k - 1)}, j' \neq j \}}$}
        \ElseIf{$j - k$ odd}{$t_j^{(k)} = t_j^{(k-1)}$}}
    \If{$\|\mathbf{t}^{(k)} - \mathbf{t}^{(k - 2)}\| < \epsilon \wedge \|\mathbf{t}^{(k - 1)} - \mathbf{t}^{(k - 3)}\| < \epsilon$}{\Return $\mathbf{t}^{(k)} = (t_1^{(k)}, \dots, t_{m-1}^{(k)})$}}
\end{algorithm}

The warping problem for two (open) piecewise linear curves has been previously discussed by Lahiri et al.\cite{lahiri}. They propose a precise matching algorithm which produces a globally optimal re-parametrisation of $\mathbf{q}$. Our algorithm can be seen as an alternative, which is much more straightforward to implement (we provide an implementation in the \texttt{R}-package \texttt{elasdics}\cite{elasdics}) but does not guarantee to find a globally optimal solution. Nevertheless, we observe convincing results in simulations (Section \ref{sec:simulation}) and we can prove local maximisation in the following sense.

\begin{theorem} \label{theo:local_max}
Every accumulation point of the sequence   $(\mathbf{t}^{(k)})_{k \in \mathbb{N}} = (t_1^{(k)}, \dots, t_{m-1}^{(k)})_{k \in \mathbb{N}}$ resulting from Algorithm \ref{algo:dist_open_curves} is a local maximiser. 
\end{theorem}

To prove this theorem we first establish that the directional derivatives exist and are non-positive for all coordinate directions. Then we show that this carries over to all directional derivatives using local convexity of the loss function. More details can be found in Appendix \ref{app:local_max}. 

If the sequence has more than one accumulation point, all of them give the same loss $\Phi(\mathbf{t})$. This means they correspond to different re-parametrisations of the second curve, but give the same distance between the two curves. This can happen as the warping problem does not guarantee unique solutions (see the example given in Appendix \ref{app:non_unique}). In practise, one can pick any maximising $\mathbf{t} =  (t_1, \dots t_{m-1})$ to obtain a locally optimal warping function. As we cannot guarantee that this locally optimal $\mathbf{t}$ is also a global maximiser, we also propose to exploit varying starting points and therefore construct multiple sequences to find a global maximum. A further advantage of Algorithm \ref{algo:dist_open_curves} is that it can be easily adapted to closed curves, which has not been explicitly addressed by Lahiri et al.\cite{lahiri}. We adjust our algorithm for open polygons via appropriately updating $t_0$ and $t_m$. 

\begin{algorithm}[ht]
\caption{Elastic distance for two closed polygons \label{algo:dist_closed_curves}}
\KwIn
{
 piecewise constant SRV curves $\mathbf{p}, \mathbf{q}$;
convergence tolerance $\epsilon > 0$\;
starting values $0 \leq t_1^{(0)} \leq \dots \leq t_{m-1}^{(0)} \leq t_m^{(0)} = t_0^{(0)} + 1$
\tcp*{e.g. relative arc length.}
}
\For{$k \in \mathbb{N}$}{
\For{$j = 1, \dots, m-1$}{
    \If{$j - k$ even}{$t_j^{(k)} = 
    \argmax_{t_j \in \left[t_{j-1}^{(k-1)}, t_{j+1}^{(k-1)}\right]}
 	\Phi\mid_{\{t_{j'} = t_{j'}^{(k - 1)}, j' \neq j \}}$}
 	\ElseIf{$j - k$ odd}{$t_j^{(k)} = t_j^{(k-1)}$}
 	\If{$k$ even}{$t_0^{(k)} = \argmax_{t_0 \in [t_{m-1}^{(k)} - 1, t_{1}^{(k)}]}
 	\Phi^*|_{\{t_{j'} = t_{j'}^{(k)}, j' \neq 0 \}}$\;
 	$t_m^{(k)} = t_0^{(k)} + 1$
 	}
}
\If{$\|\mathbf{t}^{(k)} - \mathbf{t}^{(k - 2)}\| < \epsilon \wedge \|\mathbf{t}^{(k - 1)} - \mathbf{t}^{(k - 3)}\| < \epsilon$}{\Return $\mathbf{t}^{(k)} = (t_1^{(k)}, \dots, t_{m-1}^{(k)})$}
}
\end{algorithm}

Thus, we provide algorithms to compute the elastic distance between two (open or closed) piecewise linear and continuous curves. These curves form a subspace in the space of absolutely continuous curves and are called splines of degree 1. If we would like to model smooth (that is differentiable) curves, for example for a mean function, a spline space of higher degree might be more suitable.

\subsection{Modelling spline curves or spline SRV curves}
\label{subsec:splines}
As common in functional data analysis (Ramsay and Silverman\cite{ramsay_silverman}), we like to model curves or means for samples of curves as piecewise polynomial functions. This is in particular beneficial as we target our analysis at sparsely observed curves, which cannot be evaluated at arbitrary points.  Moreover, spline models impose parsimonious models for smooth curves, which can help to avoid overfitting the observed curves given limited information.

\begin{defi}[Spline curves]
We call $\boldsymbol{\xi} = (\xi_1, \dots, \xi_d)^T: [0,1] \to \R^d$ with $d \in \mathbb{N}$ a $d$-dimensional spline curve of degree $l \in \mathbb{N}_0$ if all its components $\xi_1, \dots, \xi_d: [0,1] \to \R$ are spline curves of degree $l$ with common knot set $0 = \kappa_0 < \kappa_1 < \dots < \kappa_{K - 1} < \kappa_K = 1$ for some $K \geq 2$. That means $\xi_1, \dots, \xi_d$ are piecewise polynomial of degree $l$ between the knots $\kappa_1, \dots, \kappa_K$, as well as continuous and $(l-1)$-times continuously differentiable on the whole domain $[0,1]$ for $l \geq 1$. Denote by $\mathcal{S}^l_{K; \kappa_0, \dots, \kappa_K}$ the set of all spline curves of degree $l$ with common knot set $0 = \kappa_0 < \kappa_1 < \dots < \kappa_{K - 1} < \kappa_K = 1$.
\end{defi}

\begin{figure}[ht]
\centering
\includegraphics[scale=0.65]{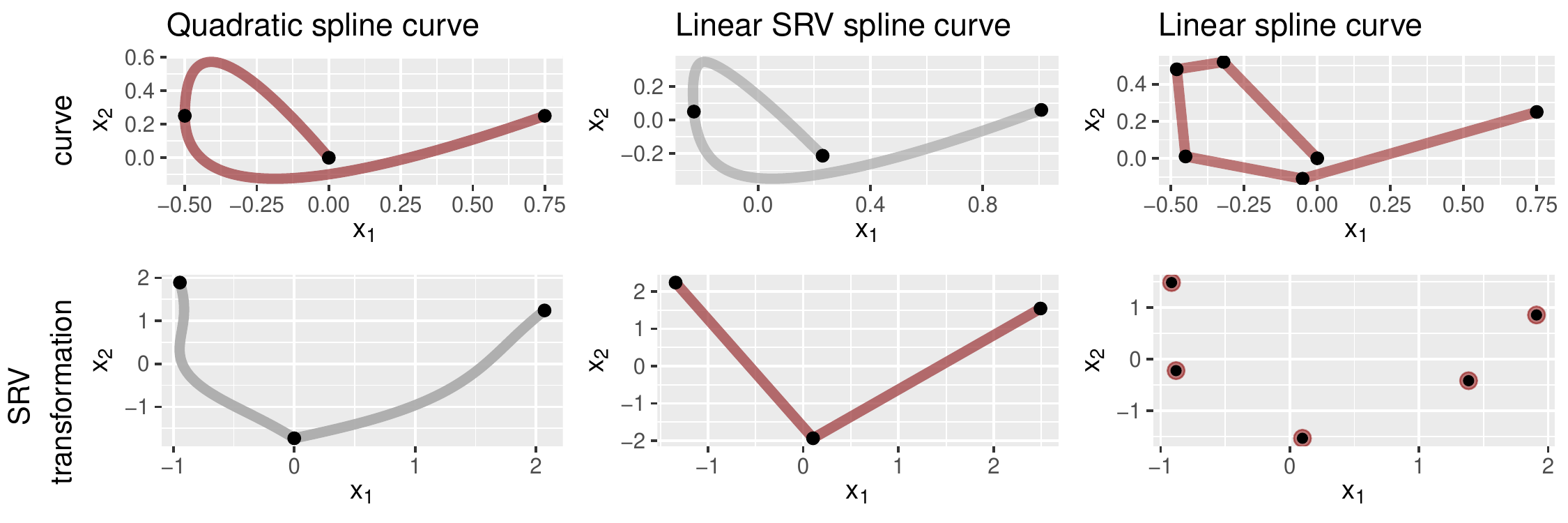}
\caption{\label{fig:spline_curves} Two-dimensional curves and corresponding SRV transformations. Spline curves are plotted as red curves with their values at knots marked as black dots; other curves are grey.  Note that the SRV curve in the bottom right panel is piecewise constant in $t$ and $t$ is not visible in the image.}
\end{figure}

We can either model the curve $\boldsymbol{\beta}$  as a $d$-dimensional spline curve, or its SRV transformation $\mathbf{p}$ (see Fig. \ref{fig:spline_curves}). If $\boldsymbol{\beta}$ is a spline of degree $l \geq 2$, the corresponding SRV curve $\mathbf{p}$ will not be a spline curve. The same holds true for the curve $\boldsymbol{\beta}$ if we model the SRV curve as a spline of degree $l \geq 1$. Only if $\boldsymbol{\beta}$ has degree $l = 1$ are both the piecewise linear curve itself and its piecewise constant SRV transformation are splines. However, if we use linear spline curves, we need a large number of knots to obtain curves that visually appear similarly smooth as if we use linear splines on SRV level and thus, we expect less parsimonious models.

To use these spline curves or spline SRV curves as model spaces for curves modulo parametrisation, we need to ensure model identifiability, that is that 
each equivalence class contains at most one spline curve. The unique spline representative then allows to identify and interpret the equivalence class of a curve modulo warping via its spline basis coefficients.
We will see in Subsection \ref{subsec:identify} that this is true for quadratic or cubic splines on curve level and for linear spline SRV curves (under mild conditions). Linear spline curves are identifiable under additional assumptions.

Therefore, we can use the space of cubic, quadratic or linear spline curves as a model space for smooth curves. However, 
using quadratic or cubic splines to model on the curve level would not imply a vector space structure on the SRV level, on which the distance is computed. We therefore propose to consider linear spline (and thus continuous) SRV curves
to model smooth curves. If $\mathbf{p}$ is the SRV transformation of $\boldsymbol{\beta}$ and $\mathbf{p}$ is continuous, we have that the back-transform $\boldsymbol{\beta}(t) = \boldsymbol{\beta}(0) + \int_0^t \mathbf{p}(s) \| \mathbf{p}(s) \| ds$ is differentiable, as the norm $\|\cdot\|$ is continuous as well. Alternatively, constant spline SRV curves can be used to model less regular, polygonal mean curves. We thus work with a linear or constant spline model on SRV level in the following.

\subsection{Elastic means for samples of curves}
\label{subsec:spline_mean}
Since the space of curves modulo parametrisation and translation does not form a Euclidean space, standard statistical techniques for describing probability distributions cannot be applied directly. In particular, taking sums or integrals requires a linear structure of the space, which means that we cannot define the expected value as an integral or the mean as a weighted average here. In order to generalise the mean as a notion of location to arbitrary metric spaces, Fréchet\cite{frechet} proposed to use its property of being the minimiser of the expected squared distances.
\begin{defi}[Fréchet mean (Fréchet\cite{frechet})]
Let $(\Omega, \mathcal{F}, P)$ be a probability space and $\mathcal{X}$ a metric space with distance function $d$, equipped with the Borel-$\sigma$-Algebra. For a random variable $X: \Omega \to \mathcal{X}$ we call every element in
\begin{align*}
\arginf_{A \in \mathcal{X}} \mathbb{E}_P \left(d(X, A)^2 \right) 
\end{align*}
an expected element of $X$. For a set of observations $x_1, \dots, x_n \in \mathcal{X}$ we define the Fréchet mean as an element in
\begin{align*}
\arginf_{A \in \mathcal{X}} \sum_{i = 1}^n d(x_i, A)^2.
\end{align*}
\end{defi}
That means Fréchet means are empirical versions of expected elements and neither of them need to exist or be unique. Consider for example a uniform distribution on the sphere where every point on the sphere is a valid Fréchet mean. This non-uniqueness can occur for the elastic distance as well, see the example given in Appendix \ref{app:non_unique}. Nevertheless, Ziezold\cite{ziezold} showed a set version of the law of large numbers for the Fréchet mean, which means that for independently and identically distributed random variables $X_1, \dots, X_n: \Omega \to \mathcal{X}$ the set of Fréchet means converges to the set of the expected elements.

As discussed in the previous subsection, we propose to use linear or constant splines on SRV level as model spaces. Hence, to compute a Fréchet mean w.r.t. the elastic distance \eqref{eq:warp_prob} for a set of curves with SRV transformations $\mathbf{q}_1, \dots, \mathbf{q}_n$, we need to solve the following minimisation problem for a given degree $l \in \{0, 1\}$ (constant or linear splines).
\begin{align} \label{eq:mean_prob_open}
\text{Minimise } &\quad  \sum_{i = 1}^n \inf_{\gamma_i} \left\| \mathbf{p} - (\mathbf{q}_i \circ {\gamma_i}) \sqrt{\dot{\gamma_i}} \right\|_{L_2}^2 \quad
\text{w.r.t. } \ \textbf{p}: [0,1] \to \R^d \text{ spline of degree $l$.} 
\end{align}

A solution $\bar{\mathbf{p}}$ to this optimisation problem is a piecewise constant respectively piecewise linear approximation of the SRV transformation of a Fréchet mean. Hence, the corresponding mean curve $\bar{\boldsymbol{\beta}}$ is either a polygon or a differentiable approximation of the Fréchet mean. If we consider the optimisation problem \eqref{eq:mean_prob_open} for only one observed curve ($n = 1$), we get as a solution a spline approximation of this curve as a special case. Similarly to the proposal of \cite{srivastava_book} for densely observed curves, we tackle the minimisation problem (\ref{eq:mean_prob_open}) with an iterative approach in Algorithm \ref{algo:smooth_mean_open}, alternating between fitting the mean and optimising the warping for each of the observations, but now using our warping approach for sparse curves and modelling the mean with a constant or linear spline.

\begin{algorithm}[H]
\caption{Elastic spline mean for open curves \label{algo:smooth_mean_open}}
\KwIn
{
convergence tolerance $\epsilon > 0$\;
SRV transformations $\mathbf{q}_i$, $i = 1, \dots, n$ of discretely observed curves $\boldsymbol{\beta}_i$, $i = 1, \dots, n$\;
initial mean $\bar{\mathbf{p}}_{new} = \arginf_{\bar{\mathbf{p}}} \sum_{i = 1}^n \left\| \bar{\mathbf{p}} - \mathbf{q}_i \right\|_{L_2}^2$
}
\While{$\| \bar{\mathbf{p}}_{old} - \bar{\mathbf{p}}_{new} \| > \epsilon$}{
$\bar{\mathbf{p}}_{old} = \bar{\mathbf{p}}_{new}$\;
$\gamma_i = \arginf_{\gamma} \left\| \bar{\mathbf{p}}_{old} - (\mathbf{q}_i \circ {\gamma}) \sqrt{\dot{\gamma}} \right\|_{L_2}^2, \quad \forall i = 1, \dots, n$ \tcp*{warping step}
$\bar{\mathbf{p}}_{new} = \arginf_{\bar{\mathbf{p}}} \sum_{i = 1}^n \left\| \bar{\mathbf{p}} - (\mathbf{q}_i \circ {\gamma_i}) \sqrt{\dot{\gamma_i}} \right\|_{L_2}^2$ \tcp*{$L_2$ spline fitting step}
}
\Return{$\bar{\mathbf{p}} = \bar{\mathbf{p}}_{new}$}
\end{algorithm}

\begin{figure}[ht]
\centering
\includegraphics[scale=0.65]{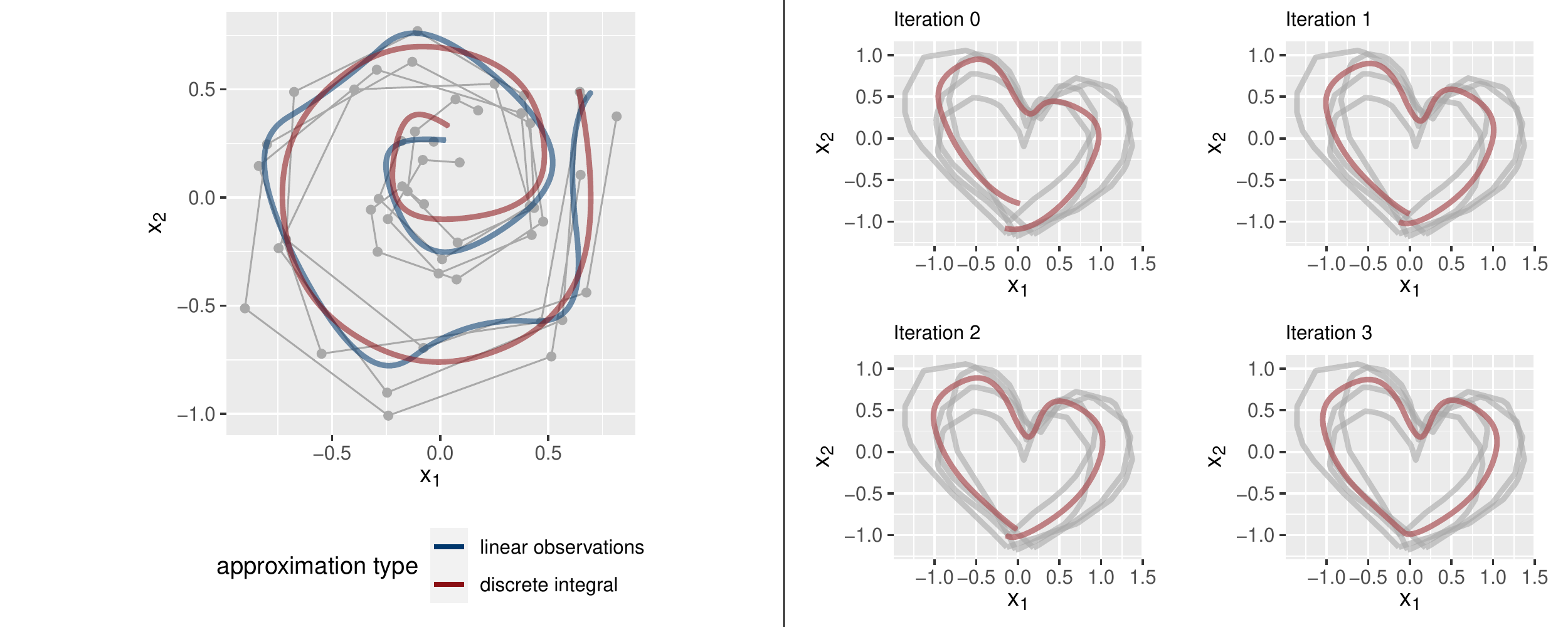}
\caption{\label{fig:elastic_means}Left: Smooth means (with 11 knots each) for four spiral curves based on linear splines on SRV level. The blue mean curve is based on assuming piecewise linear observations for the integral approximations and the red mean curve is based on the integral approximation using the mean value theorem. \newline
Right: First three iterations of the algorithm for closed mean curves on a toy dataset.}
\end{figure}

For the warping step we update the optimal warpings $\gamma_i$ of the observed curves $\boldsymbol{\beta}_i$, $i = 1, \dots n$ via interpreting them as observed polygons with piecewise constant SRV transformations $\mathbf{q}_i$, $i = 1, \dots n$,  as in Lemma \ref{lem:warp_prob_discrete}. We tackle the remaining maximisation problem (\ref{eq:warp_prob_discrete}) using a gradient descent algorithm as discussed before
if $\bar{\mathbf{p}}$ is piecewise linear and Algorithm \ref{algo:dist_open_curves} if $\bar{\mathbf{p}}$ is piecewise constant. In the $L_2$ spline fitting step the integrals
\begin{align} \label{eq:L2_integral}
\left\| \bar{\mathbf{p}} - (\mathbf{q}_i \circ {\gamma_i}) \sqrt{\dot{\gamma_i}} \right\|_{L_2}^2 &= \int_0^1 \left\| \bar{\mathbf{p}}(t) - (\mathbf{q}_i(t) \circ \gamma_i(t)) \sqrt{\dot{\gamma_i}(t)} \right\|^2 \ dt
\end{align}
in the sum need to be approximated, since the curves $\boldsymbol{\beta}_i$ are only observed on a finite grid $0 = s_{i,0} \leq s_{i,1} \leq \dots \leq s_{i,m_i} = 1$, which means the SRV curves $\mathbf{q}_1, \dots, \mathbf{q}_n$ are unobserved. One option is to assume that the SRVs $\mathbf{q}_i$ of the observed curves are piecewise constant, like we do in the warping step. Since $\bar{\mathbf{p}}$ is piecewise linear (or even piecewise constant), $(\mathbf{q}_i \circ {\gamma_i}) \sqrt{\dot{\gamma_i}}$ will be piecewise linear as well (see proof of Lemma \ref{lem:warp_prob_discrete} in the appendix), which leads to a closed form solution of the integral. If we use this approximation of the integral, the resulting mean tends to overfit the edges of the observed polygons (see for an example the mean plotted in blue on the left hand side of Figure \ref{fig:elastic_means}).

Alternatively, we derive an approximation of the integrals in the $L_2$ fitting step of Algorithm \ref{algo:smooth_mean_open} using the mean value theorem and the monotonicity of the warping. For all $j = 0, \dots, m_i - 1$, there is a $t_{i,j} \in [\gamma_i^{-1}(s_{i,j}) , \gamma_i^{-1}(s_{i,j + 1})]$ with $(\boldsymbol{\beta}_i \circ \gamma_i)'(t_{i,j}) = \frac{\boldsymbol{\beta}_i(s_{i,j+1}) - \boldsymbol{\beta}_i(s_{i,j})}{ \gamma_i^{-1}(s_{i,j+1})-\gamma_i^{-1}(s_{i,j})}$ and therefore 
\begin{align*}
(\mathbf{q}_i \circ \gamma_i(t_{i,j})) \sqrt{\dot{\gamma_i}(t_{i,j})} &= 
\frac{(\boldsymbol{\beta}_i \circ \gamma_i)'(t_{i,j})}{ \sqrt{\left\| (\boldsymbol{\beta}_i \circ \gamma_i)'(t_{i,j}) \right\|}} \\ &=
\frac{\boldsymbol{\beta}_i(s_{i,j+1}) - \boldsymbol{\beta}_i(s_{i,j})}{\sqrt{\left\| \boldsymbol{\beta}_i(s_{i,j+1}) - \boldsymbol{\beta}_i(s_{i,j})\right\|} \sqrt{\gamma_i^{-1}(s_{i,j+1})-\gamma_i^{-1}(s_{i,j}))}}.
\end{align*}
While this is exact for unkown points $t_{i,j}$, we use an approximation by assuming this mean value of the derivative $(\boldsymbol{\beta}_i \circ \gamma)'$ is attained in the middle of the interval $[\gamma_i^{-1}(s_{i,j}) , \gamma_i^{-1}(s_{i,j + 1})]$; hence we approximate $t_{i,j} \approx \frac{\gamma_i^{-1}(s_{i,j + 1}) + \gamma_i^{-1}(s_{i,j})}{2}$ for all $j = 0, \dots, m_i - 1$. Thus, for $i = 1, \dots, n$, the integral in \eqref{eq:L2_integral} is replaced by the weighted sum $ \sum_{j = 0}^{m_i - 1} \omega_{i,j} \left\| \bar{\mathbf{p}}(t_{i,j}) - (\mathbf{q}_i \circ \gamma_i(t_{i,j})) \sqrt{\dot{\gamma_i}(t_{i,j})} \right\|^2$. This leaves us with a quadratic minimisation problem w.r.t. the spline coefficients in $\bar{\mathbf{p}}$, for which we compute the solution analytically as a generalised least squares estimate. There are different options to choose the weights $\omega_{i,j}$ in this integral approximation. The weights $\omega_{i,j} = \left( \gamma_i^{-1}(s_{i,j+1})-\gamma_i^{-1}(s_{i,j})) \right)$  based on the trapezoidal rule for numerical integration give equal importance to each of the observed curves, independent of the number of  points $m_i$  observed on each of them. 
An alternative choice of $\omega_{i,j} = 1$ puts more weight on single observations on a specific curve. Consequently, curves or parts of curves with more observations have higher influence on the estimated mean than curves or parts of curves with fewer observations. The difference between this approximation (with $\omega_{i,j} = 1$) and the one based on assuming observed polygons also for the $L_2$ spline fitting step is displayed in Fig. \ref{fig:elastic_means} on the left. In this example, the estimated mean based on this discrete integral approximation (in red) is closer to a proper spiral shape. In the following we will use weights $\omega_{i,j} = 1$ unless stated otherwise.

\begin{remark}[Smooth elastic mean for closed curves]
\label{rem:mean_closed}
Algorithm \ref{algo:smooth_mean_open} can be adapted for closed curves. We replace the first step by updating the optimal parametrisations $\gamma_i$ via considering the corresponding minimisation problem for closed curves \eqref{eq:warp_prob_discrete_closed} via gradient descent or Algorithm \ref{algo:dist_closed_curves} depending on the spline degree. In the second step, that is updating the least-squares estimate for given parametrisations, we use a penalty function method to deal with the non-linear constraint of closedness for $\bar{\mathbf{p}}$ (see for example Sun and Yuan\cite{sun_optimization}). Thus, we add a cost function penalising openness with increasing weight. Precisely, in the $k$-th iteration step, we consider the loss function 
\begin{align*}
\left\| \bar{\mathbf{p}} - (\mathbf{q}_i \circ {\gamma_i}) \sqrt{\dot{\gamma_i}} \right\|_{L_2}^2 + \lambda_k \left\| \int_0^1 \bar{\mathbf{p}}(t) \| \bar{\mathbf{p}}(t)\| \ dt \right\|^2,
\end{align*}
with $\lambda_k \to \infty$ for $k \to \infty$. Since $\int_0^1 \bar{\mathbf{p}}(t) \|\bar{\mathbf{p}}(t)\| \ dt = \bar{\boldsymbol{\beta}}(1) - \bar{\boldsymbol{\beta}}(0)$, if $\bar{\mathbf{p}}$ is the SRV of $\bar{\boldsymbol{\beta}}$, the penalty term vanishes if and only if $\bar{\boldsymbol{\beta}}$ is closed.
\end{remark}

Figure \ref{fig:elastic_means} shows three iterations of this adapted algorithm for calculating a smooth mean of four, irregularly sampled, closed heart shapes. The initial mean (iteration 0) was computed as a least-squares-estimate assuming the curves were parametrised by relative arc length. The sequence $(\lambda_k)_{k \in \mathbb{N}}$ was chosen as $\lambda_k = 10^{-3}k$ for all $k \in \mathbb{N}$. 

\subsection{Identifiability of spline curves}
\label{subsec:identify}
We model curves or means for samples of curves using basis representations. If we study equivalence classes of curves modulo re-parametrisation, we have to ensure unique spline representatives in each class, meaning that elements of the quotient space are identifiable via their basis coefficients. To see why this is not self-evident, consider as a simple counterexample in $\mathbb{R}^1$ the space of quadratic polynomials  $P: [0,1] \to \mathbb{R}$, 
a subspace of the quadratic spline space. Note that $\gamma_a(x) = ax^2 + (1-a)x$ defines a feasible warping function for all $a \in ]0,1[$, since $\gamma_a$ is differentiable with $\gamma_a'(x) 
\geq 0$ and $\gamma_a(0) = 0$, $\gamma_a(1) = 1$. Hence all quadratic polynomials of the form $P(x) = p_1 \gamma_a(x) + p_0$ with $p_0, p_1 \in \mathbb{R}$ are elements of the same equivalence class, although they have varying basis coefficients $ap_1$, $(1-a)p_1$ and $p_0$ for $a \in ]0, 1[$  w.r.t. the monomial basis expansion. This counterexample
shows in particular that one-dimensional spline functions do not have unique representatives in the space of functions modulo re-parametrisation. 
As identifiability plays an important role in any spline based modelling approach, it is fortunate that in contrast to the one-dimensional case we can show that in $\R^d$ with $d \geq 2$, nearly all quadratic or cubic spline curves have unique basis representations.

\begin{theorem} \label{theo:unique_spline}
Let $d \geq 2$ and $\mathbf{Q}, \mathbf{P}: [0, 1] \to \R^d$ be quadratic or cubic spline curves, where $\mathbf{Q}$ has a non-linear image between each of its knots. Moreover let $\gamma: [0, 1] \to [0,1]$ be monotonically increasing and onto. Then
\begin{align*}
\mathbf{P} = \mathbf{Q} \circ \gamma \quad \Rightarrow \quad \gamma = id.
\end{align*}
\end{theorem}

This means nearly all equivalence classes modulo re-parametrisation contain at most one spline curve. Hence we can identify these curves modulo warping via their spline basis coefficients. Only if the spline has a linear image, are there splines with differing coefficients in its equivalence class. This is the case if and only if the splines in each coordinate direction are multiples of each other modulo translation. For more details refer to the proof of this theorem in  Appendix \ref{app:unique_spline}. Note that the we do not make any assumptions on the knots here, in particular the knots could be different for $\mathbf{Q}$ and $\mathbf{P}$. That means there is almost always a unique representative modulo warping in $\bigcup_{K, \kappa_0, \dots, \kappa_K} \mathcal{S}^l_{K; \kappa_0, \dots, \kappa_K}$ for given $l = 2,3$, i.e.\ in the union of all spline spaces with varying (also number of) knots.
Considering only quadratic or cubic splines is crucial, as the following counterexample with splines of degree four shows. Let
\begin{align*}
\mathbf{Q}(t) = \left(
\begin{matrix}
4t^4 - 2t^2 \\
4t^4
\end{matrix}
\right) \text{ and } 
\mathbf{P}(t) = \left(
\begin{matrix}
t^4 + 2t^3 - t  \\
t^4 +2t^3 + t^2
\end{matrix}
\right).
\end{align*}
Then $\gamma(t) = \sqrt{0.5(t^2 + t)}$ is a suitable warping function since it
fullfills $\mathbf{P} = \mathbf{Q}(\gamma(t))$ and is monotonically increasing and onto, but monomial coefficients differ between $\mathbf{P}$ and $\mathbf{Q}$ and are thus not identifiable modulo warping. The result for cubic spline curves also implies uniqueness of representatives for linear spline SRV curves, another useful result for identifiable modelling of elastic curves.

\begin{cor}
\label{cor:ident_srv_splines}
Let $\boldsymbol{\beta}_1, \boldsymbol{\beta}_2:[0,1] \to \R^d$ with SRV functions $\mathbf{q}_1$ and $\mathbf{q}_2$, respectively. If $\mathbf{q}_1$ and $\mathbf{q}_2$ are nowhere constant linear splines and  $\mathbf{q}_2(t) = \mathbf{q}_1(\gamma(t)) \sqrt{\dot{\gamma}(t)}$, then $\mathbf{q}_1$ = $\mathbf{q}_2$.
\end{cor}

\begin{proof}{}
Let $\mathbf{q}^2_1$ and $\mathbf{q}^2_2$ the component-wise squares of $\mathbf{q}_1$ and $\mathbf{q}_2$, respectivly. We compute
\begin{align*}
\mathbf{P}(s) := \int_0^s \mathbf{q}^2_2(t) \ dt = \int_0^s \mathbf{q}_1^2(\gamma(t)) \dot{\gamma}(t) \ dt = \int_0^{\gamma(s)} \mathbf{q}^2_1(t') \ dt' =: \mathbf{Q}(\gamma(s))
\end{align*}
for all $s \in [0, 1]$ via substituting $\gamma(t) \mapsto t'$. Here we have cubic splines $\mathbf{P}$ and $\mathbf{Q}$ on both sides. Hence we deduce $\gamma = id$ by Theorem \ref{theo:unique_spline} and consequently $\mathbf{q}_2 = \mathbf{q}_1$. Note that the cubic spline curve $\mathbf{P}(s) = \int_0^s \mathbf{q}_2^2(t) \ dt$ is linear on any interval if and only if $\mathbf{q}_2(t)$ is constant on this interval, which is excluded by the assumptions.
\end{proof}
To sum up, the space of linear SRV spline curves seems particularly suitable to model smooth curves modulo parametrisation and translation as these curves can be identified via their basis coefficients, i.e. there is a unique representation in this space, and the corresponding curves are differentiable, which leads to visually smooth curves.

\begin{remark}[Splines of lower or higher degree]
\label{rem:constant_splines}
Piecewise linear spline curves or equivalently piecewise constant SRV curves are identifiable via their spline basis coefficients, if we consider one spline space $\mathcal{S}^1_{K; \kappa_0, \dots, \kappa_K}$ but not the union of several such spaces, and assume that the curve is not differentiable at all of its knots (i.e.\ no knot is superfluous). Hence, with this weaker identifiability result, piecewise constant srv-curves are a suitable model space as well, with curves modelled as polygons instead of smooth curves. For more details see Appendix \ref{app:constant_srv_splines}. \newline
For SRV splines of higher order first note that the counterexample for splines of degree 4 could similarly be constructed for all splines with any degree that is not a prime number. If the degree of the splines is a prime number, it seems possible that one can show a similar identifiability result. This would imply identifiability for quadratic SRV curves using an analogous argument as in Corollary \ref{cor:ident_srv_splines}.
\end{remark}

Since we want to use these spline spaces for estimation of smooth or polygonal curves, we need the following result on continuity of the embedding. It allows us to interpret estimated coefficients, for instance compare the coefficients of estimated group means to investigate local differences, as it ensures convergence of the coefficients if the curves converge. Hence if we construct a sequence that converges to the mean with respect to the elastic distance, as we aim to do in Algorithm \ref{algo:smooth_mean_open}, we can conclude that the estimated spline coefficients converge to the spline coefficients of the mean as well. We show that this continuity property holds whenever we consider a (subset of a) finite dimensional spline space of the following form as a model space $\Xi$. 

\begin{defi} \label{def:xi}
Let $\Xi$ be one of the following for given fixed $K \geq 2$, $0 = \kappa_0 < \dots < \kappa_K=1$:
\begin{itemize}
    \item A subset of $\mathcal{S}^l_{K; \kappa_0, \dots, \kappa_K}$, $l= 2,3$, which consists of identifiable splines as described in Theorem \ref{theo:unique_spline}, additionally centred (i.e.\ with integral zero) to account for translation.
    \item A set of identifiable curves with linear spline SRV curves in $\mathcal{S}^1_{K; \kappa_0, \dots, \kappa_K}$ from Corollary \ref{cor:ident_srv_splines}.
    \item The set of curves with piecewise constant SRV curves in $\mathcal{S}^1_{K; \kappa_0, \dots, \kappa_K}$ from Remark \ref{rem:constant_splines}.
\end{itemize}
\end{defi}
Note that we do not consider unions of spline spaces here for simplicity in considering convergence of corresponding coefficients.
 
\begin{lemma}[Topological embedding]
\label{lem:top_emb}
Let $f: (\Xi, \| \cdot \|) \to (\mathcal{A}, d)$ be the embedding of the spline coefficients defining the functions in $\Xi$, equipped with the usual Euclidean distance $\|\cdot\|$, into the space $\mathcal{A}$ of absolutely continuous curves w.r.t. the elastic distance $d$. Then $f$ is a topological embedding, i.e. $f$ is a homeomorphism on its image. 
\end{lemma}

A proof for this statement can be found in Appendix \ref{app:top_emb}.
It shows that the distance on spline coefficients and elastic distance of curves modulo translation are topologically equivalent on suitable spline spaces. This means a sequence of curves converges with respect to the spline coefficients if and only if it converges with respect to the elastic distance.
Overall we thus have that any spline model $\Xi$ in Definition \ref{def:xi} yields an identifiable model for the Fréchet mean of observed curves, with the possibility to interpret spline coefficients, and this also holds for converging series of estimators as we aim to construct in our algorithms. 

%% file: 2_simulation.tex
\section{Simulations}
\label{sec:simulation}
We test our methods, which we made available for public use in the \texttt{R}-package \texttt{elasdics}\cite{elasdics}, on simulated data. Since there is an implementation of the SRV framework already available for \texttt{R} implemented in the package \texttt{fdasrvf}\cite{fdasrvf} based on Srivastava et al.\cite{srivastava}, we compare our results to their output whenever possible.

\subsection{Simulation: Aligning sparsely and irregularly sampled curves}
In this first simulation, we compare our methods for aligning sparsely and irregularly sampled curves to the implementation of the dynamic programming (DP) algorithm in \texttt{fdasrvf}\cite{fdasrvf}. Since this DP implementation only allows for an equal number of observed points on both curves, we restrict the simulation to this case, although we developed our methods in particular for differing numbers of observed points per curve. In Figure \ref{fig:simulate_dists} in Appendix \ref{app:plots}, we present one simulated example for open and closed curves each.

For the open setting we choose a parametrised curve $\boldsymbol{\beta}(t) = \sin(t)(\cos(12t) + 2t, \sin(12t) + t)^T$, which we use as a template for both curves. The first curve $\boldsymbol{\beta}_1$ (displayed in red in Figure \ref{fig:simulate_dists} in Appendix \ref{app:plots}) is obtained via sampling an unbalanced observation grid $t_1, \dots, t_m$ with $m \in \{10, 30, 50\}$ and adding a Gaussian random walk error (with standard deviation $sd = 0.01$) to the evaluations $\boldsymbol{\beta}_1(t_1), \dots, \boldsymbol{\beta}_1(t_m)$. The second curve $\boldsymbol{\beta}_2$ is re-sampled 30 times (displayed in grey in Figure \ref{fig:simulate_dists}) using the same sampling scheme as for $\boldsymbol{\beta}_1$.

For the closed setting we choose two butterfly shapes available in \texttt{fdasrvf}\cite{fdasrvf}. These are discretely observed curves with 100 observations each. We down-sample the curves such that $m \in \{30, 60, 90\}$ points per curve are left and such that points with high estimated curvature are more likely to be included. This way, the images of the curves are well preserved, as we are more likely to remove points on straight lines. Furthermore, we add an error term $\sin(\pi \frac{i - 1}{m - 1})\epsilon_i$ to the $i$-th remaining observation for all $i = 1, \dots, m$, where $\epsilon_i$ is distributed according to a Gaussian random walk with standard deviation $sd = 0.5$ and the modification with the sinus function ensures closedness. According to this sampling scheme, we draw one copy (plotted in red) of $\boldsymbol{\beta}_1$ from the first butterfly shape and 30 copies (plotted in grey) of $\boldsymbol{\beta}_2$ from the second butterfly shape.

For each of the settings we compare the optimal alignment for each copy of $\boldsymbol{\beta}_2$ to the corresponding $\boldsymbol{\beta}_1$ using our coordinate-wise-optimisation (CWO) algorithm with the alignment produced by the dynamic programming (DP) from \texttt{fdasrvf}\cite{fdasrvf}. When looking at the coordinates separately, we visually observe slightly better alignment for our method CWO compared to DP. This is also evident in a smaller average elastic distance, e.g. a reduction of 33\% and 26\% on average for $m=30$ in the open and closed setting, respectively. For moderate $m$ we observe an reduction of 48\% (open, $m=50$) and 13\% (closed, $m=60$).
As expected, this difference decreases if 90 points of the butterfly shapes are selected (4\% reduction on average), as in this case the points are nearly observed on a regular, fairly dense grid, which is the setting the implementation in \texttt{fdasrvf} is designed for.

A highly unbalanced distribution of observed points on the curves described above causes difficulties for the mean computation in \texttt{fdasrvf}\cite{fdasrvf} as well. Figure \ref{fig:compare_means} in Appendix \ref{app:plots} demonstrates this for sets of partially densely and partially sparsely observed curves each, for which we compute means with respect to the elastic distance. The means in red, which are computed by the \texttt{curve\_karcher\_mean} function in \texttt{fdasrvf}\cite{fdasrvf}, do not capture the image of the observed curves well (with e.g.\ a butterfly no longer recognisable). Contrarily, our methods are specifically developed for such unbalanced data, which results in visually appealing mean curves displayed in blue (e.g.\ of butterfly shape). Since the implementation in \texttt{fdasrvf} aims at computing a mean with respect to the geodesic shape distance, i.e. minimises the geodesic distance on the sub-manifold of (closed) curves with fixed curve length, the results are not completely comparable. Nevertheless, in particular for the open curves, which are of similar length, we expect the impact of this aspect to be relatively small compared to the warping. 

\subsection{Simulation: Convergence of spline mean coefficients}
\begin{figure}[!ht]\centering
\includegraphics[scale=0.65]{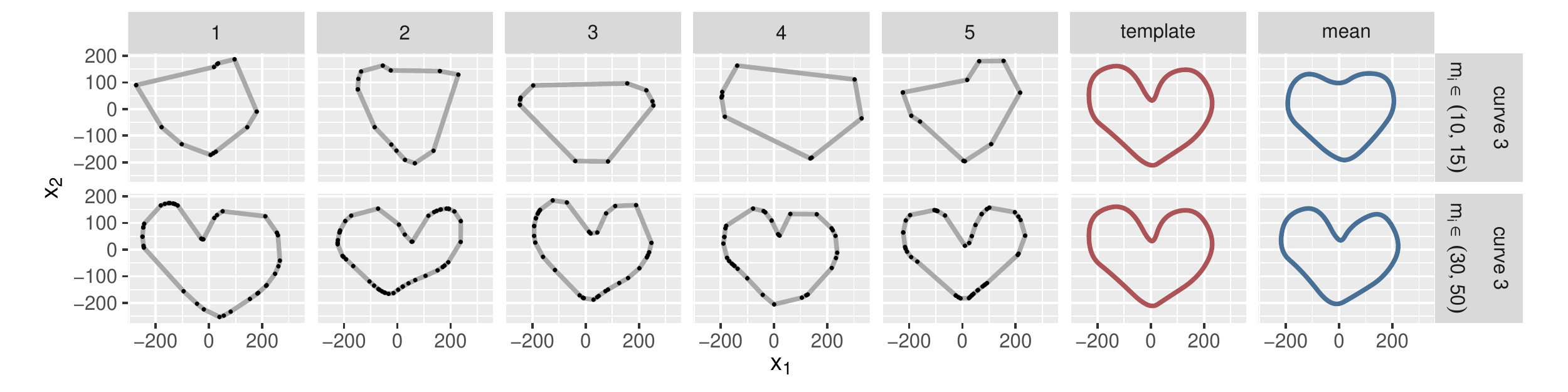}
\vspace*{-0.2cm}
\caption{\label{fig:simu_coefs_closed_data} Simulated data in grey with observed values marked as black dots and corresponding smooth elastic means over $n = 5$ observations in blue. The irregularly sampled curves are drawn from a heart-shaped template (in red) with varying number $m_i$ of observed points per curve.}
\end{figure}

The second simulation is concerned with the convergence and the identifiability of spline means and their associated coefficients, now also with varying numbers of points per curve. For a known template curve $\boldsymbol{\beta}$ with known B-spline coefficients $\xi_1, \dots, \xi_B$ we generate a set of observed curves $\boldsymbol{\beta}_1, \dots, \boldsymbol{\beta}_n$ via independently sampling the coefficients $\xi_{i,b} \sim \mathcal{N}(\xi_b, \sigma^2)$ for all $i = 1, \dots, n$, $b = 1, \dots, B$. If the template curve is closed, we additionally close the sampled curves via minimising the penalty function given in Remark \ref{rem:mean_closed} (for estimating a closed mean) in gradient direction. The points $t_{i,1}, \cdots, t_{i, m_i - 1}$ on which $\boldsymbol{\beta}_i$ is observed are sampled uniformly on $[0, 1]$, where the number of observed points $m_i$ is sampled uniformly either from $\{10, \dots, 15\}$ (very sparse and unbalanced setting) or $\{30, \dots, 50\}$ (less sparse but still unbalanced setting). Examples for curves sampled from two open template curves modelled as linear splines on SRV level with three or nine, equally spaced, inner knots, respectively, are displayed in Figure \ref{fig:simu_coefs_open_data} in Appendix \ref{app:plots}.  Here we choose the standard deviation as $\sigma = 0.3$, $\sigma = 0.4$ for the two open template curves, respectively.
Examples for curves sampled from a heart-shaped template (with standard deviation $\sigma = 4$) are displayed in Figure \ref{fig:simu_coefs_closed_data}. The closed, heart-shaped curve is modelled as linear spline on SRV level with ten, equally spaced, inner knots. 

\begin{figure}[!ht]\centering
\includegraphics[scale=0.7]{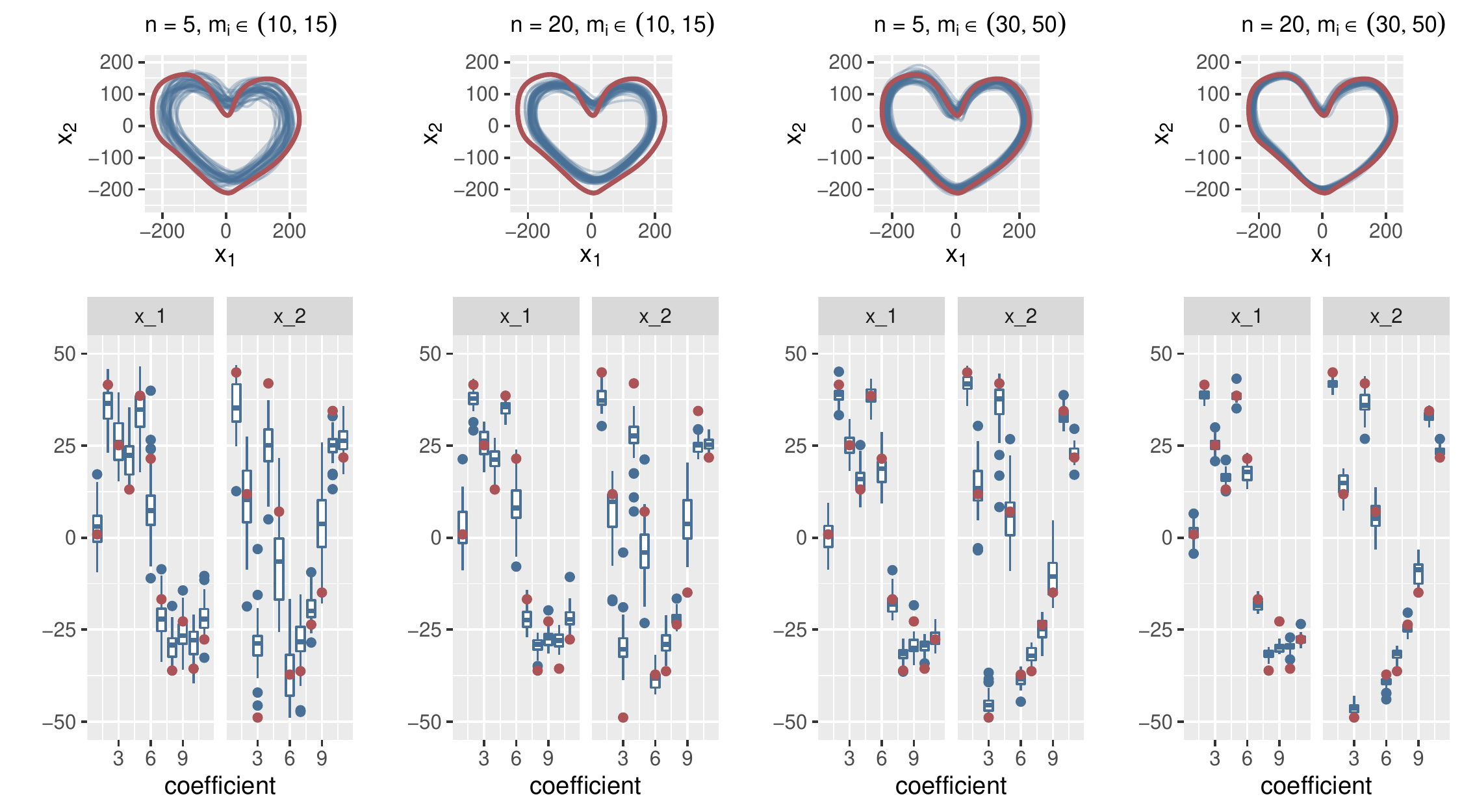}
\vspace*{-0.2cm}
\caption{\label{fig:simu_coefs_curve3} Top: Smooth means (in blue) computed for a set of $n$ curves drawn from the heart shaped template curve (in red) via sampling its B-spline coefficients and with $m_i, i = 1, \dots, n$ points observed per curve. The means are computed using linear SRV splines and the same knot set as the template (ten equally spaced inner knots)\newline
Bottom: Corresponding distribution of spline mean coefficients (in blue) and template coefficients (in red).}
\end{figure}

In this very sparse setting, the sampled curves are hardly recognisable as heart shapes (cf.\ Figure\ \ref{fig:simu_coefs_closed_data}). However, the elastic mean curve over $n = 5$ observations, estimated using the true knot set and linear SRV splines to allow comparison of estimated and true coefficients, represents the original heart surprisingly well even in this challenging setting. We repeated this simulation 40 times (Figure \ref{fig:simu_coefs_curve3}) each for varying numbers of observations $n \in \{5, 20\}$ and observed points per curve $m_i$. For $m_i \in \{10, \dots, 15\}$ observations per curve we generally obtain a heart-shaped curve, which seems smaller and with less pronounced features than the template. If we increase the number of observed curves from $n = 5$ to $n = 20$, the variance of the mean curves decreases but a certain bias due to under-sampling the curves remains. This also manifests in the coefficients of the spline means: for $n = 20$ we observe lower variance of the estimated coefficients than for $n = 5$, but the distribution of the estimated spline coefficients is still not centered at the coefficients of the template (indicated as red dots in Figure \ref{fig:simu_coefs_curve3}).

If we increase the number of points on each curve to $m_i \in \{30, \dots, 50\}$, the estimated means with respect to the elastic distance adapt closer to the template. Moreover, the variance of the estimated spline coefficients decreases as well as the distance of the estimated coefficients to the template. The reduction of variance indicates convergence of the spline coefficients for $n \to \infty$, although we do not expect them to precisely converge to the coefficients of the template, not even if $m_i \to \infty$ for all $i = 1, \dots, n$. This is because we draw the sample curves $\boldsymbol{\beta}_1, \dots, \boldsymbol{\beta}_n$ such that $\boldsymbol{\beta}$ is the mean with respect to the $L_2$ distance on SRV level, but this does in general not imply that $\boldsymbol{\beta}$ is the mean with respect to the elastic distance. Nevertheless, we expect this difference to be small, as the coefficients in the rightmost barplot are close to the red dots indicating the template's coefficients. In addition, their low variance for $n = 20$ confirms our theoretical results on identifiability of spline coefficients in our model (Corollary \ref{cor:ident_srv_splines}) and continuity of the embedding (Lemma \ref{lem:top_emb}). We observe similar behaviour of the estimated means and associated coefficients for the open curves displayed in Figure \ref{fig:simu_coefs_curve1} and Figure \ref{fig:simu_coefs_curve2} in Appendix \ref{app:plots}.

So far, we only elaborated on the convergence of correctly specified spline means, also to show convergence of corresponding spline coefficients. Since this assumption is usually questionable in real data applications, we demonstrate the behaviour of our methods in the case of model misspecification. Figure \ref{fig:simu_misspec_coefs} in Appendix \ref{app:plots} shows means with varying knots using linear SRV splines (smooth means in blue) or constant SRV splines (polygonal means in red). All means are computed for the same set of $n = 20$ heart-shaped curves, which have been sampled as described above from the third template with $m_i \in \{30, \dots, 50\}$ points per curve. For a sufficient number of knots, both the smooth and the polygonal means reproduce the original heart shape well. If we consider the number of coefficients $n_{coefs}$ as a measure for model complexity, we observe that the smooth means are closer to the template than the polygonal ones, given the same number of coefficients, with a local minimum at the correctly specified model. This shows that one can obtain more parsimonious models for smooth means using linear SRV curves. Even though the distance to the template for a polygonal mean can be reduced by using more knots, it does not seem to become as low as for the linear SRV mean.
This indicates that using linear SRV splines for modelling a smooth 'true' mean might reduce the bias due to under-sampling the curves. While we see a local minimum for the $n_{coefs}$ used to generate the data, close $n_{coefs}$ give similar results and in particular values larger than the true one give similarly good results, with the  distance generally decreasing in $n_{coefs}$. This indicates that results are not very sensitive to $n_{coefs}$ given it is sufficiently large.

%% file: 3_application.tex
\section{Applications on real data}  \label{sec:application}

\subsection{Clustering and modelling smooth means of GPS-tracks}
\begin{figure}[!ht]\centering
\includegraphics[scale=0.7]{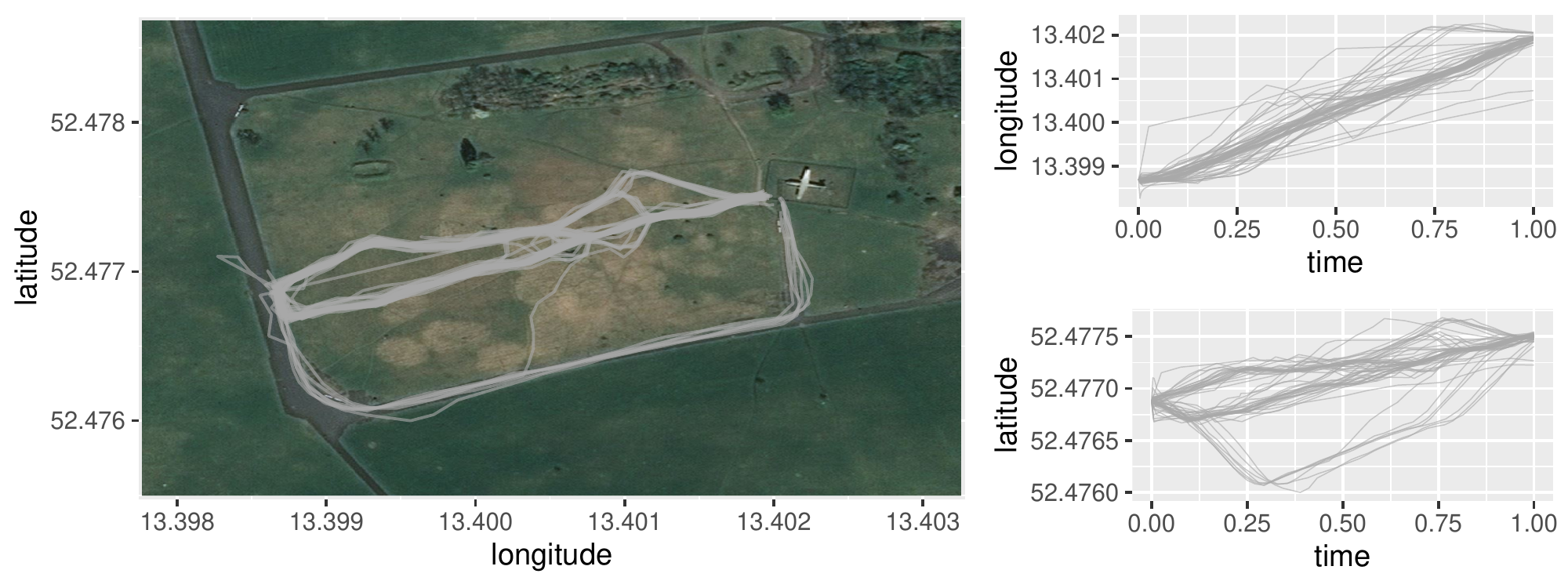}
\vspace*{-0.2cm}
\caption{\label{fig:tracks_on_map} Left: GPS paths tracked on Tempelhof Field and plotted on OpenStreetMap. Right: Longitude and latitude over relative time.}
\end{figure}

As our main goal is the development of statistical (elastic) analysis methods for discretely observed data curves, we demonstrate their practical usefulness on two datasets. The first one comprises GPS waypoints tracked on Tempelhof Field, a former airfield (up to 2008) in Berlin, which is now publicly used as a recreation area. Clustering and smooth mean estimation allow us to find new paths on Tempelhof field not yet included in OpenStreetMap.
The dataset consists of 55 paths with 15 to 45 waypoints each, recorded by members of our working group using their mobile phones for tracking. Due to the variety of   mobile devices used, the number of points per curve differs considerably, hence the data is highly irregular and quite sparsely observed (see Figure \ref{fig:tracks_on_map}).
We are solely interested in analysing the paths the participants walked on, not the trajectories over time. Separately looking at longitude and latitude over time suggests that the individuals had quite different walking patterns, namely did not move with constant speed. This implies that classical functional analysis of the trajectories is not suitable to study the paths used by the test subjects.

From the GPS data we recover the paths the individuals walked on while tracking their trajectories. This is done in two steps. First, the tracks are clustered using average linkage based on the elastic distance and the elbow criterion for stopping. Here we apply Algorithm \ref{algo:dist_open_curves} to approximate the pairwise distance between the irregularly observed open tracks.
Afterwards we compute a smooth elastic Fréchet mean for each of the four largest clusters using Algorithm \ref{algo:smooth_mean_open} and linear splines on SRV level with 10 inner knots.

\begin{figure}[!ht]\centering
\includegraphics[scale=0.7]{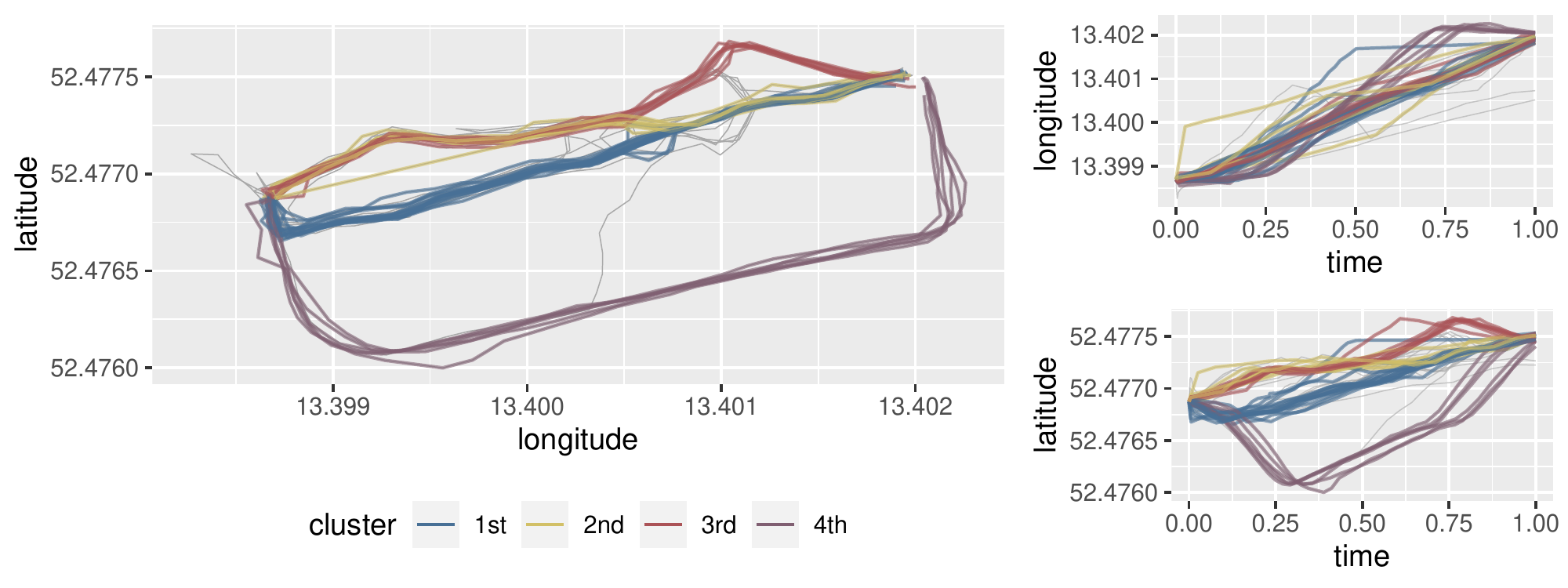}
\caption{\label{fig:tracks_cluster}Left: The observed trajectories as elements of the four largest clusters. Right: Longitude and latitude for the trajectories of the four largest clusters over relative time.}
\end{figure}

The clustering result displayed in Figure \ref{fig:tracks_cluster} on the left is visually satisfying. Looking again at longitude and latitude separately (on the right) clearly indicates that clustering based on the usual $L_2$ distance would lead to worse results. In particular, elements of the first and third largest clusters might be classified differently using a non-elastic distance.

\begin{figure}[!ht]\centering
\includegraphics[scale = 0.65]{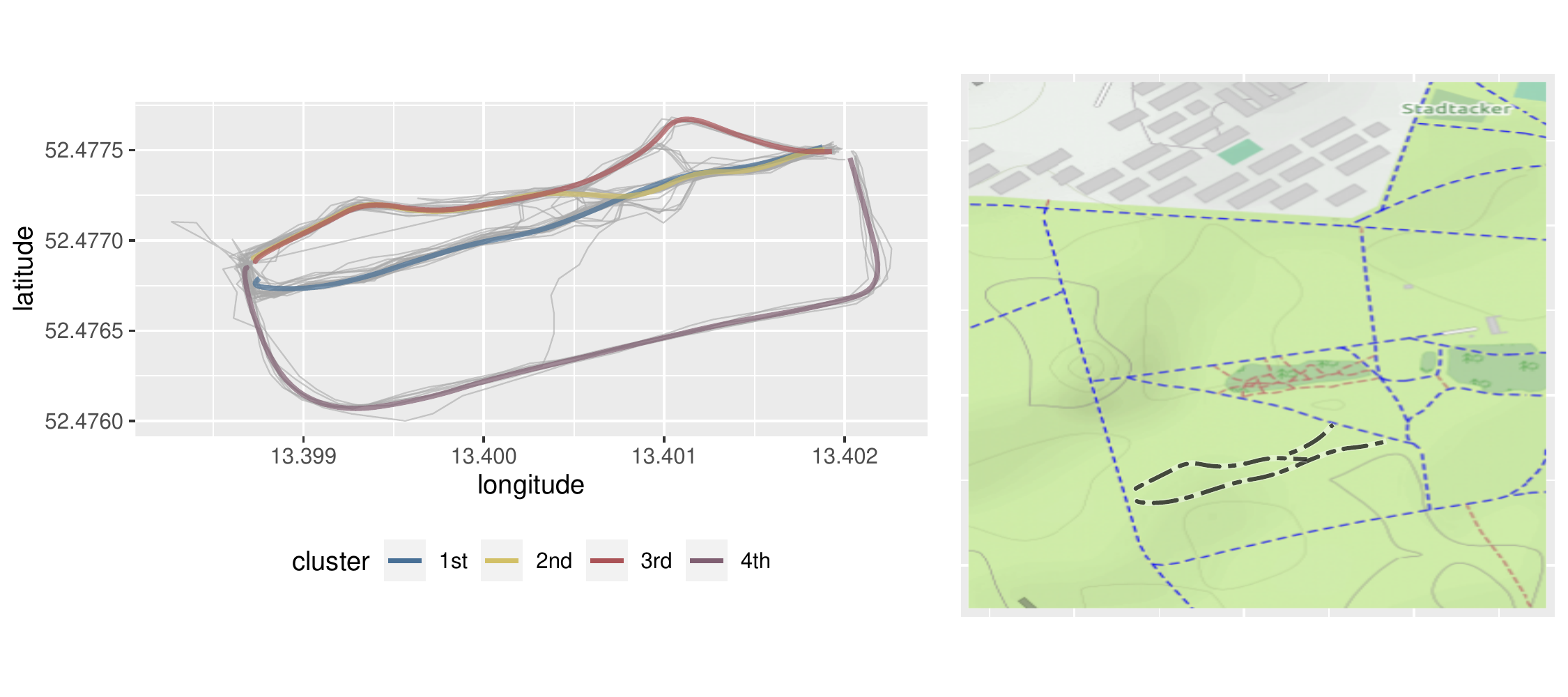}
\caption{\label{fig:tracks_mean_paths} Left: Smooth means modelled as linear SRV curves with 10 inner knots for the four largest clusters.  The mean paths have been centered at the mean center of the observed paths per cluster to account for translation.
Right: The new paths (in black) of the four largest clusters added to the existing OpenStreetMap.}
\end{figure}

The smooth mean curves for each of the four largest clusters displayed in Figure \ref{fig:tracks_mean_paths} on the left seem to describe the observed tracks well, although the number of estimated spline coefficients and therefore model parameters is low (24 coefficients per mean curve compared to 30 to 90 values per observed curve). Thus, we obtain a smooth mean curve for irregularly sampled curves based on the elastic distance that captures the data well and allows dimension reduction. 

One application of the procedure outlined above can be to identify new paths not yet included in an existing map.
The smooth mean curves can be added to an OpenStreetMap, for instance, where we only add parts of our estimated means that are notably different from already existing paths. An example of the resulting map is displayed in Figure \ref{fig:tracks_mean_paths} on the right.

\subsection{Classifying spiral curve drawings for detecting Parkinson's disease}
The Archimedes spiral-drawing test is a common, non-invasive tool for diagnosing patients with Parkinson’s disease. Usually, the drawing task is performed on paper and analysed by medical experts to identify deviations of the shape to the spiral template (Alty et al.\cite{alty}). Recently, there have been approaches (Saunders et al.\cite{saunders}, Isenkul et al.\cite{isenkul}) using digitising tablets to obtain more detailed data, not only on the image of the spiral curve
 but also on the position of the pen at each time point. 

\begin{figure}[!ht]\centering
\includegraphics[scale = 0.7]{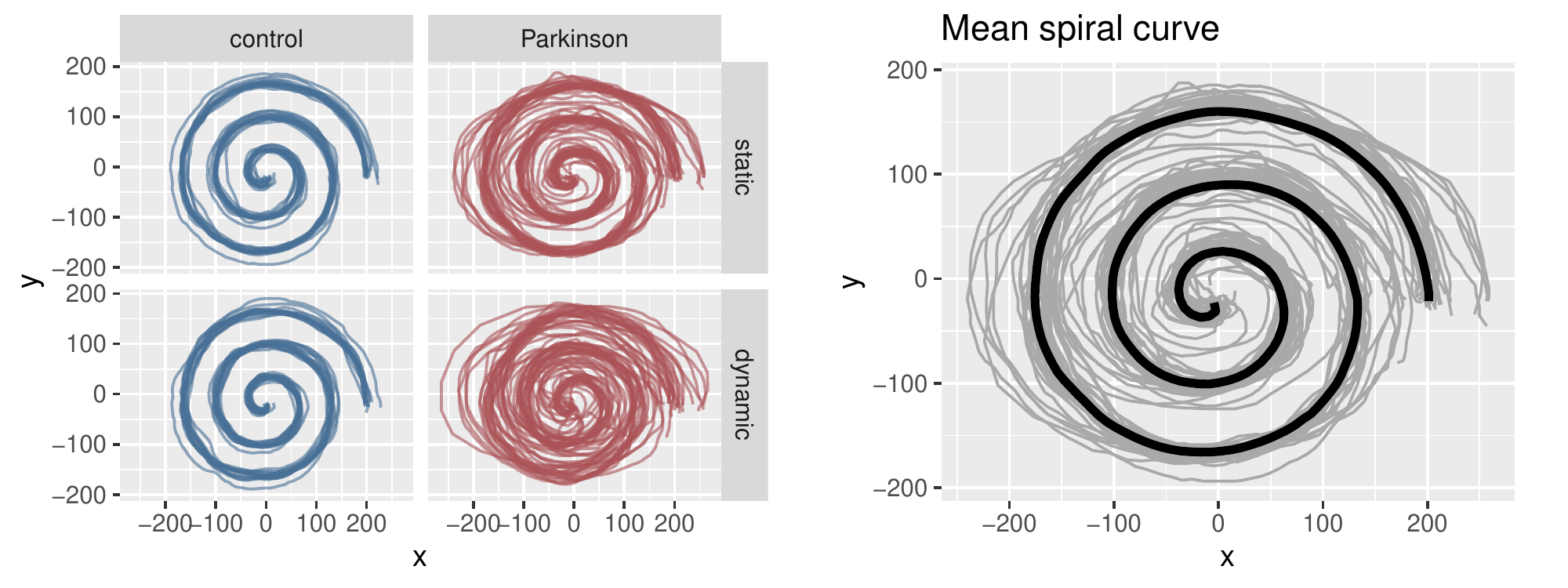}
\caption{\label{fig:parkinson_spiral_mean} Left: Spiral curves drawn by either a healthy control group or by patients with Parkinson's disease in two different settings.
Right: The mean curve (black) of all static curves (grey) computed with respect to the elastic distance.}
\end{figure}

Additional to this so-called \textit{static} spiral test, Isenkul et al.\cite{isenkul} proposed a modified, \textit{dynamic} spiral test, where the template spiral curve appears and disappears in certain time intervals, hence the spiral blinks. A group of 25 Parkinson's patients and a control group of 15 participants performed both tests and the resulting data is publicly available through \url{https://www.kaggle.com/team-ai/parkinson-disease-spiral-drawings}. Figure \ref{fig:parkinson_spiral_mean} displays the spiral curves drawn by the participants. It is visually notable that the non-impaired subjects in the control group follow the template more closely than the patients with Parkinson's disease. This difference seems to become even more severe for the dynamic spiral test.

While the authors of the original study based their analysis on differences in speed distributions of both tasks, Kurt et al.\cite{kurt} imposed pre-alignment of the spiral curves using a heuristic dynamic time warping algorithm. We will follow up on this, but use the elastic distance defined in Section \ref{sec:intro} as a proper distance between the observed curve and a template instead. Moreover, we are only looking for highly interpretable classifiers giving decision rules of the following form: Classify an individual as being at high risk of having Parkinson's disease if the distance of the curve drawn by this individual to the template exceeds a certain threshold. This procedure mimics the decision made by medical experts based on the spiral drawing and allows us to assess whether the additional information provided by time or speed is actually necessary for good classification.

For our analysis, we only use 10\% of the values per curve, which results in irregularly sampled curves with 55 to 269 points each. Based on visual inspection, the images of the curves still almost coincide with the original curves. We compute the elastic mean (see Subsection \ref{subsec:spline_mean}) of all curves drawn in the static spiral test using piecewise constant splines with 201 knots on SRV level. Afterwards, we use the resulting polygonal mean (displayed in black in Figure \ref{fig:parkinson_spiral_mean} on the right) as a template curve. Alternatively, a parametrised version of the original template curve could be directly used in practice, if available.

\begin{figure}[!ht]\centering
\includegraphics[scale = 0.7]{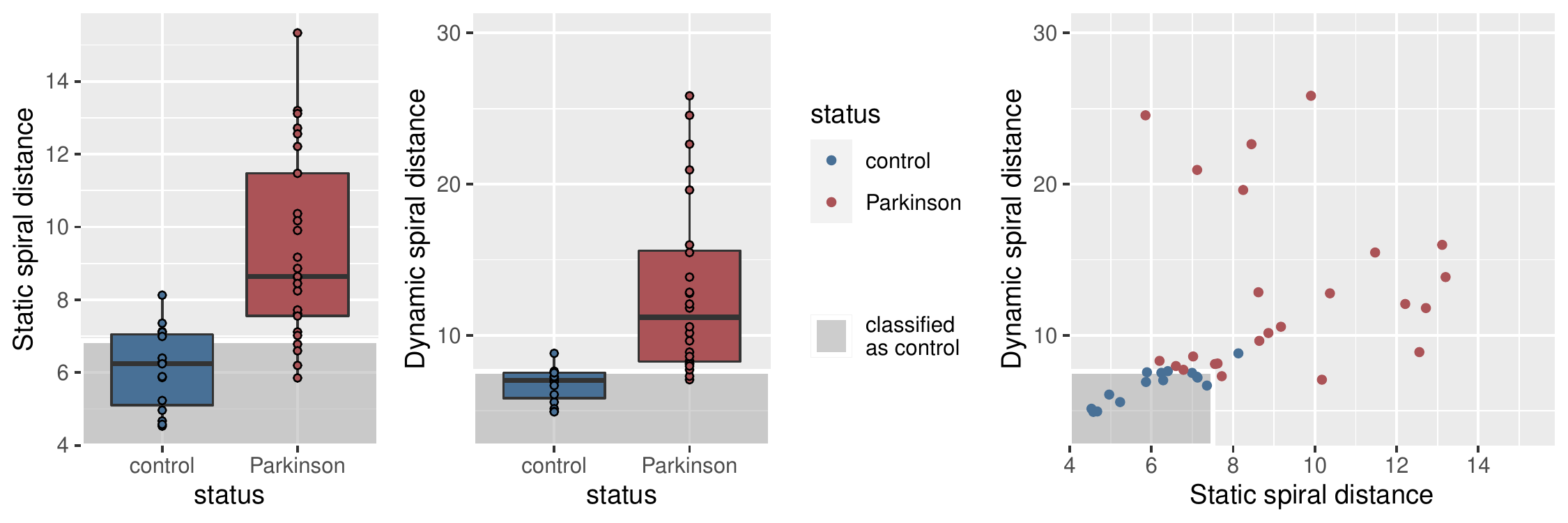}
\caption{\label{fig:parkinson_dists_classify} Left: Distance of the curves drawn by the participants to the mean spiral curve for both settings.\newline
Right: Distance of the curve in the static setting compared to the distance of the curve in the dynamic setting. \newline
Note that one observation for a Parkinson's patient with an extreme distance greater than 35 in the dynamic setting is not displayed.\newline
The grey areas indicate the decision rule based on the zero-one loss with in-sample accuracies of 77.5\%, 92.5\% and 97.5\% for the classifiers based on the static test, dynamic test or both tests, respectively.}
\end{figure}

Figure \ref{fig:parkinson_dists_classify} shows the elastic distances of the curves drawn by the participants to the template curve. As expected, this distance is generally greater for Parkinson's patients than for the control group in both settings. Moreover, looking at the scatter plot on the right of Figure \ref{fig:parkinson_dists_classify}, there seems to be a strong positive correlation between the distance in the static test and the distance in the dynamic test for healthy individuals (in blue). For Parkinson's patients, this trend is not strongly present.

We propose intuitive decision rules of the form: Classify as status 'Parkinson' if the distance of the curve drawn by the test subject to the mean curve exceeds a threshold. Here we either analyse the curves in the static or the dynamic spiral test (Figure \ref{fig:parkinson_dists_classify} on the left and in the middle). The grey areas in Figure \ref{fig:parkinson_dists_classify} indicate the corresponding decision rule. Alternatively, we classify as status 'Parkinson' if any of the distances in the two tests exceeds a respective threshold (as indicated in grey in Figure \ref{fig:parkinson_dists_classify} on the right). To estimate those thresholds, we directly optimise the zero-one loss (also called misclassification loss) as this is feasible for a small dataset and a small set of possible decision functions. For the classifier with one single variable, that is the distance in either the static or dynamic setting, we would expect similar decision boundaries for alternative classifiers like logistic regression or support vector machines with linear kernel and the hinge-loss. 

To evaluate our classifiers we use leave-one-out cross-validation for which we obtain 72.5\% accuracy for the static setting, 90.0\% accuracy for the dynamic setting and 92.5\% accuracy for the classifier based on static and dynamic spiral drawings. Since we observe in-sample only one misclassified observation for the classifier based on both distances, including additional features like the difference or the quotient of the two distances is not advisable. Nevertheless, if more data were available, those variables might improve the classification further. To see that an elastic analysis of the observed curves is favourable we compare our results to classification based on the usual $L_2$-distance. For this analysis, we re-parametrise the curves according to their relative arc length to account for different speed patterns but do not align them in an elastic manner. For these $L_2$-distances we obtain accuracies of 55.0\% for the static setting, 80.0\% for the dynamic setting and 77.5\% for the classifier based on both distances. Hence the elastic distance performs better for all three classifiers. 

\begin{figure}[!ht]\centering
\includegraphics[scale = 0.7]{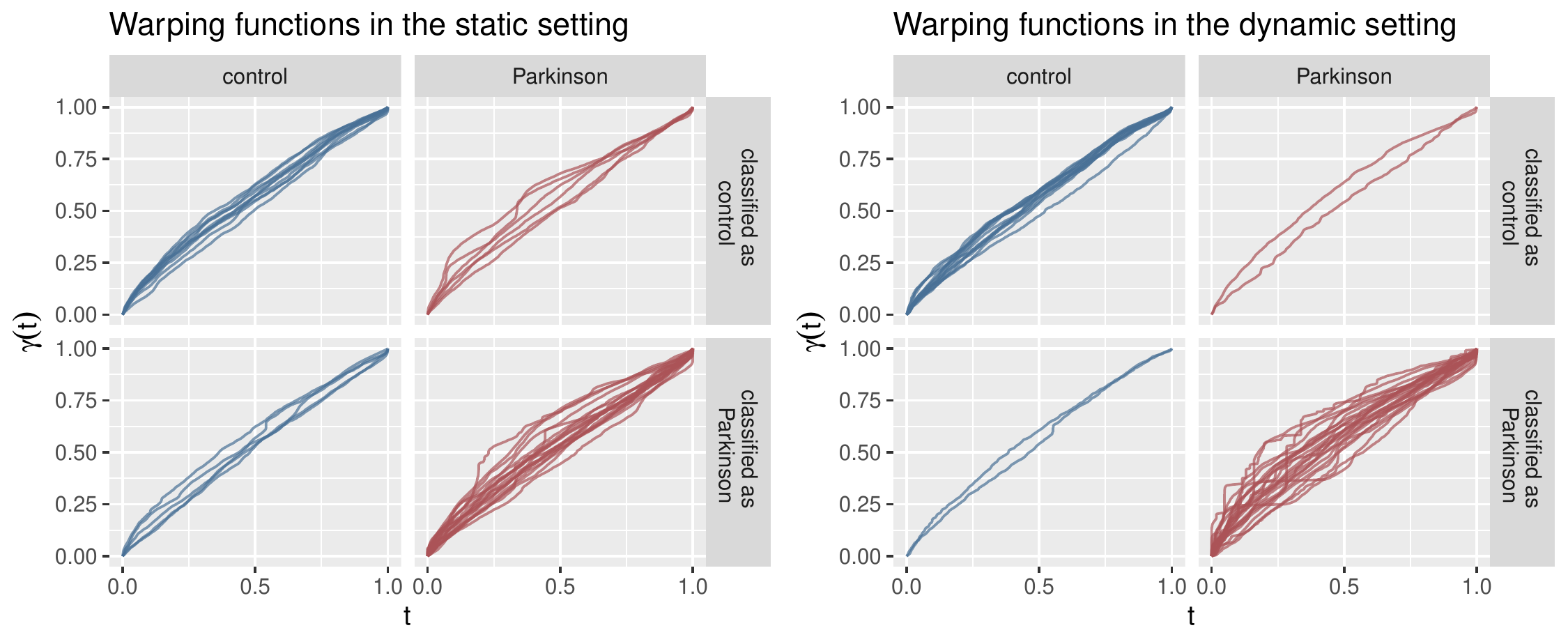}
\caption{\label{fig:parkinson_warping} Optimal warping in both settings separated by the actual status and the predicted status using the classifiers based on only the corresponding distance each and leave-one-out cross-validation.}
\end{figure}

Elastic alignment of the observed curves to a template allows us to separate phase and amplitude variation. Our classifiers depend on the elastic distance, which means we rely only on the amplitude variation. To see if the phase, that is the temporal pattern, yields additional information compared to only the image, we look in Figure \ref{fig:parkinson_warping} at the warping functions separated according to the classification result. This comparison of real-time parametrisation to the parametrisation after alignment to the mean curve shows whether the speed patterns of patients with Parkinson's disease are dissimilar to those of healthy individuals.

Looking at the general pattern of the warping functions in both settings, we observe more deviation from a smooth speed pattern in the group of Parkinson's patients than in the control group. To decide whether this yields additional information to the elastic distance of the curve to the template, we further inspect the warping curves which belong to misclassified subjects. There are two Parkinson's patients with conspicuous speed patterns we misclassify as 'control' in the static setting. Their speed pattern shows starting and stopping motions, which is not present in the curves of any of the healthy control subjects. Contrarily, we do not observe any noticeably different speed pattern for the misclassified individuals in the dynamic setting. Here the image of the curve seems to capture all available information on the status of the participant.

In conclusion, the elastic distance of the curve drawn by the patient to a template curve is an intuitive measurement of performance for both the static and the dynamic spiral drawing test. Using this feature for classification, we mimic and objectify the medical diagnosis process of a doctor. Our classifier in particular performs well for the dynamic spiral test, as the struggle of the Parkinson's patients to follow the curve is captured in the image of the curve here. If more data were available, maybe even from patients with differing but related neurological conditions like essential tremor, it might also be beneficial to analyse the whole aligned curves and not only use there distance to the template, or to additionally analyse the warping functions, which our approach allows to separate from the images. 

%% file: 4_discussion.tex
\section{Discussion}
\label{sec:discussion}
The SRV framework has been developed to analyse curves in $\mathbb{R}^d$ without taking their parametrisation into account (elastic analysis). Analogously to developments in functional data analysis, these methods were at first targeted at densely observed curves. Since curves are usually observed on a discrete grid in statistical practice, existing methods (as in Srivastava and Klassen\cite{srivastava_book}) relied on discretising the warping functions and on interpolation of the curves. However, this approach has limitations if curves
 are sparsely observed. The main contribution of our work is to address the discrete and often sparse nature of observed curves explicitly, and to go beyond the pairwise alignment of curves to develop statistical elastic analysis methods for samples of irregularly or sparsely observed curves in $\mathbb{R}^d$. 

To do so, we proposed to interpret observed curves as polygons with constant speed parametrisation between the corners to make the alignment problem accessible, either between two such curves or between an observed curve and a model-based curve such as a mean curve. We suggested using splines on SRV level for modelling the mean, either piecewise linear splines, leading to smooth mean curves, or piecewise constant splines, implying polygonal mean curves, and developed corresponding estimation algorithms for open and closed curves. Our approach for elastic mean estimation does not need to interpolate the discretely observed curves as other approaches developed for densely observed curves do, but directly approximates the integral that appears in the corresponding optimisation problem. 
However, since polygons underestimate the curvature of the real unobserved curves, the polygonal assumption does lead to a kind of shrinkage bias for the estimated elastic mean for sparsely observed curves. While this bias towards curves with smaller curvature decreases with increasing observations per curve, it would be of interest to develop correction methods for (very) sparse settings in future work.

We have shown that the SRV splines modulo parametrisation used for modelling the elastic mean are in general identifiable via their coefficients and we confirmed this result in simulations. While we did not explicitly address the choice of the optimal number of knots for such splines, 
a further simulation has shown that the estimation of the mean curve is not sensitive to the specific spline degree and choice of knots, given the number of knots is sufficiently large.
It may be interesting in the future to investigate penalised estimation with a large number of spline basis functions, although the interpretation of coefficients and identifiability modulo parametrisation  would need to be studied in this setting.

Another appealing direction for further research is to include our methods for sparsely and irregularly sampled curves in existing approaches for functional shape analysis. Here the curves have to be aligned with respect to scaling and/or rotation in addition to the alignment with respect to parametrisation and translation. Since this is usually done iteratively, it seems promising to combine this with the iterative  warping and mean fitting steps in our methods. Furthermore, elastic mean estimation for irregularly and/or sparsely sampled curves can be seen as a first step towards elastic regression models for such data.  That means our methods might be useful building blocks for modelling curves or shapes depending on continuous and/or discrete covariates.

%% file: proofs_computations.tex
\subsection{Proofs and Computations}
In this part of the appendix we provide proofs to all statements presented in Section \ref{sec:methods}.
\subsubsection{Proof of Lemma \ref{lem:warp_prob_discrete}}
\label{app:warp_prob_discrete}
\begin{proof2}
To calculate the elastic distance between two square root velocity curves $\mathbf{p,q}:[0,1] \to \R^d$ one has to consider the following minimisation problem.
\begin{align*}
\textbf{Minimise } &\quad \int_0^1 \| \mathbf{p}(t) - \mathbf{q}(\gamma(t))\sqrt{\dot{\gamma}(t)} \|^2 \ dt \\
\textbf{w.r.t. } &\quad \gamma:[0,1] \to [0, 1] \text{ monotonically increase increasing, onto and differentiable.}
\end{align*}
The objective function can be written as
\begin{align*}
\int_0^1 \| \mathbf{p}(t) - \mathbf{q}(\gamma(t))\sqrt{\dot{\gamma}(t)} \|^2 \ dt 
&= \int_0^1 \|\mathbf{p}(t) \|^2 \ dt - 2 \int_0^1 \langle \mathbf{p}(t), \mathbf{q}(\gamma(t)) \rangle \sqrt{\dot{\gamma}(t)}  \ dt +  \int_0^1 \| \mathbf{q}(\gamma(t)) \|^2  \dot{\gamma}(t) \ dt \\
&= \| \mathbf{p} \|_{L_2}^2 -2\int_0^1 \langle \mathbf{p}(t), \mathbf{q}(\gamma(t)) \rangle \sqrt{\dot{\gamma}(t)}  \ dt + \| \mathbf{q} \|_{L_2}^2.
\end{align*}
Hence the minimisation problem stated above is equivalent to
\begin{align*}
\textbf{Maximise } &\quad \int_0^1 \langle \mathbf{p}(t), \mathbf{q}(\gamma(t)) \rangle \sqrt{\dot{\gamma}(t)}  \ dt \\
\textbf{w.r.t. } &\quad \gamma:[0,1] \to [0, 1]\  \text{monotonically increasing, onto and differentiable.}
\end{align*}

We assume that $\mathbf{q}$ is the square root velocity curve of a polygon (for example a polygon with observations at its corners). Hence $\mathbf{q}$ is piecewise constant, which means there exist time points $0 = s_0 < s_1 <  \dots < s_{m -1} < s_m = 1$ such that
$\mathbf{q}|_{[s_j, s_{j+1}]} = \mathbf{q}_j \in \R^d$  for all $j = 0, \dots, m -1$.
Since $\gamma$ is increasing and onto, this gives time points $0 = t_0 < \dots < t_m = 1$ such that $\gamma(t_j) = s_j$ for all $j = 1, \dots, m$. Hence the optimisation problem becomes equivalently
\begin{align*}
\textbf{Maximise } &\quad \sum_{j = 0}^{m - 1} \int_{t_j}^{t_{j+1}} \langle \mathbf{p}(t), \mathbf{q}_j \rangle \sqrt{\dot{\gamma}(t)}  \ dt \\
\textbf{w.r.t. } &\quad \gamma:[0,1] \to [0, 1]\  \text{monotonically increasing, onto, differentiable}\\
&\quad \text{and } \gamma(t_j) = s_j \ \forall j = 1, \dots, m - 1.
\end{align*}
We can split this optimisation problem into an outer maximisation over $t_1, \dots, t_{m-1}$ and an inner one, where for fixed $j = 0, \dots, m-1$, the following maximisation problem needs to be solved.
\begin{align}
\label{eq:split_opti}
\textbf{Maximise } &\quad \int_{t_j}^{t_{j+1}} \langle \mathbf{p}(t), \mathbf{q}_j \rangle \sqrt{\dot{\gamma}(t)}  \ dt \\
\textbf{w.r.t. } &\quad \dot{\gamma}:[t_j, t_{j+1}] \to \R_0^+ \text{ and } \int_{t_j}^{t_{j+1}} \dot{\gamma}(t) \ dt = s_{j+1} - s_j. \nonumber
\end{align}
We obtain an upper bound for these objective functions using the Cauchy-Schwarz inequality. We have
\begin{align} \nonumber
\int_{t_j}^{t_{j+1}} \langle \mathbf{p}(t), \mathbf{q}_j \rangle \sqrt{\dot{\gamma}(t)}  \ dt  
&\leq \int_{t_j}^{t_{j+1}} \langle \mathbf{p}(t), \mathbf{q}_j \rangle_+ \sqrt{\dot{\gamma}(t)} \ dt \\ \nonumber
&\overset{\small C.S.}{\leq} \sqrt{ \int_{t_j}^{t_{j+1}} \langle \mathbf{p}(t), \mathbf{q}_j \rangle_+^2 \ dt} \sqrt{\int_{t_j}^{t_{j+1}}  \dot{\gamma}(t) \ dt} \\ \label{ineq:warping}
&= \sqrt{ (s_{j+1} - s_j) \int_{t_j}^{t_{j+1}} \langle \mathbf{p}(t), \mathbf{q}_j \rangle^2_+ \ dt}
\end{align}
To show this upper bound is actually the supremum over all feasible functions $\dot{\gamma}$ we consider two distinct cases.

\begin{itemize}
\item[i)] If $\int_{t_j}^{t_{j+1}} \langle \mathbf{p}(t), \mathbf{q}_j \rangle_+^2 \ dt > 0$ we can choose 
\begin{align}
\label{eq:warp_fun}
\dot{\gamma}(t) = \frac{(s_{j+1} - s_j)\langle \mathbf{p}(t), \mathbf{q}_j \rangle_+^2}{\int_{t_j}^{t_{j+1}} \langle \mathbf{p}(t), \mathbf{q}_j \rangle_+^2 \ dt}.
\end{align} 
This choice of $\dot{\gamma}$ is feasible as it attains only non-negative values and $\int_{t_j}^{t_{j + 1}} \dot{\gamma}(t) \ dt = s_{j + 1} - s_{j}$ for all $j = 0, \dots, m-1$. We calculate
\begin{align*}
\int_{t_j}^{t_{j+1}} \langle \mathbf{p}(t), \mathbf{q}_j \rangle \sqrt{\dot{\gamma}(t)}  \ dt 
&= \int_{t_j}^{t_{j+1}} \langle \mathbf{p}(t), \mathbf{q}_j \rangle \frac{ \sqrt{s_{j+1} - s_j}\langle \mathbf{p}(t), \mathbf{q}_j \rangle_+}{\sqrt{\int_{t_j}^{t_{j+1}} \langle \mathbf{p}(t), \mathbf{q}_j \rangle_+^2 \ dt}}  \ dt \\
&=\frac{\sqrt{s_{j+1} - s_j}}{\sqrt{\int_{t_j}^{t_{j+1}} \langle \mathbf{p}(t), \mathbf{q}_j \rangle_+^2 \ dt}} \int_{t_j}^{t_{j+1}} \langle \mathbf{p}(t), \mathbf{q}_j \rangle \langle \mathbf{p}(t), \mathbf{q}_j \rangle_+  \ dt \\
&= \sqrt{s_{j+1} - s_j} \sqrt{\int_{t_j}^{t_{j+1}} \langle \mathbf{p}(t), \mathbf{q}_j \rangle_+^2 \ dt},
\end{align*}
where the last equality is due to $\langle \mathbf{p}(t), \mathbf{q}_j \rangle \langle \mathbf{p}(t), \mathbf{q}_j \rangle_+ = \langle \mathbf{p}(t), \mathbf{q}_j \rangle_+^2$, since $\langle \mathbf{p}(t), \mathbf{q}_j \rangle < 0$ implies $\langle \mathbf{p}(t), \mathbf{q}_j \rangle_+ = 0$. Hence $\dot{\gamma}$ is a maximising function.

\item[ii)] If $\int_{t_j}^{t_{j+1}} \langle \mathbf{p}(t), \mathbf{q}_j \rangle_+^2 \ dt = 0$ the objective function is bounded above by 0 due to \eqref{ineq:warping} and we construct a sequence $(\dot{\gamma_k})_{k \in \N}$ of feasible functions to reach that upper bound. For all $k \in \N$ let
\begin{align*}
\dot{\gamma_k} = (s_{j + 1} - s_j) k \1_{[t_j, t_j + \frac{1}{k}]} \geq 0. 
\end{align*}
Hence we have for sufficiently large $k \in \N$
\begin{align*}
 \int_{t_j}^{t_{j + 1}} \dot{\gamma_k}(t) \ dt = (s_{j + 1} - s_{j})  \int_{t_j}^{t_j + \frac{1}{k}} k \ dt = s_{j + 1} - s_{j},
\end{align*}
which shows that the functions $\dot{\gamma}_k$ are feasible for $k \geq \frac{1}{t_{j + 1} - t_j}$.

Since $\| \mathbf{p} \|_\infty < \infty $ we have for sufficiently large $k \in \N$
\begin{align*}
\left| \int_{t_j}^{t_{j+1}} \langle \mathbf{p}(t), \mathbf{q}_j \rangle \sqrt{\dot{\gamma_k}(t)}  \ dt \right| 
&\leq  \int_{t_j}^{t_{j+1}} \left| \langle \mathbf{p}(t), \mathbf{q}_j \rangle \right| \sqrt{\dot{\gamma_k}(t)}  \ dt \\
&\leq  \int_{t_j}^{t_{j+1}} \| \mathbf{p}(t) \| \| \mathbf{q}_j \| \sqrt{\dot{\gamma_k}(t)}  \ dt \\
&\leq \| \mathbf{p} \|_\infty \| \mathbf{q}_j \| \int_{t_j}^{t_j + \frac{1}{k}}  \sqrt{(s_{j + 1} - s_j) k}  \ dt \\
&= \| \mathbf{p} \|_\infty \| \mathbf{q}_j \| \sqrt{s_{j + 1} - s_{j}} \frac{\sqrt{k}}{k} \quad\overset{k \to \infty}{\xrightarrow{\hspace{1.2cm}}} 0.
\end{align*}
This shows that $(\dot{\gamma_k})$ is a maximising sequence of warping functions since
\begin{align*}
0 \geq \int_{t_j}^{t_{j+1}} \langle \mathbf{p}(t), \mathbf{q}_j \rangle \sqrt{\dot{\gamma_k}(t)} \ dt \overset{k \to \infty}{\longrightarrow} 0.
\end{align*}
In this cases, we do not find a maximising warping function $\gamma$ but a sequence of maximising warping functions $\gamma_k$.
\end{itemize}
In both cases i) and ii), the inner optimisation \eqref{eq:split_opti} takes the value $\sqrt{ (s_{j+1} - s_j) \int_{t_j}^{t_{j+1}} \langle \mathbf{p}(t), \mathbf{q}_j \rangle^2_+ \ dt}$ for given $j = 0, \dots, m-1$. The overall optimisation thus becomes the outer optimisation over the sum of these terms with respect to $t_1, \dots, t_{m-1}$, i.e. takes the form \eqref{eq:warp_prob_discrete}.
\end{proof2}

\subsubsection{Gradient of the loss function in Lemma \ref{lem:warp_prob_discrete}}
\label{app:gradient}
The simplified loss function given in Lemma \ref{lem:warp_prob_discrete},
\begin{align*}
\Phi(\mathbf{t}) = \Phi(t_1, \dots, t_{m-1}) = \sum_{j = 0}^{m-1} \sqrt{ (s_{j+1} - s_j) \int_{t_j}^{t_{j+1}} \langle \mathbf{p}(t), \mathbf{q}_j \rangle^2_+ \ dt}
\end{align*}
is differentiable if $\mathbf{p}$ is at least continuous. In this case the partial derivatives can be computed as
\begin{align*}
\frac{\partial}{\partial t_j} \Phi(\mathbf{t}) 
&= \frac{\partial}{\partial t_j} 
\sum_{k = 0}^{m-1} \sqrt{ (s_{k+1} - s_k) \int_{t_k}^{t_{k+1}} \langle \mathbf{p}(t), \mathbf{q}_k \rangle^2_+ \ dt} \\
&= \frac{\partial}{\partial t_j} \sqrt{ (s_{j} - s_{j-1}) \int_{t_{j-1}}^{t_{j}} \langle \mathbf{p}(t), \mathbf{q}_{j - 1} \rangle^2_+ \ dt} +
\frac{\partial}{\partial t_j}
\sqrt{ (s_{j+1} - s_j) \int_{t_j}^{t_{j+1}} \langle \mathbf{p}(t), \mathbf{q}_j \rangle^2_+ \ dt} \\
&= \frac{\frac{1}{2}(s_{j} - s_{j-1}) \langle \mathbf{p}(t_j), \mathbf{q}_{j - 1} \rangle^2_+}
{\sqrt{ (s_{j} - s_{j-1}) \int_{t_{j-1}}^{t_{j}} \langle \mathbf{p}(t), \mathbf{q}_{j - 1} \rangle^2_+ \ dt}} -
\frac{\frac{1}{2} (s_{j+1} - s_j) \langle \mathbf{p}(t_j), \mathbf{q}_j \rangle^2_+}
{\sqrt{ (s_{j+1} - s_j) \int_{t_j}^{t_{j+1}} \langle \mathbf{p}(t), \mathbf{q}_j \rangle^2_+ \ dt}}  \\
&=\frac{1}{2} \left( \frac{ \sqrt{s_{j} - s_{j-1}} \langle \mathbf{p}(t_j), \mathbf{q}_{j - 1} \rangle^2_+}
{\sqrt{ \int_{t_{j-1}}^{t_{j}} \langle \mathbf{p}(t), \mathbf{q}_{j - 1} \rangle^2_+ \ dt}} -
\frac{ \sqrt{s_{j+1} - s_j} \langle \mathbf{p}(t_j), \mathbf{q}_j \rangle^2_+}
{\sqrt{ \int_{t_j}^{t_{j+1}} \langle \mathbf{p}(t), \mathbf{q}_j \rangle^2_+ \ dt}} \right)
\end{align*}
for all $j = 1, \dots, m-1$. If $\mathbf{p}$ is piecewise linear, $t \mapsto \langle \mathbf{p}(t), \mathbf{q}_j \rangle^2_+$ is piecewise quadratic and one can compute the integral in the denominator exactly.

\subsubsection{Closed form solution for the coordinate wise maximisation needed in Algorithm \ref{algo:dist_open_curves}}
\label{app:closed_form_max}
For fixed $j \in \{1, \dots, m-1\}$ and fixed $0 = t_0 \leq \dots \leq t_{j-1} \leq t_{j+1} \leq \dots \leq t_m = 1$ we need to solve
\begin{align} \label{eq:warp_prob_coord}
\textbf{Maximise } &\quad L(t_j) =  \sqrt{ (s_{j} - s_{j-1}) \int_{t_{j-1}}^{t_j} \langle \mathbf{p}(t), \mathbf{q}_{j -1} \rangle^2_+ \ dt} +
\sqrt{ (s_{j+1} - s_j) \int_{t_j}^{t_{j+1}} \langle \mathbf{p}(t), \mathbf{q}_j \rangle^2_+ \ dt} \\
\textbf{w.r.t } &\quad t_{j-1} \leq t_j \leq t_{j + 1}. \nonumber
\end{align}

Since $\mathbf{p}$ is assumed to be piecewise constant on $[t_{j-1}, t_{j+1}]$ there exists $t_{j-1} = r_0 < \dots < r_l = t_{j+1}$ such that $\mathbf{p}|_[r_\iota, r_{\iota +1}[ = \mathbf{p}_\iota \in \R^d$ for all $\iota = 0, \dots, l-1$. Hence the objective function restricted to $[r_\iota, r_{\iota + 1}[$ can be written as
\begin{align*}
L|_{[r_\iota, r_{\iota + 1}]}(t_j) =& \sqrt{ (s_{j} - s_{j-1}) \left((t_j - r_{\iota})\langle \mathbf{p}_\iota, \mathbf{q}_{j -1} \rangle^2_+ + \sum_{k = 0}^{\iota - 1} (r_{k + 1} - r_{k})\langle \mathbf{p}_k, \mathbf{q}_{j -1} \rangle^2_+ \right)}\\
&+
\sqrt{ (s_{j+1} - s_j) \left((r_{\iota + 1} - t_j)\langle \mathbf{p}_\iota, \mathbf{q}_{j} \rangle^2_+ + \sum_{k = \iota + 1}^{l - 1} (r_{k + 1} - r_{k})\langle \mathbf{p}_k, \mathbf{q}_{j} \rangle^2_+ \right)}.
\end{align*}
This shows that for all $\iota = 0, \dots, l-1$ there are constant values 
\begin{align*}
A_{\iota 1} &= (s_{j} - s_{j-1})\langle \mathbf{p}_\iota, \mathbf{q}_{j -1} \rangle^2_+ \\
A_{\iota 2} &= (s_{j+1} - s_{j})\langle \mathbf{p}_\iota, \mathbf{q}_{j} \rangle^2_+ \\
B_{\iota 1} &= (s_{j} - s_{j-1}) \left(r_{\iota}\langle \mathbf{p}_\iota, \mathbf{q}_{j -1} \rangle^2_+ - \sum_{k = 0}^{\iota - 1} (r_{k + 1} - r_{k})\langle \mathbf{p}_k, \mathbf{q}_{j -1} \rangle^2_+ \right) \\
B_{\iota 2} &= (s_{j + 1} - s_{j}) \left(r_{\iota + 1}\langle \mathbf{p}_\iota, \mathbf{q}_{j} \rangle^2_+ + \sum_{k = \iota + 1}^{l - 1} (r_{k + 1} - r_{k})\langle \mathbf{p}_k, \mathbf{q}_{j} \rangle^2_+ \right)
\end{align*}
such that
\begin{align*}
L|_{[r_\iota, r_{\iota + 1}]}(t_j) = \sqrt{A_{\iota 1}t_j - B_{\iota 1}} + \sqrt{B_{\iota2} - A_{\iota2}t_j}
\end{align*}
with $A_{\iota1}t_j - B_{\iota1} \geq 0$ and $B_{\iota2} - A_{\iota2}t_j \geq 0$ for all $t_j \in [r_\iota, r_{\iota + 1}]$. Without loss of generality we assume $A_{\iota1}, A_{\iota2} > 0$ since otherwise the objective function is monotonic, hence attains its maximum on the boundary and this case can be included separately below. Thus $L|_{[r_\iota, r_{\iota + 1}]}$ is twice continuously differentiable on $]r_\iota, r_{\iota + 1}[$ with
\begin{align*}
\frac{\partial}{\partial t_j} L|_{[r_\iota, r_{\iota + 1}]}(t_j) &= 
\frac{1}{2} \left( \frac{A_{\iota1}}{\sqrt{A_{\iota1}t_j - B_{\iota1}}} - \frac{A_{\iota2}}{\sqrt{B_{\iota2} - A_{\iota2}t_j}} \right), \\
\frac{\partial^2}{\partial t_j^2} L|_{[r_\iota, r_{\iota + 1}]}(t_j) &= 
-\frac{1}{4} \left( \frac{A_{\iota1}^2}{\sqrt{A_{\iota1}t_j - B_{\iota1}}^3} + \frac{A_{\iota2}^2}{\sqrt{B_{\iota2} - A_{\iota2}t_j}^3} \right) < 0.
\end{align*}
Therefore, every maximiser $t_j$ within $]r_\iota, r_{\iota + 1}[$ fullfills
\begin{align*}
\frac{A_{\iota1}}{\sqrt{A_{\iota1}t_j - B_{\iota1}}} = \frac{A_{\iota2}}{\sqrt{B_{\iota2} - A_{\iota2}t_j}}
\quad \Leftrightarrow &\quad A_{\iota1}^2 (B_{\iota2} - A_{\iota2}t_j) = A_{\iota2}^2 (A_{\iota1}t_j - B_{\iota1}) \\
\Leftrightarrow &\quad t_j = \frac{A_{\iota1}^2B_{\iota2} + A_{\iota2}^2 B_{\iota1}}{A_{\iota1}A_{\iota2}^2 + A_{\iota1}^2A_{\iota2}}.
\end{align*}
We conclude that every solution to the coordinate wise maximisation problem (\ref{eq:warp_prob_coord}) is contained in the set \begin{align*}
\bigcup_{\iota = 0}^{l} \{r_\iota\} \cup \bigcup_{\iota = 0}^{l -1} \lbrace \frac{A_{\iota1}^2B_{\iota2} + A_{\iota2}^2 B_{\iota1}}{A_{\iota1}A_{\iota2}^2 + A_{\iota1}^2A_{\iota2}} \rbrace
\end{align*}
and can compare function values of $L$ over this set to find the maximiser.

\subsubsection{Proof of Theorem \ref{theo:local_max}}
\label{app:local_max}
\begin{proof2} Let $\Phi$ be defined as in Equation (\ref{eq:warp_prob_discrete}),
\begin{align*} 
\Phi(\mathbf{t}) = \Phi(t_1, \dots, t_{m-1}) = \sum_{j = 0}^{m-1} \sqrt{ (s_{j+1} - s_j) \int_{t_j}^{t_{j+1}} \langle \mathbf{p}(t), \mathbf{q}_j \rangle^2_+ \ dt},
\end{align*}
with $\mathbf{p}$ being piecewise constant. Furthermore let $(\mathbf{t}^{(\iota)})_{\iota \in \mathbb{N}} = t^{(1)}, t^{(2)}, \dots $ be a sequence resulting from Algorithm \ref{algo:dist_open_curves} and $\mathbf{t}^*$ an accumulation point of $(\mathbf{t}^{(\iota)})_{\iota \in \mathbb{N}}$.

We proof this main result in three steps. First, we show that the accumulation point $\mathbf{t}^* = (t_1^*, \dots, t_{m-1}^*)$ is a maximiser of $\Phi$ restricted to coordinate directions. Then we conclude that $\Phi$ is semi-differentiable at $\mathbf{t}^*$ for every direction $\mathbf{u} \in \mathbb{R}^{m-1}$. Last we use Lemma \ref{lem:local_concave} below, which establishes local concavity of the loss function, to see that $\mathbf{t}^*$ is a local maximum of $\Phi$.

Since $\mathbf{t}^*$ is an accumulation point, there is a subsequence $(\mathbf{t}^{(\iota_k)})_{k \in \mathbb{N}}$ with $\displaystyle \lim_{k \to \infty} \mathbf{t}^{(\iota_k)} = \mathbf{t}^*$. Denote by
\begin{align*}
\Phi_{odd}^{(k)} &:= \Phi|_{ \{ t_j = t_j^{(\iota_k)}, \ j \text{ even} \} } \\
\Phi_{even}^{(k)} &:= \Phi|_{ \{ t_j = t_j^{(\iota_k)}, \ j \text{ odd} \} }
\end{align*}
the restrictions of $\Phi$ at the current sequence value with either fixed odd or even coordinate entries. $\Phi$ is continuous, hence we have point-wise limits
\begin{align*}
\lim_{k \to \infty} \Phi_{odd}^{(k)}  &= \Phi|_{ \{ t_j = t_j^*, \ j \text{ even} \} } =: \Phi_{odd}^*, \\
\lim_{k \to \infty} \Phi_{even}^{(k)}  &= \Phi|_{ \{ t_j = t_j^*, \ j \text{ odd} \} } =: \Phi_{even}^*,
\end{align*}
with $\Phi_{odd}^*, \Phi_{even}^*$ being the restrictions to odd and even coordinate directions at the accumulation point $\mathbf{t}^*$. Since at each step we either update all odd or all even entries, $\Phi_{odd}^{(k)}$ and $\Phi_{even}^{(k)}$ attain their maximum at either the current or the next sequence value. That is
\begin{align*}
\left\| \Phi_{odd}^{(k)} \right\|_\infty, \left\| \Phi_{even}^{(k)} \right\|_\infty \in 
\lbrace \Phi(\mathbf{t}^{(\iota_k)}), \Phi(\mathbf{t}^{(\iota_k + 1)})  \rbrace
\end{align*}
for all $k \in \mathbb{N}$. Thus, $\Phi_{odd}^*$ and $\Phi_{even}^*$ are bounded as well:
\begin{align*}
\left\| \Phi_{odd}^* \right\|_\infty = \lim_{k \to \infty} \left\| \Phi_{odd}^{(k)} \right\|_\infty \leq \lim_{k \to \infty} \Phi(\mathbf{t}^{(\iota_k + 1)}) = \Phi(\mathbf{t}^*),
\end{align*}
since $\displaystyle \lim_{\iota \to \infty} \Phi(t^{(\iota)}) = \Phi(\mathbf{t}^*)$. We can conclude this as coordinate-wise maximisation produces a monotonically increasing sequence $\Phi(\mathbf{t}^{(\iota + 1)}) \geq \Phi(\mathbf{t}^{(\iota)})$ for all $\iota \in \mathbb{N}$ and the subsequence $\Phi(\mathbf{t}^{(\iota_k )})$ converges to $\Phi(\mathbf{t}^*)$ due to $\Phi$ being continuous, which implies the whole sequence converges.
Analogously we have $\displaystyle \left\| \Phi_{even}^* \right\|_\infty  \leq \Phi(\mathbf{t}^*)$, hence $\mathbf{t}^*$ is a maximiser of $\Phi$ restricted to any coordinate direction (i.e.\ $t_j^*$ maximises $\Phi(t_1^*, \dots, t_j, \dots, t_{m-1}^*)$ over $t_j$ for all $j = 1, \dots, m-1$).

To show that this implies that $\Phi$ is partially semi-differentiable at $\mathbf{t}^*$ first note that $\Phi$ is partially semi-differentiable at every point $\mathbf{t} = (t_1, \dots, t_{m-1})$ with 
$(s_{j+1} - s_j) \int_{t_j}^{t_{j+1}} \langle \mathbf{p}(t), \mathbf{q}_j \rangle^2_+ \ dt > 0$ for all $j = 1, \dots, m-1$, since the square-root function is differentiable for strictly positive values and $\int_{t_j}^{t_{j+1}} \langle \mathbf{p}(t), \mathbf{q}_j \rangle^2_+ \ dt$ is piecewise linear, thus semi-differentiable.\\
Assume there is a $j \in \{1, \dots, m-1\}$ with $(s_{j+1} - s_j) \int_{t^*_j}^{t^*_{j+1}} \langle \mathbf{p}(t), \mathbf{q}_j \rangle^2_+ \ dt = 0$. We show that $\Phi$ is still partially semi-differentiable at $\mathbf{t}^*$ in direction $t_j$. A similar argument shows differentiability in direction $t_{j + 1}$.\\
Let $L$ be the relevant part of the loss function $\Phi$ in direction $t_j$.
\begin{align*}
L(t_j) = \sqrt{ (s_{j} - s_{j-1}) \int_{t^*_{j-1}}^{t_j} \langle \mathbf{p}(t), \mathbf{q}_{j -1} \rangle^2_+ \ dt} +
\sqrt{ (s_{j+1} - s_j) \int_{t_j}^{t^*_{j+1}} \langle \mathbf{p}(t), \mathbf{q}_j \rangle^2_+ \ dt}
\end{align*}
We need to show that both, left and right derivatives of $L$ at $t_j^*$ exist.
\begin{itemize}
\item If $(s_{j} - s_{j-1}) \int_{t^*_{j-1}}^{t^*_{j}} \langle \mathbf{p}(t), \mathbf{q}_{j-1} \rangle^2_+ \ dt = 0$,\\
we have $L(t_j^*) = 0$. This implies $L(t_j) = 0$ for all $t_j \in [t_{j -1}^*, t_{j + 1}^*]$ since $t_j^*$ is a maximiser (in $t_j$ coordinate direction) and $L$ is non-negative. Therefore $L = 0$ which means $L$ is differentiable on its whole domain.
\item If $(s_{j} - s_{j-1}) \int_{t^*_{j-1}}^{t^*_{j}} \langle \mathbf{p}(t), \mathbf{q}_{j-1} \rangle^2_+ \ dt > 0$,\\
the left summand of $L$ is strictly positive in a neighbourhood of $t_j^*$ and consequently semi-differentiable in a neighbourhood of $t_j^*$. The right summand $H(t_j) := \sqrt{ (s_{j+1} - s_j) \int_{t_j}^{t^*_{j+1}} \langle \mathbf{p}(t), \mathbf{q}_j \rangle^2_+ \ dt}$ is differentiable at $t_j^*$ since it is 0 in a neighbourhood of $t_j^*$. This is due to $t_j \mapsto (s_{j+1} - s_j) \int_{t_j}^{t^*_{j+1}} \langle \mathbf{p}(t), \mathbf{q}_j \rangle^2_+ \ dt$ being piecewise linear, non-negative and monotonically decreasing. Since it attains 0 at $t_j^*$, it is also 0 in a right neighbourhood of $t_j^*$. If $H$ were strictly positive in a neighbourhood left of $t_j^*$, its left derivative would tend to $- \infty$ at $t^*_j$ as $H(t_j^*)$ = 0 and the derivative of the square-root tends to $\infty$ for values tending linearly to 0. But $\frac{\partial_-}{\partial t_j} H(t_j^*) = -\infty$ would imply $\frac{\partial_-}{\partial t_j} L(t_j^*) = -\infty$, which contradicts $t_j^*$ being a maximiser.
\end{itemize}

Taking all those cases into account we conclude that $\Phi$ is partially semi-differentiable at the accumulation point $\mathbf{t}^*$ produced by coordinate-wise maximisation. Since we already know that $t^*_j$ is the coordinate-wise maximiser of $\Phi$ for all $j = 1, \dots, m-1$ in coordinate directions, the left-sided partial derivatives need to be non-negative, the right-sided partial derivatives non-positive.

To show that this implies that $\mathbf{t}^*$ is a local maximiser, consider sets $U = \bigtimes_{j = 1}^{m-1} U_j \cap \{0 \leq t_1 \leq \dots \leq t_{m-1} \leq 1\}$ such that $\mathbf{t}^* \in U$ and $\mathbf{p}$ is constant on the interior of the interval $U_j \neq \emptyset$ for all $j = 1, \dots, m-1$.

We prove that $\mathbf{t}^*$ is the maximiser of $\Phi|_U$ by contradiction. Assume there is a $\mathbf{u} \in U$ such that $\Phi(\mathbf{u}) > \Phi(\mathbf{t}^*)$. 
Let $\mathbf{\alpha}(s) = s\mathbf{u} + (1-s)\mathbf{t}^*$ for all $s \in [0,1]$. Since the square-root is improperly differentiable on $[0, \infty[$, with the derivative at 0 being $\infty$, this implies that $\Phi \circ \mathbf{\alpha}$ is improperly differentiable on $[0, 1]$ with 
\begin{align*}
\left( \Phi \circ \mathbf{\alpha} \right)'(s) = \langle \frac{\partial \Phi}{\partial \mathbf{t}} (\mathbf{\alpha}(s)), \mathbf{u} - \mathbf{t}^* \rangle = \sum_{j = 1}^{m-1} (u_j - t_j^*)\frac{\partial \Phi}{\partial t_j} (\mathbf{\alpha}(s)).
\end{align*}
Considering the limit $s \searrow 0$ yields
\begin{align*}
 \lim_{s \searrow 0} \frac{\partial \Phi}{\partial t_j} (\mathbf{\alpha}(s)) =
\begin{cases}
\frac{\partial_+ \Phi}{\partial t_j} (t_j^*) \leq 0 & \text{if } u_j - t_j^* > 0, \\
\frac{\partial_- \Phi}{\partial t_j} (t_j^*) \geq 0 & \text{if } u_j - t_j^* < 0.
\end{cases}
\end{align*}
Hence the right-sided derivative will be attained if $u_j - t_j^*$ is positive and the left-sided derivative if $u_j - t_j^*$ is negative. This implies $(u_j - t_j^*) \lim_{s \searrow 0} \frac{\partial \Phi}{\partial t_j} (\alpha(s)) \leq 0$ for all $j = 1, \dots, m - 1$ and therefore, 
\begin{align*}
\left( \Phi \circ \mathbf{\alpha} \right)'(0) = \lim_{s \searrow 0} \left( \Phi \circ \mathbf{\alpha} \right)'(s) = \sum_{j = 1}^{m-1} (u_j - t_j^*) \lim_{s \searrow 0} \frac{\partial \Phi}{\partial t_j} (\mathbf{\alpha}(s)) \leq 0.
\end{align*}
But since $U$ is a convex set, $\Phi$ is concave on the interior of $U$ (see Lemma \ref{lem:local_concave}) and therefore on $U$ as it is continuous and we compute
\begin{align*}
\left( \Phi \circ \mathbf{\alpha} \right)'(0) &= \lim_{s \searrow 0} \frac{(\Phi \circ \mathbf{\alpha})(s) - (\Phi \circ \mathbf{\alpha})(0)}{s} \\
&= \lim_{s \searrow 0} \frac{\Phi(s\mathbf{u} + (1 - s)\mathbf{t}^*) - \Phi(\mathbf{t}^*)}{s} \\
&\geq \lim_{s \searrow 0} \frac{s\Phi(\mathbf{u}) + (1 - s)\Phi(\mathbf{t}^*) - \Phi(\mathbf{t}^*)}{s} \\
&= \Phi(\mathbf{u}) - \Phi(\mathbf{t}^*) > 0,
\end{align*}
which contradicts $\left( \Phi \circ \mathbf{\alpha} \right)'(0) \leq 0$.

Thus, $\mathbf{t}^*$ is a maximum of $\Phi|_U$. This means it is a maximum on the union of such $U$'s, whose interior is a relatively open neighbourhood of $\mathbf{t}^*$ with respect to the relative topology on $\{0 \leq t_1 \leq \dots \leq t_{m-1} \leq 1\}$. Hence $\mathbf{t}^*$ is a local maximiser of $\Phi$.
\end{proof2}

\begin{lemma} \label{lem:local_concave}
Let $\Phi$ be the loss function defined in Equation (\ref{eq:warp_prob_discrete}), $\mathbf{p}$ piecewise constant and $U \subset \mathbb{R}^{m-1}$ a convex set such that $\mathbf{p}(t_j)$ is constant for all $j  = 1, \dots, m$ and all $(t_1, \dots, t_{m-1}) \in U$. Then $\Phi|_U$ is concave. 
\end{lemma}

\begin{proof}{} Note that $\Phi$ is twice continuously differentiable on the interior $\mathring{U}$ of $U$.
We show that all second directional derivatives $\partial^2_{\mathbf{u} \mathbf{u}} \Phi$ are non positive. This implies the Hessian $H$ is negative semi-definite, since $\mathbf{u}^T H \mathbf{u} =\partial^2_{\mathbf{u} \mathbf{u}} \Phi$ for all $\mathbf{u} \in \mathbb{R}^{m-1}$. Hence $\Phi|_U$ is concave.\\ 
To show the second derivative at $\mathbf{t} = (t_1, \dots, t_{m-1}) \in \mathring{U}$ is non-positive in any direction let $\alpha \in \mathbb{R}$ and $\mathbf{u} = (u_1, \dots, u_{m-1}) \in \mathbb{R}^{m-1}$. Define 
\begin{align*}
Q_j(\alpha) = (s_{j+1} - s_j) \int_{t_j + \alpha u_j}^{t_{j+1} + \alpha u_{j + 1}} \langle \mathbf{p}(t), \mathbf{q}_j \rangle^2_+ \ dt.
\end{align*}
$Q_j$ is linear around $\alpha = 0$ and therefore differentiable with constant derivative $Q_j'(\alpha) =: c_j \in \mathbb{R}$. 
If $Q_j(0) \neq 0$ for all $j \in 1, \dots, m-1$ we compute the directional derivative of the loss function as 
\begin{align*}
\partial_\mathbf{u} \Phi (\mathbf{t}) = \left. \frac{\partial}{\partial \alpha} \sum_{j = 1}^{m-1} \sqrt{Q_j(\alpha)} \right|_{\alpha = 0}
= \frac{1}{2} \left. \sum_{j = 1}^{m-1}\frac{Q_j'(\alpha)}{ \sqrt{Q_j(\alpha)}} \right|_{\alpha = 0}
= \frac{1}{2} \left. \sum_{j = 1}^{m-1} \frac{c_j}{ \sqrt{Q_j(\alpha)}} \right|_{\alpha = 0},
\end{align*}
and the second derivative becomes
\begin{align*}
\partial^2_{\mathbf{u}\mathbf{u}} \Phi (\mathbf{t}) = \left. \frac{\partial^2}{\partial \alpha^2} \sum_{j = 1}^{m-1} \sqrt{Q_j(\alpha)} \right|_{\alpha = 0}
= \left. - \frac{1}{4}\sum_{j = 1}^{m-1} \frac{c_j^2}{ \sqrt{Q_j(\alpha)}^3} \right|_{\alpha = 0} \leq 0.
\end{align*}
If $Q_j(0) = 0$ for some $j = , \dots, m-1$, we have in particular $\langle \mathbf{p}(t_j), \mathbf{q}_j \rangle^2_+ = 0$ and  $\langle \mathbf{p}(t_{j+1}), \mathbf{q}_j \rangle^2_+ = 0$, which means $Q_j$ is zero in a neighbourhood of $\alpha = 0$. Hence the second derivative of $\sqrt{Q_j(\alpha)}$ is zero as well and does not contribute to the sum.
\end{proof}

\subsubsection{Optimal warping and Fr\`echet means are not unique}
\label{app:non_unique}
We give an example that illustrates that both the optimal warping function minimising the elastic distance and the Fr\`echet mean for a set of curves with respect to the elastic distance are not necessarily unique. Consider two piecewise linear curves $\boldsymbol{\beta_1}$ and $\boldsymbol{\beta_2}$ with respective piecewise constant SRV curves $\mathbf{p}$ and $\mathbf{q}$ given as
\begin{align*}
\mathbf{p}(t) =
\begin{cases}
(-3,0)^T & \text{if } t \in [0, 0.25[ \\
(2, -4)^T & \text{if } t \in [0.25, 0.5[ \\
(-4, 2)^T & \text{if } t \in [0.5, 0.75[ \\
(0, -3)^T & \text{if } t \in [0.75, 1]
\end{cases}
\quad \text{and} \quad
\mathbf{q}(t) =
\begin{cases}
(-3,1)^T & \text{if } t \in [0, 0.5[ \\
(1, -3)^T & \text{if } t \in [0.5, 1]
\end{cases}
\end{align*}

\begin{figure}[ht]
\centering
\includegraphics[scale=0.65]{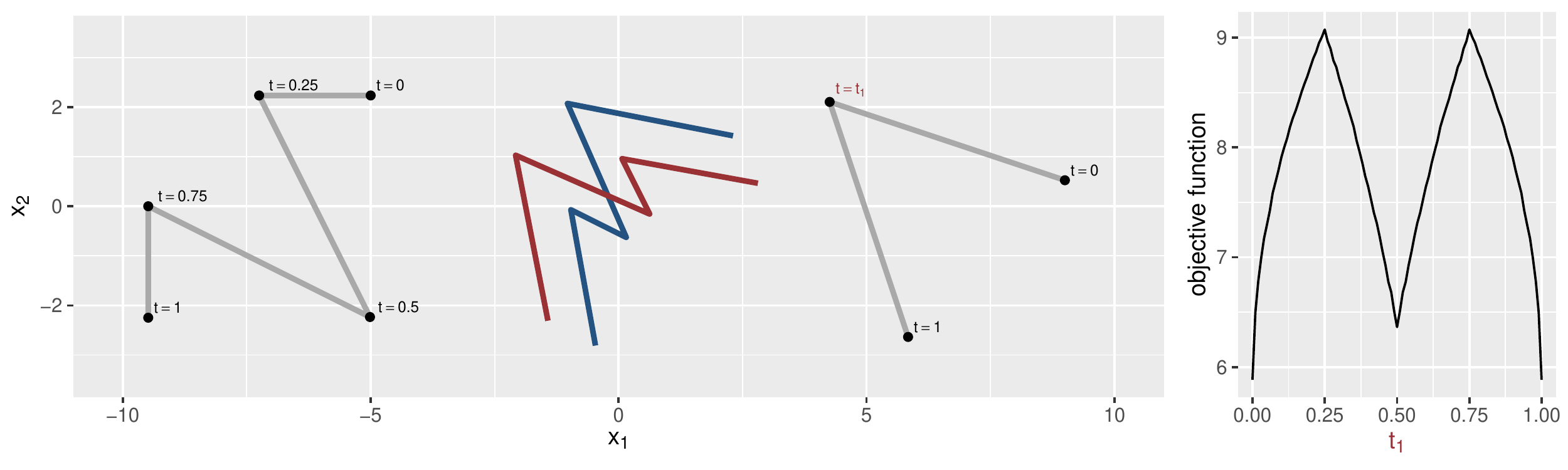}
\caption{\label{fig:non_unique}Left: Two piecewise linear curves in grey with Fr\`echet mean curves in red and blue.
Right: Objective function with two modes. Both maximisers $t_1 = 0.25$ and $t_1 = 0.75$ correspond to optimal warping functions.}
\end{figure}
The corresponding curves $\boldsymbol{\beta_1}$ and $\boldsymbol{\beta_2}$ are displayed in Figure \ref{fig:non_unique} on the left. The objective function $\Phi$, which needs to be maximised in order to find the optimal warping of the second curve to the first, only depends on one parameter $t_1$ and is given as
\begin{align*}
\Phi(t_1) = 
\sqrt{ 0.5 \int_{0}^{t_1} \langle \mathbf{p}(t), \begin{pmatrix} -3 \\ 1 \end{pmatrix} \rangle^2_+ \ dt} + 
\sqrt{ 0.5 \int_{t_1}^{1} \langle \mathbf{p}(t), \begin{pmatrix} 1 \\ -3 \end{pmatrix} \rangle^2_+ \ dt},
\end{align*}
where 
\begin{align}
\label{eq:sym_scalar_prod}
\langle \mathbf{p}(t), \begin{pmatrix} -3 \\ 1 \end{pmatrix} \rangle^2_+ = \begin{cases}
9^2 & \text{if } t \in [0, 0.25[ \\
0 & \text{if } t \in [0.25, 0.5[ \\
14^2 & \text{if } t \in [0.5, 0.75[ \\
0 & \text{if } t \in [0.75, 1]
\end{cases} \text{ and }
\langle \mathbf{p}(t), \begin{pmatrix} 1 \\ -3 \end{pmatrix} \rangle^2_+ = \begin{cases}
0 & \text{if } t \in [0, 0.25[ \\
14^2 & \text{if } t \in [0.25, 0.5[ \\
0 & \text{if } t \in [0.5, 0.75[ \\
9^2 & \text{if } t \in [0.75, 1]
\end{cases}.
\end{align}
With this we compute
\begin{align*}
\Phi(1-t_1) &= 
\sqrt{ 0.5 \int^{1- t_1}_{0} \langle \mathbf{p}(t), \begin{pmatrix} -3 \\ 1 \end{pmatrix} \rangle^2_+ \ dt} + 
\sqrt{ 0.5 \int_{1 - t_1}^1 \langle \mathbf{p}(t), \begin{pmatrix} 1 \\ -3 \end{pmatrix} \rangle^2_+ \ dt} \\
&= 
\sqrt{ 0.5 \int_{t_1}^{1} \langle \mathbf{p}(1-t), \begin{pmatrix} -3 \\ 1 \end{pmatrix} \rangle^2_+ \ dt} + 
\sqrt{ 0.5 \int_{0}^{t_1} \langle \mathbf{p}(1-t), \begin{pmatrix} 1 \\ -3 \end{pmatrix} \rangle^2_+ \ dt} \\
&\overset{\eqref{eq:sym_scalar_prod}}{=} 
\sqrt{ 0.5 \int_{t_1}^{1} \langle \mathbf{p}(t), \begin{pmatrix} 1 \\ -3 \end{pmatrix} \rangle^2_+ \ dt} + 
\sqrt{ 0.5 \int_{0}^{t_1} \langle \mathbf{p}(t), \begin{pmatrix} -3 \\ 1 \end{pmatrix} \rangle^2_+ \ dt} \\
&= \Phi(t_1),
\end{align*}
which shows that $\Phi$ is symmetric around $0.5$.
Looking at the gradient of $\Phi$ given in Appendix \ref{app:gradient} we observe $\Phi'(t_1) > 0$ if $t_1 \in ]0, 0.25[ \cup ]0.5, 0.75[$ and $\Phi'(t_1) < 0$ if $t_1 \in ]0.25, 0.5[ \cup ]0.75, 1[$, which implies that both $t_1 = 0.25$ and $t_1 = 0.75$ are local maximisers and therefore global maximisers due to $\Phi$ being symmetric. For illustration of the objective function please refer to the right part of Figure \ref{fig:non_unique}. 

The two maximisers of $\Phi$ correspond to two different optimal warping functions $\gamma_1$ and $\gamma_2$ of $\boldsymbol{\beta_2}$ to $\boldsymbol{\beta_1}$.
For $t_1 = 0.25$ we obtain $\dot{\gamma}_1$ according to \eqref{eq:warp_fun} in Appendix \ref{app:warp_prob_discrete} as 
\begin{align*}
    \dot{\gamma}_1(t) = 
    \begin{cases}
    \frac{0.5\langle \mathbf{p}(t), ( -3, 1)^T \rangle^2_+}{\int_{t_j}^{t_{j+1}} \langle \mathbf{p}(t), ( -3, 1)^T \rangle^2_+ \ dt} &  \text{if } t \in [0, 0.25[ \\
    \frac{0.5\langle \mathbf{p}(t), ( 1, -3)^T \rangle^2_+}{\int_{t_j}^{t_{j+1}} \langle \mathbf{p}(t), ( 1, -3)^T \rangle^2_+ \ dt} &  \text{if } t \in [0.25, 1]
    \end{cases}
    = \begin{cases}
    \frac{0.5 \cdot 9^2}{0.25 \cdot 9^2} &  \text{if } t \in [0, 0.25[ \\
    \frac{0.5\langle \mathbf{p}(t), ( 1, -3)^T \rangle^2_+}{0.25 \cdot 14^2 + 0.25 \cdot 9^2} &  \text{if } t \in [0.25, 1]
    \end{cases}
\end{align*}

Therefore, $\dot \gamma_1(t)$ for $t_1 = 0.25$ and analogously $\dot \gamma_2(t)$ for $t_1 = 0.75$ are piecewise constant with
\begin{align*}
\dot\gamma_1(t) = 
\begin{cases}
2 & \text{if } t \in [0, 0.25[ \\
c_1 & \text{if } t \in [0.25, 0.5[ \\
0 & \text{if } t \in [0.5, 0.75[ \\
c_2 & \text{if } t \in [0.75, 1]
\end{cases} \text{ and } \dot\gamma_2(t) = 
\begin{cases}
c_2 & \text{if } t \in [0, 0.25[ \\
0 & \text{if } t \in [0.25, 0.5[ \\
c_1 & \text{if } t \in [0.5, 0.75[ \\
2 & \text{if } t \in [0.75, 1]
\end{cases},
\end{align*}
where the constant values are given as $c_1 = \frac{2 \cdot 14^2}{14^2 + 9^2}$ and $c_2 = \frac{2 \cdot 9^2}{14^2 + 9^2}$. Here, the form of the derivative of the second optimal warping function $\gamma_2$ of the second curve to the first curve is due to symmetry of this particular problem. Thus, both SRV curves
\begin{align*}
\mathbf{q}(\gamma_1(t)) \sqrt{\dot \gamma_1(t)} =
\begin{cases}
\sqrt{2}(-3,1)^T & \text{if } t \in [0, 0.25[ \\
\sqrt{c_1}(1, -3)^T & \text{if } t \in [0.25, 0.5[ \\
0 & \text{if } t \in [0.5, 0.75[ \\
\sqrt{c_2}(1, -3)^T & \text{if } t \in [0.75, 1]
\end{cases} \text{ and }
\mathbf{q}(\gamma_2(t)) \sqrt{\dot \gamma_2(t)} =
\begin{cases}
\sqrt{c_2}(-3,1)^T & \text{if } t \in [0, 0.25[ \\
0 & \text{if } t \in [0.25, 0.5[ \\
\sqrt{c_1}(-3, 1)^T & \text{if } t \in [0.5, 0.75[ \\
\sqrt{2}(1, -3)^T & \text{if } t \in [0.75, 1]
\end{cases}
\end{align*}
are SRV transformations of optimally aligned curves.
This also means that both $L_2$-means of $\mathbf{p}$ and the SRV transformations $(\mathbf{q} \circ \gamma_{i})\sqrt{\dot{\gamma_i}}$, $i=1,2$ of either optimally aligned $\boldsymbol{\beta}_2$ are SRV transformations of Fr\`echet means of $\boldsymbol{\beta}_1$ and $\boldsymbol{\beta}_2$ (in red and blue in Figure \ref{fig:non_unique}).

To see this, let $\boldsymbol{\bar{\beta}}$ be a curve with SRV tranformation 
\begin{align*}
    \mathbf{\bar{p}} \in \biggl\{ \frac{1}{2}\mathbf{p} + \frac{1}{2} (\mathbf{q} \circ \gamma_i) \sqrt{\dot \gamma_i} \ \biggl| \ i = 1,2 \biggr\}.
\end{align*}

We compute for $i = 1,2$
\begin{align*}
    d([\boldsymbol{\beta}_1], [\boldsymbol{\beta_2}]) &\leq d([\boldsymbol{\beta}_1], [\boldsymbol{\bar{\beta}}]) + d([\boldsymbol{\bar{\beta}}], [\boldsymbol{\beta_2}]) \\
    &= \inf_\gamma \|  \frac{1}{2}\mathbf{p} + \frac{1}{2} (\mathbf{q} \circ \gamma_i) \sqrt{\dot \gamma_i}  - (\mathbf{p} \circ \gamma) \sqrt{\dot \gamma}\|_{L_2} + \inf_\gamma \|  \frac{1}{2}\mathbf{p} + \frac{1}{2} (\mathbf{q} \circ \gamma_i) \sqrt{\dot \gamma_i}  - (\mathbf{q} \circ \gamma_i \circ \gamma) \sqrt{\dot \gamma_i} \sqrt{\dot \gamma} \|_{L_2} \\
    &\overset{\gamma = \text{id}}{\leq} \frac{1}{2} \| (\mathbf{q} \circ \gamma_i) \sqrt{\dot \gamma_i}  - \mathbf{p} \|_{L_2} + \frac{1}{2} \| \mathbf{p} -  (\mathbf{q} \circ \gamma_i) \sqrt{\dot \gamma_i}   \|_{L_2} =  \|\mathbf{p} -  (\mathbf{q} \circ \gamma_i) \sqrt{\dot \gamma_i}   \|_{L_2} \\
    &= d([\boldsymbol{\beta}_1], [\boldsymbol{\beta_2}]),
\end{align*} 
which shows that all inequalities have to be equalities and, therefore, $\gamma$ the identity function. This also implies that $\boldsymbol{\bar{\beta}}$ is optimally aligned to $\boldsymbol{\beta}_1$ and $\boldsymbol{\beta}_2 \circ \gamma_i$, $i = 1,2$ and $d([\boldsymbol{\beta}_1], [\boldsymbol{\bar{\beta}}]) = d([\boldsymbol{\bar{\beta}}], [\boldsymbol{\beta_2}]) = \frac{1}{2} d([\boldsymbol{\beta}_1], [\boldsymbol{\beta_2}])$. 
Hence, for every other curve $\tilde{\boldsymbol{\beta}}$ it holds that
\begin{align*}
d([\tilde{\boldsymbol{\beta}}], [\boldsymbol{\beta_1}])^2 + d([\tilde{\boldsymbol{\beta}}], [\boldsymbol{\beta_2}])^2
&\geq 2 \left( \frac{d([\tilde{\boldsymbol{\beta}}], [\boldsymbol{\beta_1}]) + d([\tilde{\boldsymbol{\beta}}], [\boldsymbol{\beta_2}])}{2} \right)^2 \\
&\geq \frac{1}{2} d([\boldsymbol{\beta_1}], [\boldsymbol{\beta_2}])^2 \\
&= d([\bar{\boldsymbol{\beta}}], [\boldsymbol{\beta_1}])^2 + d([\bar{\boldsymbol{\beta}}], [\boldsymbol{\beta_2}])^2,
\end{align*}
where the first inequality is due to the square being convex and the second due to the triangle inequality. This shows that every $\bar{\boldsymbol{\beta}}$ is a minimiser of the sum of squared distances and therefore a Fr\`echet mean. Hence, both $\frac{1}{2}\mathbf{p} + \frac{1}{2} (\mathbf{q} \circ \gamma_i) \sqrt{\dot \gamma_i}$, $i = 1, 2$ are equivalently valid SRV transformations of Fr\`echet mean curves.

\subsubsection{Proof of Theorem \ref{theo:unique_spline}}
\label{app:unique_spline}
\begin{proof2}
Let $\mathbf{Q} = \left(Q_1, Q_2, \dots, Q_d \right)$.
Without loss of generality we assume $d = 2$. For $d > 2$ perform a coordinate transformation such that $(Q_1, Q_2)$ has a non-linear image between its knots and consider the first two coordinates.

Hence we assume $\mathbf{P} = \mathbf{Q} \circ \gamma$ with $\deg(\mathbf{P}), \deg(\mathbf{Q}) \in \{2,3\}$ and $\mathbf{Q}$ has non-linear image between its knots. First, we show that $\gamma$ is piecewise polynomial, which implies $\gamma$ is piecewise linear since $\deg(\gamma) \geq 2$ would imply $\deg(\mathbf{P}) = \deg(\mathbf{Q} \circ \gamma) \geq 4$.

Let $I \subseteq [0,1]$ be an interval such that $\mathbf{P}|_I$ and $\mathbf{Q}|_{\gamma(I)}$ are polynomials of degree $ \in \{2, 3\}$. That means we can denote
\begin{align*}
\mathbf{P}(t) &=
\left(
\begin{matrix}
P_1(t) \\
P_2(t)
\end{matrix}
\right) =
\left(
\begin{matrix}
p_{10} + p_{11}t + p_{12}t^2 + p_{13}t^3 \\
p_{20} + p_{21}t + p_{22}t^2 + p_{23}t^3
\end{matrix}
\right) \quad \text{ for all } t \in I,\\
\mathbf{Q}(t) &=
\left(
\begin{matrix}
Q_1(t) \\
Q_2(t)
\end{matrix}
\right) =
\left(
\begin{matrix}
q_{10} + q_{11}t + q_{12}t^2 + q_{13}t^3 \\
q_{20} + q_{21}t + q_{22}t^2 + q_{23}t^3
\end{matrix}
\right) \quad \text{ for all } t \in \gamma(I).
\end{align*}
We compute
\begin{align} \label{eq:unique_poly} 
q_{13}P_2(t) - q_{23} P_1(t) &= q_{13}Q_2(\gamma(t)) - q_{23}Q_1(\gamma(t)) \nonumber \\ 
&= q_{13}q_{20} - q_{23}q_{10} + \left(q_{13}q_{21} - q_{23}q_{11} \right)\gamma(t) +
\left(q_{13}q_{22} - q_{23}q_{12} \right)\gamma(t)^2.
\end{align}
Note that either $\left|\begin{matrix}
q_{13} & q_{12} \\
q_{23} & q_{22} 
\end{matrix} \right| = q_{13}q_{22} - q_{23}q_{12} \neq 0$ or $ \left|\begin{matrix}
q_{13} & q_{11} \\
q_{23} & q_{21} 
\end{matrix} \right| =  q_{13}q_{21} - q_{23}q_{11} \neq 0$, because otherwise $\left(\begin{matrix} q_{12} \\ q_{22} \end{matrix} \right)$ and
$\left(\begin{matrix} q_{11} \\ q_{21} \end{matrix} \right)$ are multiples of 
$\left(\begin{matrix} q_{13} \\ q_{23} \end{matrix} \right)$, which means $\mathbf{Q}$ has  a linear image on $\gamma(I)$. Thus we need to consider two cases.
\begin{itemize}
\item[i)]
If $q_{13}q_{22} - q_{23}q_{12} = 0$, this implies $(q_{13}q_{21} - q_{23}q_{11}) \neq 0$ and the claim follows via solving Equation (\ref{eq:unique_poly}) for $\gamma(t)$.

\item[ii)]
If $ c_1 := q_{13}q_{22} - q_{23}q_{12} \neq 0$ there exists a polynomial $\tilde{P}_1$ with $\deg(\tilde{P}_1) \leq 3$ and a constant $c_2 \in \R$ such that $\gamma(t) = \sqrt{\tilde{P}_1} + c_2$ (derive this from Equation (\ref{eq:unique_poly}) by completing the square). Thus we observe that
\begin{align*}
q_{12}P_2(t) - q_{22} P_1(t) &= q_{12}Q_2(\gamma(t)) - q_{22}Q_1(\gamma(t))\\
&= q_{12}(q_{20} + q_{21} \gamma(t) + q_{23} \gamma(t)^3) - q_{22}(q_{10} + q_{11} \gamma(t) + q_{13} \gamma(t)^3)\\ 
&= q_{12}q_{20} - q_{22}q_{10} + \left(q_{12}q_{21} - q_{22}q_{11} \right)(\sqrt{\tilde{P}_1} + c_2) - c_1 \left( \sqrt{\tilde{P}_1} + c_2 \right)^3\\
&= c_3 + c_4 c_2 + c_4\sqrt{\tilde{P}_1} - c_1 \left(\tilde{P}_1  \sqrt{\tilde{P}_1} + 3 c_2 \tilde{P}_1 + 3 c_2^2 \sqrt{\tilde{P}_1} + c_2^3 \right) \\
&= c_3 + c_4 c_2 - c_1(c_2^3 + 3c_2\tilde{P}_1) + \left(c_4 - c_1(3c_2^2 + \tilde{P}_1) \right) \sqrt{\tilde{P}_1}
\end{align*}
with additional constants $c_3 := q_{12}q_{20} - q_{22}q_{10}$ and $c_4 := q_{12}q_{21} - q_{22}q_{11}$. Thus,

\begin{align*}
\left( q_{12}P_2(t) - q_{22} P_1(t) - c_3 - c_4 c_2 + c_1(c_2^3 + 3c_2\tilde{P}_1) \right)^2 = \left(c_4 - c_1(3c_2^2 + \tilde{P}_1) \right)^2 \tilde{P}_1,
\end{align*}
which shows that either $c_4 - c_1(3c_2^2 + \tilde{P}_1) = 0$ which implies $\tilde{P}_1$ is constant (since $c_1 \neq 0$) or every (complex) root of $\tilde{P}_1$ has even multiplicity, which implies that $\sqrt{\tilde{P}_1}$ and therefore $\gamma(t) = \sqrt{\tilde{P}_1} + c_2$ are polynomial.
\end{itemize}
Together this shows that $\gamma$ is polynomial and therefore linear on $I$. Hence $\gamma: [0, 1] \to [0, 1]$ is piecewise linear, that means $\gamma$ is differentiable everywhere but at a finite number of breakpoints $0 = t_0 < t_1 < \dots < t_m = 1$. Thus, the $k$-th derivative of $\mathbf{P}$, $k < \deg(\mathbf{P})$, can be computed as
\begin{align*}
\frac{d^k}{d t^k} \mathbf{P}(t) = \frac{d^k}{d t^k} (\mathbf{Q} \circ \gamma)(t) = \left( \left( \frac{d^k}{d t^k} \mathbf{Q} \right) ( \gamma(t)) \right) \left( \frac{d}{dt} \gamma(t) \right)^k,
\end{align*}
since $\frac{d}{dt} \gamma(t)$ is piecewise constant. Assume $\gamma(t)$ is not differentiable at $t_j, j = 1, \dots, m -1$. Hence the (weak) derivative $\frac{d}{dt} \gamma(t)$ is not continuous at $t_j$ and we need to have
\begin{align*}
\mathbf{Q}^{(k)}(\gamma(t_j)) = \left( \frac{d^k}{d t^k} \mathbf{Q} \right) ( \gamma(t_j)) = 0 \quad \text{ for all } k < \deg(\mathbf{P}),
\end{align*}
since $\mathbf{P}$ is $(\deg(\mathbf{P})-1)$-times continuously differentiable on $[0,1]$. Using a Taylor expansion of $\mathbf{Q}$ around $\gamma(t_j)$, which is identical to $\mathbf{Q}$ on $[\gamma(t_j), \gamma(t_{j+1})]$ since $\mathbf{Q}$ is piecewise polynomial, we obtain:
\begin{align*}
\mathbf{Q}(s) = \left(
\begin{matrix}
\frac{Q_1^{(l)}(\gamma(t_j))}{l!} (s - \gamma(t_j))^l + Q_1(\gamma(t_j)) \\
\frac{Q_2^{(l)}(\gamma(t_j))}{l!} (s - \gamma(t_j))^l + Q_2(\gamma(t_j))
\end{matrix}
\right) = \left(
\begin{matrix}
\frac{Q_1^{(l)}(\gamma(t_j))}{l!} \\ 
\frac{Q_2^{(l)}(\gamma(t_j))}{l!}
\end{matrix}
\right) (s - \gamma(t_j))^l + 
\left(
\begin{matrix}
Q_1(\gamma(t_j)) \\
Q_2(\gamma(t_j))
\end{matrix}
\right)
\end{align*}
for all $s \in [\gamma(t_j), \gamma(t_{j+1})]$. Here we denote $l = \deg(\mathbf{P})$. This would mean that $\mathbf{Q}$ has a linear image between $\gamma(t_j)$ and $\gamma(t_{j+1})$ in this case, which contradicts the assumptions. Hence $\gamma$ needs to be differentiable on $[0, 1]$, which implies $\gamma$ is linear. Since it is monotonically increasing and onto we conclude $\gamma = id$.
\end{proof2}

\subsubsection{Identifiability of constant SRV splines}
\label{app:constant_srv_splines}

\begin{figure}[ht]
\centering
\includegraphics[scale=0.65]{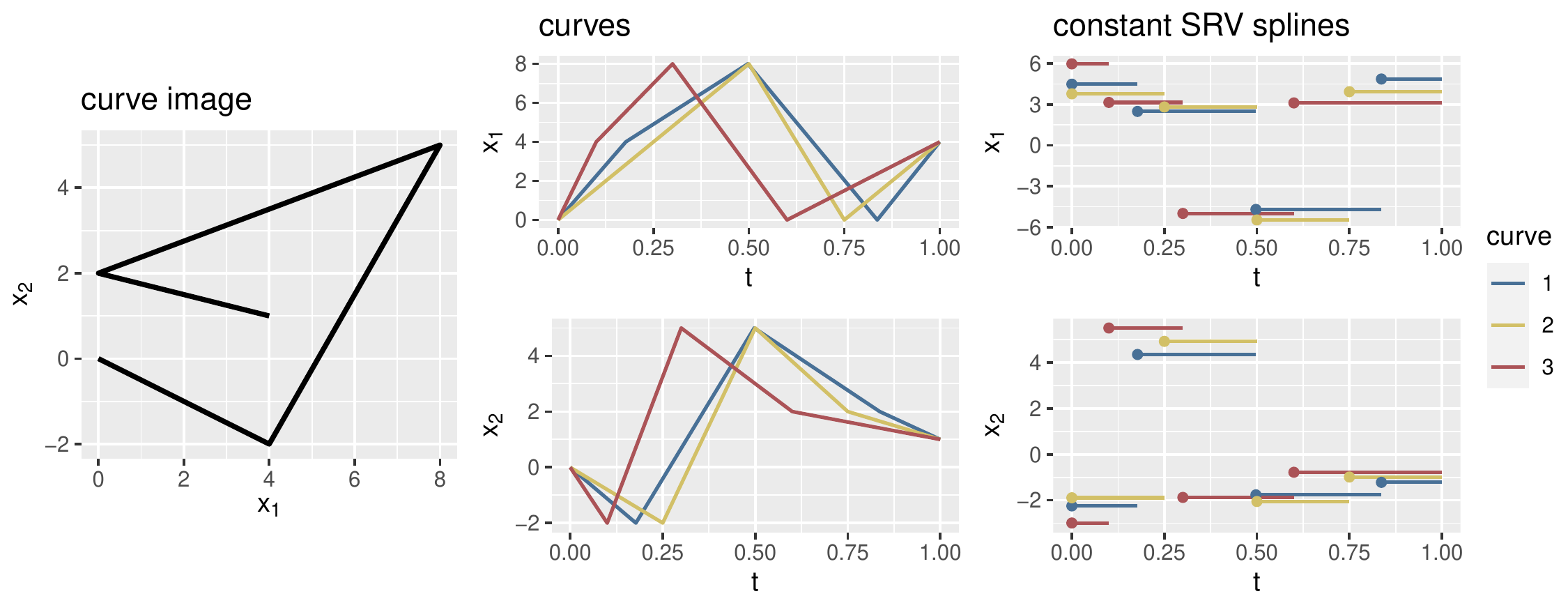}
\caption{\label{fig:constant_srv_spline} Three constant SRV splines (right) with corresponding linear spline curves (middle). All three of them have the same image displayed in black on the left.}
\end{figure}
Piecewise constant SRV curves with varying knots are not identifiable. This means multiple constant SRV splines or equivalently linear spline curves can have the same image, as for example the curves displayed in Figure \ref{fig:constant_srv_spline}. 

Fixing the set of knots determines the velocity between the knots and therefore the SRV transformation. Only if a knot is superfluous, i.e. the assignment of the knots to the corners of the polygonal image is not unique, there is more than one spline curve in each equivalence class for this case.

\subsubsection{Proof of Lemma \ref{lem:top_emb}}
\label{app:top_emb}
\begin{proof2}
The embedding $f$ is injective due to the previous results on identifiability (Theorem \ref{theo:unique_spline}, Corollary \ref{cor:ident_srv_splines}, Remark \ref{rem:constant_splines}) and continuous as the SRV transformation is continuous (Bruveris\cite{bruveris}) and $\inf_{\gamma \in \Gamma} \|\mathbf{p} - (\mathbf{q} \circ \gamma)\sqrt{\dot{\gamma}}\|_{L_2} \leq \|\mathbf{p} - \mathbf{q} \|_{L_2}$ for all $\mathbf{p}, \mathbf{q} \in L_2$.

The only part left to show is that $f^{-1}$ (which exists if we restrict the co-domain of $f$ to its image) is continuous as well. To prove this, let $(\boldsymbol{\xi}_n)_{n \in \mathbb{N}} \subseteq \Xi$ with $\boldsymbol{\beta}_n = f(\boldsymbol{\xi}_n)$ for all $n \in \mathbb{N}$ and $d(\boldsymbol{\beta}_n, \boldsymbol{\beta}) \overset{n \to \infty}{\rightarrow} 0$ for the elastic distance. Hence we have to show $\boldsymbol{\xi}_n \overset{n \to \infty}{\rightarrow} \boldsymbol{\xi}$ for $\boldsymbol{\xi} := f^{-1}(\boldsymbol{\beta})$.

Denote by $\mathbf{p}_n$ the SRV transformation of $\boldsymbol{\beta}_n$ for all $n \in  \mathbb{N}$ and by $\mathbf{q}$ the SRV transformation of $\boldsymbol{\beta}$. Then
\begin{align*}
d(\boldsymbol{\beta}_n, \boldsymbol{\beta}) = \inf_{\gamma \in \Gamma} \|\mathbf{p}_n - (\mathbf{q} \circ \gamma)\sqrt{\dot{\gamma}}\|_{L_2} \geq \inf_{\gamma \in \Gamma} \left( \|\mathbf{p}_n \|_{L_2} - \|(\mathbf{q} \circ \gamma)\sqrt{\dot{\gamma}}\|_{L_2} \right) = 
\|\mathbf{p}_n \|_{L_2} - \|\mathbf{q} \|_{L_2},
\end{align*}
which shows that $\|\dot{\boldsymbol{\beta}}_n \|_{L_2} = \|\mathbf{p}_n \|^2_{L_2}$ is bounded, as $d(\boldsymbol{\beta}_n, \boldsymbol{\beta})$ is bounded as a convergent sequence. Since $\|\dot{\boldsymbol{\beta}}_n \|_{L_2}$ or $\|\mathbf{p}_n \|_{L_2}$ induces a norm on $\Xi$, which is a subset of a finite vector space, $\|\boldsymbol{\xi}_n \|$ is bounded as well, as all norms are equivalent on finite vector spaces.\\
Consider an arbitrary subsequence of $(\boldsymbol{\xi}_n)_{n \in \mathbb{N}}$. Since this subsequence is bounded in $(\Xi, \|\cdot\|)$ as well, it contains a convergent subsequence $(\boldsymbol{\xi}_{n_k})_{k \in \mathbb{N}}$. Let $\boldsymbol{\xi}^* := \lim_{k \to \infty}\boldsymbol{\xi}_{n_k}$. 
Since the embedding $f$ is continuous, we have 
$f(\boldsymbol{\xi}^*) = \lim_{k \to \infty} f(\boldsymbol{\xi}_{n_k} ) = 
\lim_{k \to \infty} \boldsymbol{\beta}_{n_k} = \boldsymbol{\beta} = f(\boldsymbol{\xi})$ and therefore $\boldsymbol{\xi}^* = \boldsymbol{\xi}$ as $f$ is injective. Hence, every subsequence has a subsequence which converges to $\boldsymbol{\xi}$ with respect to $\|\cdot\|$. Thus, $(\boldsymbol{\xi}_n)_{n \in \mathbb{N}}$ converges to $\boldsymbol{\xi}$ in $(\Xi, \|\cdot\|)$ and $f^{-1}$ is hence continuous.
\end{proof2}

%% file: supplem_plots.tex
\subsection{Supplementary plots}
\label{app:plots}
In this part of the appendix we show plots for the simulations in Section \ref{sec:simulation}.

\begin{figure}[!ht]\centering
\includegraphics[scale=0.65]{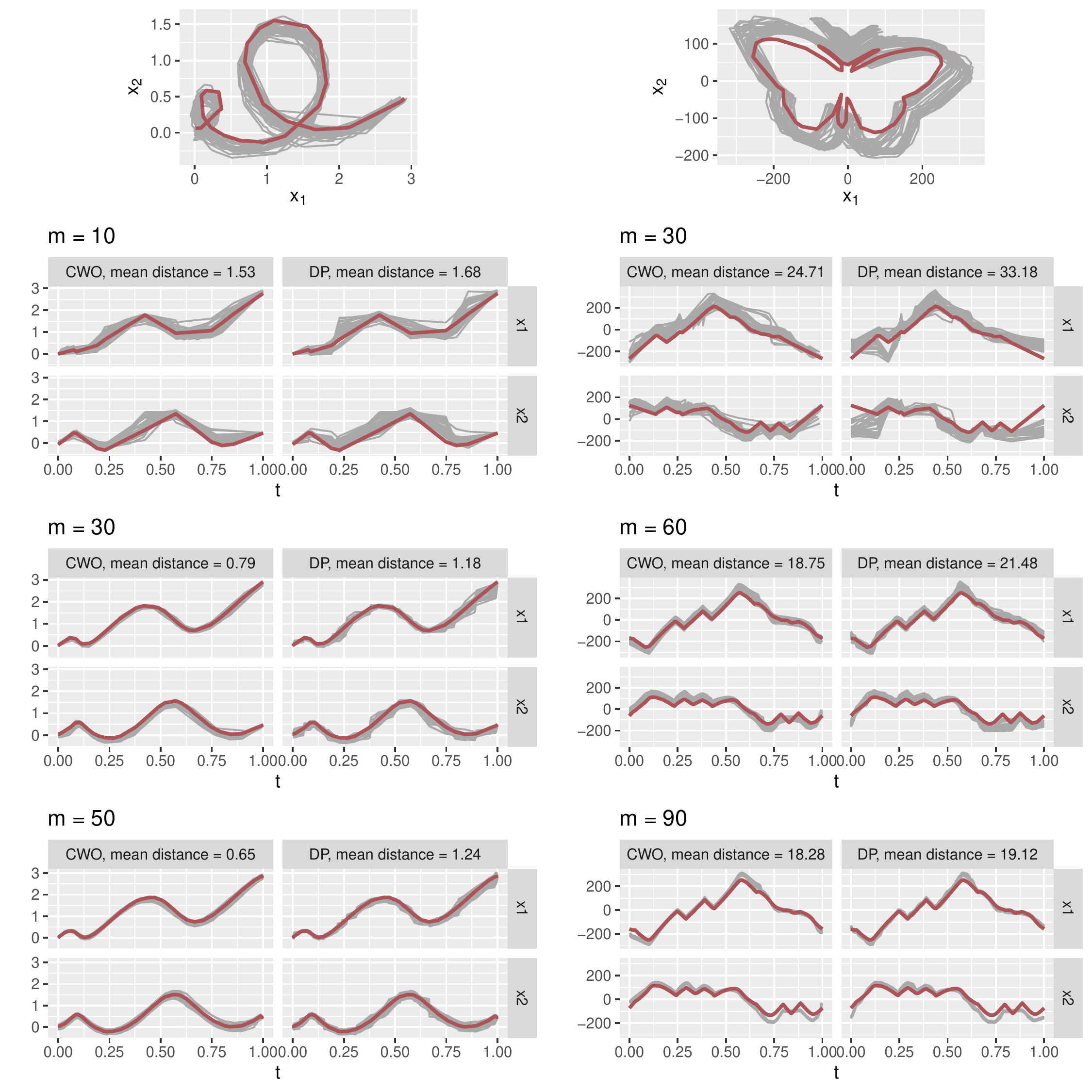}
\vspace*{-0.2cm}
\caption{\label{fig:simulate_dists} Comparison of the optimal alignment produced by our method CWO and the one computed with DP. The 40 grey curves are sampled with $m$ points per curve and aligned to the red curve.The first row shows the sampled curves in the moderately sparse setting ($m = 30$ or $m = 60$ points per curve for the open or closed curve, respectively) The optimal alignments found by both methods are depicted in the lower rows, with the resulting mean elastic distances given in the headings. To make the alignment visually comparable, the aligned curves are evaluated at the observation grid of the red curve for DP. }
\end{figure}

\begin{figure}[!ht]\centering
\includegraphics[scale=0.65]{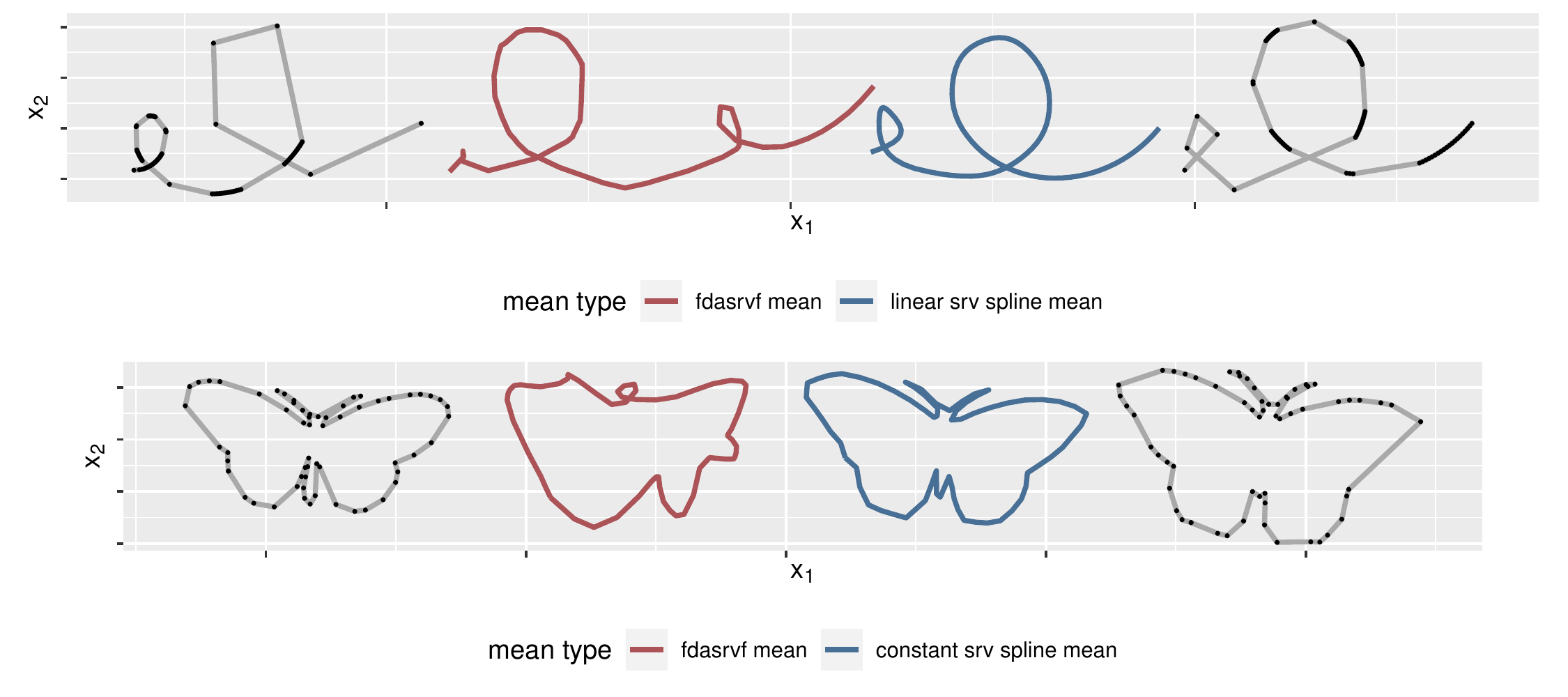}
\vspace*{-0.2cm}
\caption{\label{fig:compare_means} Elastic means for irregularly sampled curves. The observed curves are displayed in grey with black dots at the observed points. The red mean curves are computed with the \texttt{fdasrvf} package, the blue mean curves are computed using our methods and linear splines with 13 equally spaced inner knots or constant splines on SRV level with 68 equally spaced inner knots for the open curves and the closed butterfly shaped curves, respectively.}
\end{figure}

\begin{figure}[!ht]\centering
\includegraphics[scale=0.65]{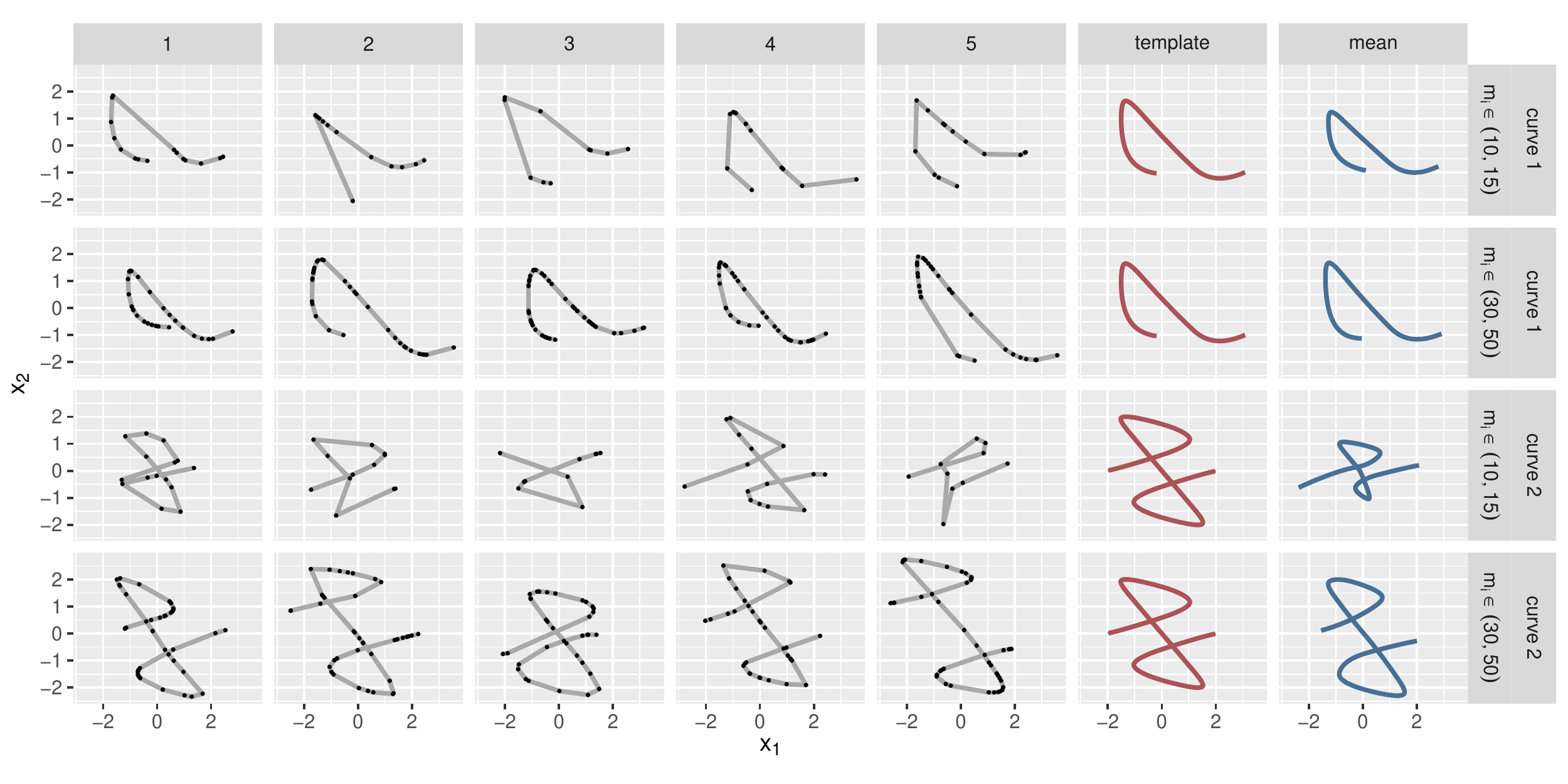}
\vspace*{-0.2cm}
\caption{\label{fig:simu_coefs_open_data} Simulated data in grey with observed values marked as black dots and corresponding smooth elastic means over $n = 5$ observations in blue. The irregularly sampled curves are drawn from two different templates (in red) with varying number $m_i$ of observed points per curve.}
\end{figure}

\begin{figure}[!ht]\centering
\includegraphics[scale=0.65]{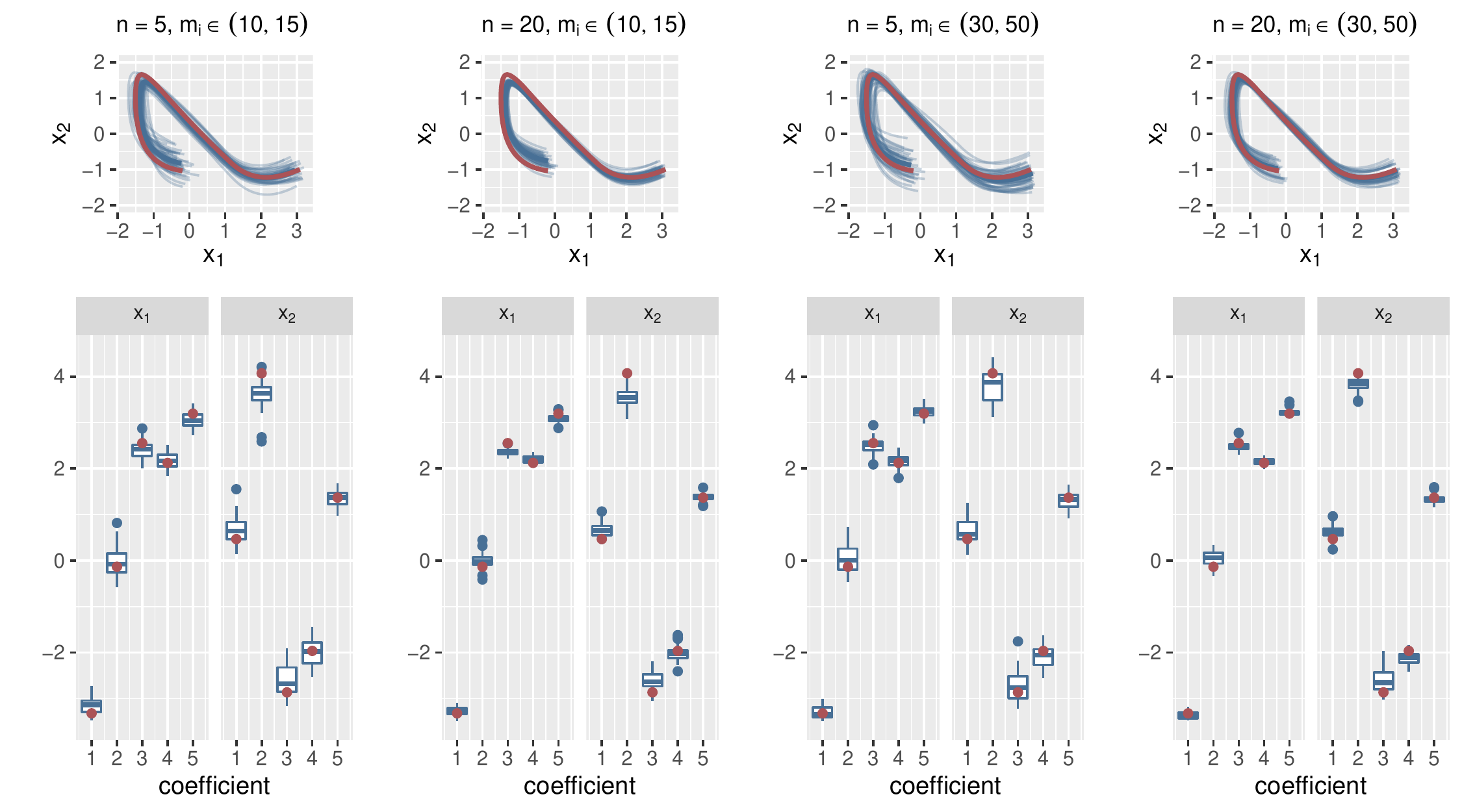}
\vspace*{-0.2cm}
\caption{\label{fig:simu_coefs_curve1}
Top: Smooth means (in blue) computed for a set of $n$ curves drawn from the open template curve (in red) via sampling its B-spline coefficients with $m_i, i = 1, \dots, n$ points observed per curve. The means are computed using linear SRV splines and the same knot set as the template (three equally spaced inner knots)\newline
Bottom: Corresponding distribution of spline mean coefficients (in blue) and template coefficients (in red).}
\end{figure}

\begin{figure}[!ht]\centering
\includegraphics[scale=0.65]{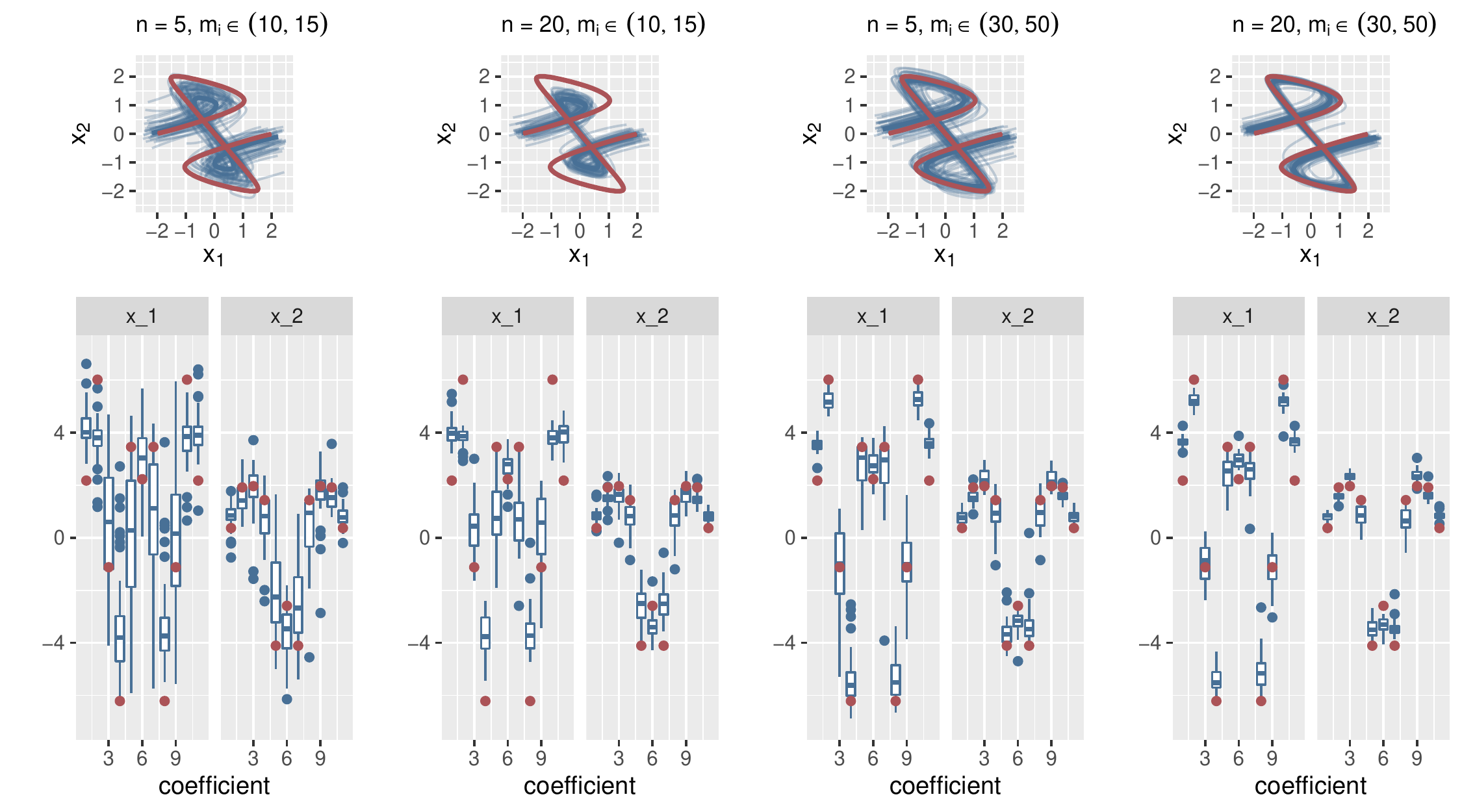}
\vspace*{-0.2cm}
\caption{\label{fig:simu_coefs_curve2}
Top: Smooth means (in blue) computed for a set of $n$ curves drawn from the open template curve (in red) via sampling its B-spline coefficients with $m_i, i = 1, \dots, n$ points observed per curve. The means are computed using linear SRV splines and the same knot set as the template (nine equally spaced inner knots)\newline
Bottom: Corresponding distribution of spline mean coefficients (in blue) and template coefficients (in red).}
\end{figure}

\begin{figure}[!ht]\centering
\includegraphics[scale=0.65]{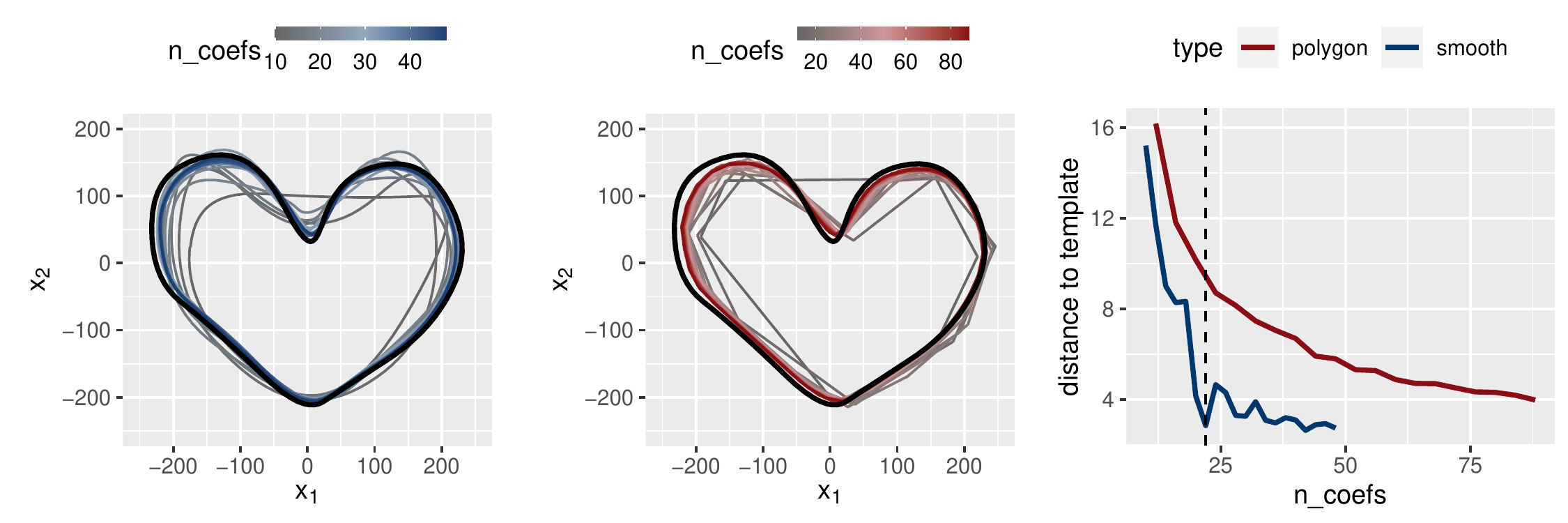}
\vspace*{-0.2cm}
\caption{\label{fig:simu_misspec_coefs} Left: Smooth means based on linear splines on SRV level (depicted in blue)  with varying number of knots and therefore coefficients computed on a sample of 20 curves with $m_i \in \{30, 50\}$ points per curve. The template is displayed in black.\newline
Middle: Polygonal means (depicted in red) with varying number of coefficients computed for the same sample of curves (from the same template in black) as the smooth means on the left.\newline
Right: Elastic Distance of the mean curves to the template curve given the number of coefficients in the mean model. The vertical dashed line indicates the true linear spline model with 22 coefficients.}
\end{figure}

%% file: paper_sparse_curves.bbl
\begin{thebibliography}{10}

\bibitem{alty}
Jane Alty, Jeremy Cosgrove, Deborah Thorpe, and Peter Kempster.
\newblock How to use pen and paper tasks to aid tremor diagnosis in the clinic.
\newblock {\em Practical Neurology}, 17, 08 2017.

\bibitem{bernal}
J.~{Bernal}, G.~{Dogan}, and C.~R. {Hagwood}.
\newblock Fast dynamic programming for elastic registration of curves.
\newblock In {\em 2016 IEEE Conference on Computer Vision and Pattern
  Recognition Workshops (CVPRW)}, pages 1066--1073, 2016.

\bibitem{bruveris}
Martins Bruveris.
\newblock Optimal reparametrizations in the square root velocity framework.
\newblock {\em SIAM Journal on Mathematical Analysis}, 48, 07 2015.

\bibitem{cheng}
Wen Cheng, Ian~L. Dryden, and Xianzheng Huang.
\newblock Bayesian registration of functions and curves.
\newblock {\em Bayesian Anal.}, 11(2):447--475, 06 2016.

\bibitem{dryden}
I.L. Dryden and K.V. Mardia.
\newblock {\em Statistical Shape Analysis: With Applications in R}.
\newblock Wiley Series in Probability and Statistics. Wiley, 2016.

\bibitem{frechet}
Maurice Fr{\'e}chet.
\newblock Les {\'e}l{\'e}ments al{\'e}atoires de nature quelconque dans un
  espace distanci{\'e}.
\newblock In {\em Annales de l'institut Henri Poincar{\'e}}, volume~10, pages
  215--310, 1948.

\bibitem{greven}
Sonja Greven and Fabian Scheipl.
\newblock A general framework for functional regression modelling.
\newblock {\em Statistical Modelling}, 17(1-2):1--35, 2017.

\bibitem{isenkul}
Muhammed Isenkul, Betul Sakar, and Olcay Kursun.
\newblock Improved spiral test using digitized graphics tablet for monitoring
  {Parkinson’s} disease.
\newblock In {\em Proc. of the Int’l Conf. on e-Health and Telemedicine},
  pages 171--5, 2014.

\bibitem{keogh}
Eamonn Keogh and Chotirat Ratanamahatana.
\newblock Exact indexing of dynamic time warping.
\newblock {\em Knowledge and Information Systems}, 7:358--386, 01 2005.

\bibitem{kurt}
{\.I}lke Kurt, Sezer Ulukaya, and O{\u{g}}uzhan Erdem.
\newblock Classification of {Parkinson’s} disease using dynamic time warping.
\newblock In {\em 2019 27th Telecommunications Forum (TELFOR)}, pages 1--4.
  IEEE, 2019.

\bibitem{kurtek}
Sebastian Kurtek, Anuj Srivastava, Eric Klassen, and Zhaohua Ding.
\newblock Statistical modeling of curves using shapes and related features.
\newblock {\em Journal of the American Statistical Association}, 107, 09 2012.

\bibitem{laborde}
Jose Laborde, Daniel Robinson, Anuj Srivastava, Eric Klassen, and Jinfeng
  Zhang.
\newblock Rna global alignment in the joint sequence–structure space using
  elastic shape analysis.
\newblock {\em Nucleic acids research}, 41, 04 2013.

\bibitem{lahiri}
Sayani Lahiri, Daniel Robinson, and Eric Klassen.
\newblock Precise matching of {PL} curves in {$R^N$} in the square root
  velocity framework.
\newblock {\em Geometry, Imaging and Computing}, 2, 01 2015.

\bibitem{lu}
Yi~Lu, Radu Herbei, and Sebastian Kurtek.
\newblock Bayesian registration of functions with a {Gaussian} process prior.
\newblock {\em Journal of Computational and Graphical Statistics},
  26(4):894--904, 2017.

\bibitem{marron}
J.~S. Marron, James~O. Ramsay, Laura~M. Sangalli, and Anuj Srivastava.
\newblock Functional data analysis of amplitude and phase variation.
\newblock {\em Statist. Sci.}, 30(4):468--484, 11 2015.

\bibitem{matuk}
James Matuk, Karthik Bharath, Oksana Chkrebtii, and Sebastian Kurtek.
\newblock Bayesian framework for simultaneous registration and estimation of
  noisy, sparse and fragmented functional data, 2019.

\bibitem{stats}
{R Core Team}.
\newblock {\em R: A Language and Environment for Statistical Computing}.
\newblock R Foundation for Statistical Computing, Vienna, Austria, 2020.

\bibitem{ramsay_silverman}
J.~Ramsay, J.~Ramsay, and B.W. Silverman.
\newblock {\em Functional Data Analysis}.
\newblock Springer Series in Statistics. Springer, 2005.

\bibitem{ramsay_li}
J.~O. Ramsay and Xiaochun Li.
\newblock Curve registration.
\newblock {\em Journal of the Royal Statistical Society: Series B (Statistical
  Methodology)}, 60(2):351--363, 1998.

\bibitem{sakoe}
H.~Sakoe and Seibi Chiba.
\newblock Dynamic programming algorithm optimization for spoken word
  recognition.
\newblock {\em IEEE Transactions on Acoustics, Speech, and Signal Processing},
  26:159--165, 1978.

\bibitem{saunders}
Rachel Saunders-Pullman, Carol Derby, Kaili Stanley, Alicia Floyd, Susan
  Bressman, Richard~B. Lipton, Amanda Deligtisch, Lawrence Severt, Qiping Yu,
  Mónica Kurtis, and Seth~L. Pullman.
\newblock Validity of spiral analysis in early {Parkinson's} disease.
\newblock {\em Movement Disorders}, 23(4):531--537, 2008.

\bibitem{srivastava_book}
A.~Srivastava and E.P. Klassen.
\newblock {\em Functional and Shape Data Analysis}.
\newblock Springer Series in Statistics. Springer New York, 2016.

\bibitem{srivastava_functional_data}
A.~Srivastava, W.~Wu, S.~Kurtek, E.~Klassen, and J.~S. Marron.
\newblock Registration of functional data using {Fisher-Rao} metric.
\newblock {\em arXiv: Statistics Theory}, 2011.

\bibitem{srivastava}
Anuj Srivastava, Eric Klassen, Shantanu~H. Joshi, and Ian~H. Jermyn.
\newblock Shape analysis of elastic curves in euclidean spaces.
\newblock {\em IEEE Trans. Pattern Anal. Mach. Intell.}, 33(7):1415--1428,
  2011.

\bibitem{elasdics}
Lisa Steyer.
\newblock {\em elasdics: Elastic Analysis of Sparse, Dense and Irregular
  Curves}, 2021.
\newblock R package version 0.1.1.

\bibitem{strait}
Justin Strait, Sebastian Kurtek, Emily Bartha, and Steven~N. MacEachern.
\newblock Landmark-constrained elastic shape analysis of planar curves.
\newblock {\em Journal of the American Statistical Association},
  112(518):521--533, 2017.

\bibitem{sun_optimization}
W.~Sun and Y.X. Yuan.
\newblock {\em Optimization Theory and Methods: Nonlinear Programming}.
\newblock Springer Optimization and Its Applications. Springer US, 2006.

\bibitem{fdasrvf}
J.~Derek Tucker.
\newblock {\em fdasrvf: Elastic Functional Data Analysis}, 2020.
\newblock R package version 1.9.4.

\bibitem{yao}
Fang Yao, Hans-Georg Müller, and Jane-Ling Wang.
\newblock Functional data analysis for sparse longitudinal data.
\newblock {\em Journal of the American Statistical Association},
  100(470):577--590, 2005.

\bibitem{ziezold}
Herbert Ziezold.
\newblock {\em On Expected Figures and a Strong Law of Large Numbers for Random
  Elements in Quasi-Metric Spaces}, pages 591--602.
\newblock Springer Netherlands, Dordrecht, 1977.

\end{thebibliography}
